\documentclass[conference]{IEEEtran}
\IEEEoverridecommandlockouts
\usepackage{cite}
\usepackage{amsmath,amssymb,amsfonts}
\usepackage{graphicx}
\usepackage{textcomp}
\usepackage{xcolor}
\def\BibTeX{{\rm B\kern-.05em{\sc i\kern-.025em b}\kern-.08em
    T\kern-.1667em\lower.7ex\hbox{E}\kern-.125emX}}

\usepackage{algorithm}
\usepackage{algpseudocode}
\usepackage{multirow}
\usepackage{xspace}
\usepackage{subcaption}
\usepackage{hyperref}
\usepackage{placeins}

\newcommand{\NumAlgorithmsImplemented}{17\xspace}
\newcommand{\NumAlgorithms}{15\xspace}
\newcommand{\NumDatasets}{16\xspace}
\newcommand{\FrameworkName}{SAGA\xspace}
\newcommand{\AAName}{PISA\xspace}
\newcommand{\AADescription}{\textbf{P}roblem-instance \textbf{I}dentification using \textbf{S}imulated \textbf{A}nnealing\xspace}

\begin{document}

\title{PISA: An Adversarial Approach To Comparing Task Graph Scheduling Algorithms\thanks{This work was supported in part by Army Research Laboratory under Cooperative Agreement W911NF-17-2-0196.}}

\author{\IEEEauthorblockN{1\textsuperscript{st} Jared Coleman}
\IEEEauthorblockA{\textit{Department of Computer Science} \\
\textit{Loyola Marymount University}\\
Los Angeles, USA \\
jared.coleman@lmu.edu}
\and
\IEEEauthorblockN{2\textsuperscript{nd} Bhaskar Krishnamachari}
\IEEEauthorblockA{\textit{Department of Electrical Engineering} \\
\textit{University of Southern California}\\
Los Angeles, USA \\
bkrishna@usc.edu}}

\maketitle

\begin{abstract}
    Scheduling a task graph representing an application over a heterogeneous network of computers is a fundamental problem in distributed computing. It is known to be not only NP-hard but also not polynomial-time approximable within a constant factor. As a result, many heuristic algorithms have been proposed over the past few decades. Yet it remains largely unclear how these algorithms compare to each other in terms of the quality of schedules they produce. We identify gaps in the traditional benchmarking approach to comparing task scheduling algorithms and propose a simulated annealing-based adversarial analysis approach called \AAName to address them. We also introduce \FrameworkName, a new open-source library for comparing task scheduling algorithms. We use \FrameworkName to benchmark \NumAlgorithms algorithms on \NumDatasets datasets and \AAName to compare the algorithms in a pairwise manner. Algorithms that appear to perform similarly on benchmarking datasets are shown to perform very differently on adversarially chosen problem instances. Interestingly, the results indicate that this is true even when the adversarial search is constrained to selecting among well-structured, application-specific problem instances. 
    We present the first known lower bounds on the performance of many of the algorithms considered in this paper compared to other popular scheduling algorithms.
    This work represents an important step towards a more general understanding of the performance boundaries between task scheduling algorithms on different families of problem instances.
\end{abstract}


\section{Introduction}
Task graph scheduling is a fundamental problem in distributed computing.
It plays a crucial role in optimizing performance across various domains, including IoT systems, scientific workflows, and load balancing in distributed networks.
Essentially, the goal is to assign computational tasks to different compute nodes in such a way that minimizes/maximizes some performance metric (e.g., total execution time, energy consumption, throughput, etc.).
In this paper, we will focus on the offline (also called static or compile-time) task scheduling problem concerning \textit{heterogeneous} task graphs and compute networks with the objective of \textit{minimizing makespan} (total execution time) under the \textit{related machines} model\footnote{In the related machines model, if the same task executes faster on some compute node $n_1$ than on node $n_2$, then $n_1$ must execute \textit{all} tasks faster than $n_2$ ($n_1$ is strictly faster than $n_2$).
Observe this model cannot describe multi-modal distributed systems, where certain classes of tasks (e.g., GPU-heavy tasks) might run better/worse on different types of machines (e.g., those with or without GPUs). The related machines model as it pertains to the task scheduling problem we study in this paper is described further in Section~\ref{sec:problem}.}.
As this problem is NP-hard~\cite{theory:npcomplete} and also not polynomial-time approximable within a constant factor~\cite{inapproximable}, many heuristic algorithms have been proposed.
Despite the overwhelming number of algorithms now available in the literature, open-source implementations of them are scarce.
Even worse, the datasets on which they were evaluated are often not available and the implementations that \textit{do} exist are not compatible with each other (different programming languages, frameworks, problem instance formats, etc.).
In order to address this technological shortcoming and enable this study, we built \FrameworkName~\cite{framework_repo}, a Python framework for running, evaluating, and comparing task scheduling algorithms.

\FrameworkName's modularity and extensibility makes it easy to benchmark algorithms on various datasets (\FrameworkName currently includes datasets of randomized problem instances and datasets based on real-world scientific workflow and IoT/edge computing applications).
The underlying theme throughout this paper, however, is that the traditional benchmarking approach to comparing task scheduling algorithms is insufficient.
Benchmarking is only as useful as the underlying datasets are representative and, in practice, peculiarities of heuristic algorithms for task scheduling make it difficult to tell just what broader family of problem instances a dataset is really representative of.
For this reason, benchmarking results for task scheduling algorithms can be misleading.
In this paper, we will show examples and provide methods for the automatic discovery of in-family problem instances --- that is, ones that are similar to other problem instances in the dataset --- on which algorithms that appear to perform well on the original dataset perform very poorly.
In fact, for \textit{every} scheduling algorithm we consider in this paper (\NumAlgorithms total), our proposed simulated-annealing based adversarial instance finder (\AAName) finds problem instances on which it performs at least twice as bad as another of the algorithms.
For most of the algorithms (10 of \NumAlgorithms), it finds a problem instance on which the algorithm performs at least \textit{five times} worse than another algorithm!

The main contributions of this paper are as follows:
\begin{enumerate}
    \item Introduces \FrameworkName, an open-source, modular, and extensible Python package with implementations of many well-known task scheduling algorithms and tools for benchmarking and adversarial analysis.
    \item Reports benchmarking results of \NumAlgorithms well-known scheduling algorithms on \NumDatasets datasets.
    \item Proposes \AAName, a novel simulated annealing-based adversarial analysis method for comparing algorithms that identifies problem instances for which a given algorithm maximally under-performs another.
    \item Identifies gaps in the traditional benchmarking approach: \AAName finds problem instances where algorithms that appear to do well on benchmarking datasets perform poorly on adversarially chosen problem instances. We also show that this is true even when \AAName is restricted to searching over well-structured, application-specific problem instances. 
\end{enumerate}

The rest of the paper is organized as follows.
We formally define the problem in Section~\ref{sec:problem}.
In Section~\ref{sec:related_work}, we provide a brief history of the task scheduling problem and related work.
In Section~\ref{sec:framework}, we introduce \FrameworkName, a Python library for running, evaluating, and comparing task scheduling algorithms.
Then, in Section~\ref{sec:benchmarking}, we present the results of benchmarking \NumAlgorithms algorithms on \NumDatasets datasets.
In Section~\ref{sec:adversarial_analysis}, we present the main contribution of this paper: \AAName, a simulated annealing-based adversarial analysis method for comparing task scheduling algorithms.
In Section~\ref{sec:sa_results} we present the results of comparing the \NumAlgorithms algorithms in a pairwise manner using this method.
We take a closer look at one example comparing HEFT and CPoP (two well-known task scheduling algorithms~\cite{scheduler:heft}) in Section~\ref{sec:case_study}, demonstrating how this framework might be used to discover the reasons why an algorithm performs well on some problem instances and poorly on others.
In Section~\ref{sec:app-specific}, we demonstrate how our proposed technique can be tailored to application-specific scenarios where certain properties of the task graph and/or network are known ahead of time.
In Section~\ref{sec:app-specific}, we show that this approach is not limited to finding only pathological (perhaps unrealistic) problem instances, but can also be used to find adversarial instances that are well-structured and application-specific.
We conclude the paper in Section~\ref{sec:conclusion} with a short discussion on the implications of this work and directions for future research.

\section{Problem Definition}\label{sec:problem}
Let us denote the task graph as $G = (T,D)$, where $T$ is the set of tasks and $D$ contains the directed edges or dependencies between them. 
An edge $(t,t^\prime) \in D$ implies that the output from task $t$ is required input for task $t^\prime$. 
Thus, task $t^\prime$ cannot start executing until it has received the output of task $t$. 
This is often referred to as a precedence constraint. 
The compute cost of given task $t \in T$ is represented by $c(t) \in \mathbb{R}^+$ and the size of the data exchanged between two inter-dependent tasks, $(t, t^\prime) \in D$, is $c(t, t^\prime) \in \mathbb{R}^+$. 
Let $N = (V, E)$, a complete undirected graph, denote the compute node network. 
$V$ is the set of nodes and $E$ is the set of edges. 
The compute speed of a node $v \in V$ is $s(v) \in \mathbb{R}^+$ and the communication strength between nodes $(v, v^\prime) \in E$ is $s(v,v^\prime) \in \mathbb{R}^+$. 
Under the \textit{related machines} model~\cite{graham}, the execution time of a task $t \in T$ on a node $v \in V$ is $ \frac{c(t)}{s(v)}$, and the data communication time between tasks $(t, t^\prime) \in D$ from node $v$ to node $v^\prime$ (i.e., $t$ executes on $v$ and $t^\prime$ executes on $v^\prime$) is $\frac{c(t, t^\prime)}{s(v, v^\prime)}$.

The goal is to schedule the tasks on different compute nodes in such a way that minimizes the makespan (total execution time) of the task graph.
For a given problem instance $(N, G)$ which represents a network/task graph pair, a \textit{schedule} is a set of tuples of the form $(t, v, r)$ where $t \in T$ is the scheduled task, $v \in V$ is the node on which task $t$ is scheduled to run, and $r \in \mathbb{R}^+$ is the start time.
A valid schedule $S$ must satisfy the following properties
\begin{itemize}
    \item All tasks must be scheduled exactly once:
    \begin{gather*}
        \forall t \in T, \exists (t, v, r) \in S \\
        \forall (t, v, r), (t^\prime, v^\prime, r^\prime) \in S,t = t^\prime 
         \Rightarrow (v, r) = (v^\prime, r^\prime)
    \end{gather*}
    \item A task cannot start executing until all of its dependencies have finished executing and their outputs have been received at the node on which the task is scheduled:
    \begin{align*}
        \forall (t, v, r), (t^\prime, v^\prime, r^\prime) &\in S, (t, t^\prime) \in D \\
        &\Rightarrow r + \frac{c(t)}{s(v)} + \frac{c(t, t^\prime)}{s(v, v^\prime)} \leq r^\prime
    \end{align*}
\end{itemize}

\begin{figure*}[h!]
    \begin{subfigure}{0.5\textwidth}
        \centering
        \includegraphics[width=\linewidth]{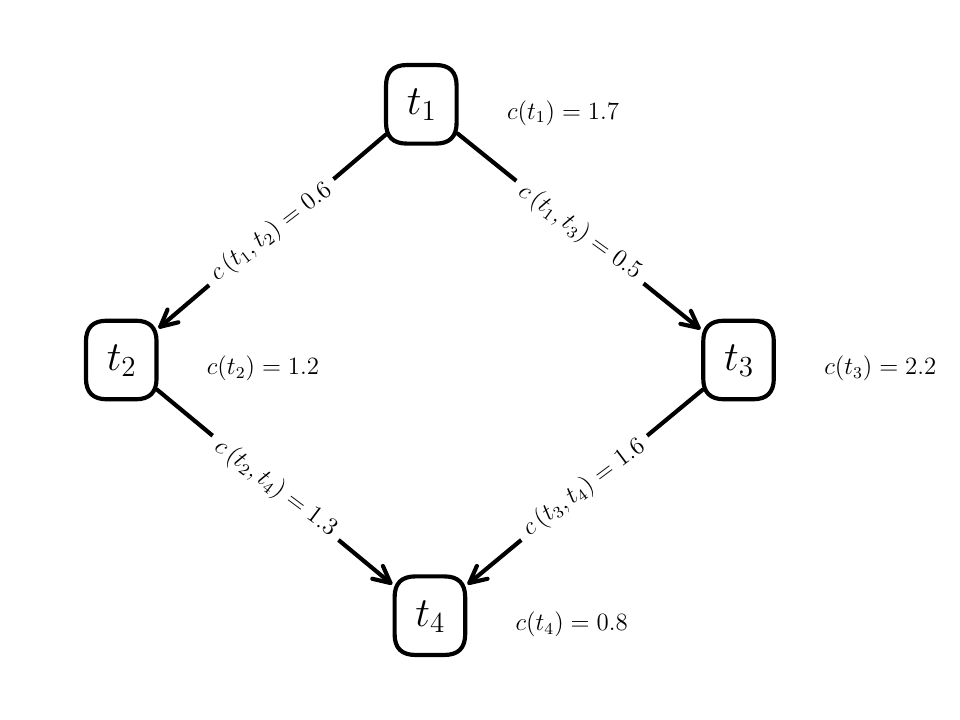}
        \caption{Task Graph}
        \label{fig:example:task_graph}
    \end{subfigure}%
    \begin{subfigure}{0.5\textwidth}
        \centering
        \includegraphics[width=\linewidth]{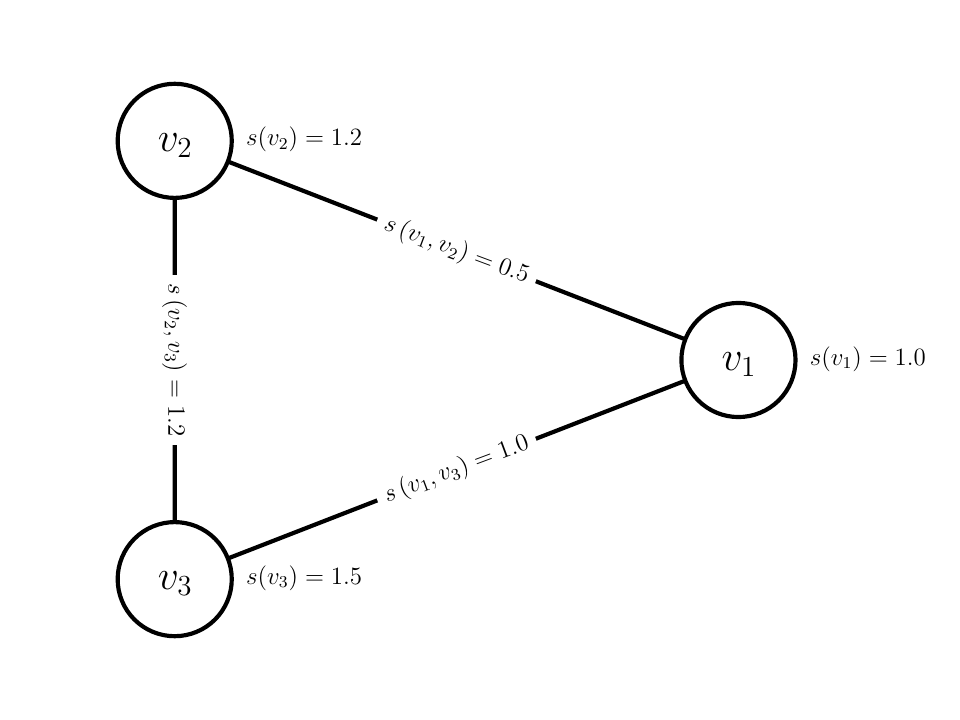}
        \caption{Network}
        \label{fig:example:network}
    \end{subfigure}%
    
    \vspace{0.25cm}%
    
    \begin{subfigure}{\textwidth}
        \centering
        \includegraphics[width=0.8\linewidth]{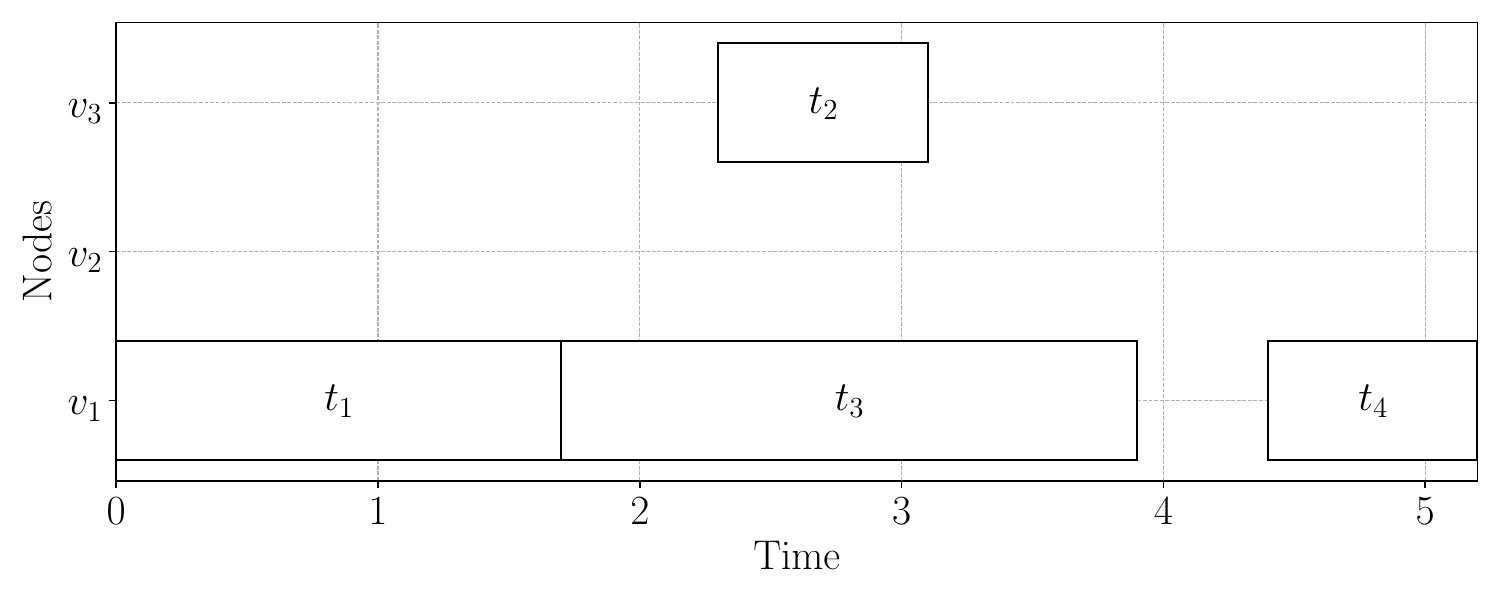}
        \caption{Schedule}
        \label{fig:example:heft}
    \end{subfigure}
    \caption{Example problem instance and schedule.}
    \label{fig:example}
\end{figure*}

Figure~\ref{fig:example} depicts an example problem instance (task graph and network) and solution (schedule) as a Gantt chart.
We define the makespan of a schedule $S$ as the time at which the last task finishes executing:
$$
    m(S) = \max_{(t, v, r) \in S} r + \frac{c(t)}{s(v)}
$$

Because the problem of minimizing makespan is NP-hard for this model~\cite{theory:npcomplete}, many heuristic algorithms have been proposed.
Traditionally, these heuristic algorithms are evaluated on a set of problem instances and compared against other algorithms based on their \textit{makespan ratio}, which for a given problem instance $(N, G)$ is the makespan of the schedule produced by the algorithm divided by the minimum makespan of the schedules produced by the baseline algorithms.
Let $S_{\mathcal{A},N,G}$ denote the schedule produced by algorithm $\mathcal{A}$ on problem instance $(N, G)$.
Then the makespan ratio of an algorithm $\mathcal{A}$ against a set of baseline algorithms $\mathcal{A}_1, \mathcal{A}_2, \ldots$ for a problem instance $(N, G)$ can be written
\begin{align*}
    \frac{m\left(S_{\mathcal{A},N,G}\right)}{\min\left\{
        m\left(S_{\mathcal{A}_1,N,G}\right), m\left(S_{\mathcal{A}_2,N,G}\right), m\left(S_{\mathcal{A}_3,N,G}\right), \ldots
    \right\}} .
\end{align*}

Makespan ratio is a well-known concept in the task scheduling literature (sometimes referred to by other names such as ``schedule length ratio'')~\cite{scheduler:heft,compare:list_vs_cluster}.
It is common to collect the makespan ratios of an algorithm against a set of baseline algorithms for a dataset of problem instances.
System designers use this technique, called \textit{benchmarking}, to decide which algorithm(s) best support(s) their application.
We highlight the shortcomings of this approach by presenting examples (and providing methods for the automatic discovery) of in-family problem instances on which algorithms that appear to perform well on the original dataset perform very poorly.

\section{Related Work}\label{sec:related_work}
A comprehensive survey by Ronald Graham in 1979 classified multiple variants of the task scheduling problem and proposed a structured approach to analyze their complexity classes~\cite{graham_survey}.
Though the variant of task scheduling discussed in this paper wasn't directly addressed in their study, the established reduction framework paved the way for its NP-completeness determination.
The survey addressed many facets of task graph scheduling, such as processor heterogeneity, precedence constraints, and makespan minimization, but did not consider inter-processor communication. 
This gap was addressed a decade later, in 1989, by Hwang et al., who introduced the concept of task graph scheduling with inter-processor communication (albeit only considering homogeneous processors)~\cite{scheduler:etf}.
They proposed the ETF (Earliest Task First) algorithm and proved that it produces schedules with makespan of at most $(2-1/n )\omega_{\text{opt}}^{(i)}+C$ where $\omega_{\text{opt}}^{(i)}$ is the optimal makespan without considering inter-processor communication delays and $C$ is the worst-case communication requirement over a terminal chain of tasks (one which determines the makespan).

Over time, as distributed computing came into the mainstream, many \textit{heuristic} scheduling algorithms emerged.
Many of the most popular fall under the \textit{list scheduling} paradigm, originally proposed by Graham.
In general, list scheduling algorithms involve two steps:
\begin{enumerate}
    \item Compute a priority $p(t)$ for each task $t$ such that tasks have higher priority than their dependent tasks.
    \item Greedily schedule tasks in order of their computed priority (from highest to lowest) to run on the node that minimizes some predefined cost function.
\end{enumerate}
In other words, list scheduling algorithms first decide on a valid topological sort of the task graph, and then use it to schedule tasks greedily according to some objective function. 
ETF, mentioned above, is a list scheduling algorithm.
Two of the most well-known heuristic algorithms for task graph scheduling, HEFT (Heterogeneous Earliest Finish Time) and CPoP (Critical Path on Processor)~\cite{scheduler:heft}, are also list scheduling algorithms.
Many other heuristic algorithms have been proposed and experimentally evaluated on different datasets of task graphs and networks (see benchmarking/comparison/survey papers~\cite{comp:benchmarking_meta,scheduler:eleven,compare:robustness,compare:benchmarking_schedulers,compare:list_vs_cluster}).
Other paradigms studied in the literature are cluster-scheduling~\cite{compare:list_vs_cluster} and meta-heuristic~\cite{scheduler:metaheuristics} algorithms.
Cluster scheduling involves dividing the task graph into groups of tasks to execute on a single node.
Meta-heuristic approaches (e.g., simulated annealing or genetic algorithms) have been shown to work well in some situations, but generally take longer to run and can be difficult to tune~\cite{compare:metaheuristics}.

In this paper, we propose a simulated annealing-based~\cite{simulated_annealing} approach to find problem instances on which an algorithm performs maximally poorly compared to a given baseline algorithm.
This approach is inspired by the recent work of Namyar et al. who use simulated annealing and other techniques to find adversarial instances for heuristics that solve convex optimization problems~\cite{compare:adversarial}.
To the best of our knowledge, this is the first work to propose an automated method for adversarial analysis of task scheduling algorithms.

\section{The \FrameworkName Framework}\label{sec:framework}
We built \FrameworkName to overcome the notable lack of publicly available datasets and task scheduling algorithm implementations.
\FrameworkName is a Python library for running, evaluating, and comparing task scheduling algorithms.
It currently contains implementations of \NumAlgorithmsImplemented algorithms using a common interface.
It also includes interfaces for generating, saving, and loading datasets for benchmarking.
Finally, it includes an implementation of the main contribution of this paper: \AAName, a simulated annealing-based adversarial analysis method for comparing algorithms.
For more information on \FrameworkName, we refer the reader to the open-source implementation~\cite{framework_repo}.
Table~\ref{tab:schedulers} lists the \NumAlgorithmsImplemented algorithms currently implemented in \FrameworkName.
\renewcommand{\arraystretch}{1.5}
\begin{table}[h]
    \centering
    \caption{Schedulers implemented in \FrameworkName}\label{tab:schedulers}
    \begin{tabular}{|r|l|l|}
        \hline
        \textbf{Abbreviation} & \textbf{Algorithm} & \textbf{Reference} \\
        \hline
        BIL & Best Imaginary Level & \cite{scheduler:bil} \\
        \hline
        BruteForce & Brute Force & - \\
        \hline
        CPoP & Critical Path on Processor & \cite{scheduler:heft} \\
        \hline
        Duplex & Duplex & \cite{scheduler:eleven} \\
        \hline
        ETF & Earliest Task First & \cite{scheduler:etf} \\
        \hline
        FastestNode & Fastest Node & - \\
        \hline
        FCP & Fast Critical Path & \cite{scheduler:fcp} \\
        \hline
        FLB & Fast Load Balancing & \cite{scheduler:fcp} \\
        \hline
        GDL & Generalized Dynamic Level & \cite{scheduler:gdl} \\
        \hline
        HEFT & Heterogeneous Earliest Finish Time & \cite{scheduler:heft} \\
        \hline
        MaxMin & MaxMin & \cite{scheduler:eleven} \\
        \hline
        MCT & Minimum Completion Time & \cite{scheduler:mct} \\
        \hline
        MET & Minimum Execution Time & \cite{scheduler:mct} \\
        \hline
        MinMin & MinMin & \cite{scheduler:eleven} \\
        \hline
        OLB & Opportunistic Load Balancing & \cite{scheduler:mct} \\
        \hline
        SMT & SMT-driven Binary Search & - \\
        \hline
        WBA & Workflow-Based application & \cite{scheduler:wba} \\
        \hline
    \end{tabular}
\end{table}
To orient the reader, we provide a brief description of each of these algorithms along with their scheduling complexity, the model they were designed for, performance guarantees (if any), datasets they have been evaluated on, and other algorithms they have been evaluated against.

\subsection{Scheduling Algorithm Descriptions}\label{sec:alg_descriptions}
To orient the reader, we provide a brief description of each of the algorithms implemented in \FrameworkName.

\textbf{BIL} (Best Imaginary Level) is a list scheduling algorithm designed for the unrelated machines model.
In this model, the runtime of a task on a node need not be a function of task cost and node speed.
Rather, it can be arbitrary.
This model is more general even than the \textit{related} machines model we study in this paper.
BIL's scheduling complexity is $O\left(|T|^2 \cdot |V| \log |V| \right)$ and was proven to be optimal for linear graphs.
The authors report a makespan speedup over the GDL (Generalized Dynamic Level) scheduler on randomly generated problem instances.
The exact process for generating random problem instances is not described except to say that CCRs (communication to computation ratios --- average communication time over average execution time) of $1/2$, $1$, and $2$ were used.

\textbf{CPoP} (Critical Path on Processor) was proposed in the same paper as \textbf{HEFT} (Heterogeneous Earliest Finish Time).
Both are list scheduling algorithms with scheduling complexity $O\left( |T|^2 |V| \right)$.
No formal bounds for HEFT and CPoP are known.
HEFT works by first prioritizing tasks based on their \textit{upward rank}, which is essentially the length of the longest chain of tasks (considering average execution and communication times).
Then, it greedily schedules tasks in this order to the node that minimizes the task completion time given previously scheduled tasks.
CPoP is similar but uses a slightly different priority metric.
The biggest difference between the two is that CPoP always schedules \textit{critical path} tasks (those on the longest chain in the task graph) to execute on the fastest node.
Both algorithms were evaluated on random graphs (the process for generating graphs is described in the paper) and on real task graphs for Gaussian Elimination and FFT applications.
They were compared against different scheduling algorithms: Mapping Heuristic (similar to HEFT without insertion)~\cite{scheduler:mapping_heuristic}, Dynamic-Level Scheduling~\cite{scheduler:dynamic_level}, and Levelized Min Time\footnote{We could not find the original paper that proposes this algorithm.}.
Schedule Length Ratios (makespan scaled by the sum of minimum computation costs of all tasks), speedup, and schedule generation times are reported.

The \textbf{MinMin}, \textbf{MaxMin}, and \textbf{Duplex} list scheduling algorithms may have been proposed many times independently, but exceptionally clear definitions can be found in a paper that compares them to OLB, MET, MCT, and a few meta-heuristic algorithms (e.g., genetic algorithms and simulated annealing)~\cite{scheduler:eleven}.
MinMin schedules tasks by iteratively selecting the task with the smallest minimum completion time (given previously scheduled tasks) and assigning it to the corresponding node.
MaxMin, on the other hand, schedules tasks by iteratively selecting the task with the \textit{largest} minimum completion time and assigning it to the the corresponding node.
Duplex simply runs both MinMin and MaxMin and returns the schedule with the smallest makespan.
In the paper, the authors evaluate these algorithms on independent non-communicating tasks with uniformly random costs and heterogeneous compute nodes with uniformly random speeds.
MinMin (and therefore also Duplex) is shown to generate schedules with low makespan compared to the other algorithms while relatively high makespans for MaxMin are reported.

\textbf{ETF} (Earliest Task First) is one of the few algorithms we discuss in this paper that has formal bounds.
It is also a list-scheduling algorithm with runtime $O(|T| |V|^2)$ and was designed for heterogeneous task graphs but homogeneous compute networks.
It works by iteratively scheduling the task with the earliest possible start time  given previously scheduled tasks (usually --- details omitted for the sake of simplicity can be found in the original paper).
Note how this is different from HEFT and CPoP, which schedule according to the earliest possible \textit{completion} time of the task.
This important difference is what allows the authors to prove a formal bound of $\omega_{\text{ETF}} \leq (2-1/n)\omega_{\text{opt}}^{(i)}+C$ where $\omega_{\text{opt}}^{(i)}$ is the optimal schedule makepsan without communication and $C$ is the total communication requirement over some terminal chain of tasks.

\textbf{FCP} (Fast Critical Path) and \textbf{FLB} (Fast Load Balancing) both have a runtime of $O(|T| \log\left(|V|) + |D|\right)$ and were designed for heterogeneous task graphs and heterogeneous node speeds but homogeneous communication strengths. 
The algorithms were evaluated on three types of task graphs with different structures based on real applications (LU decomposition, Laplace equation solver, and a stencil algorithm) with varied CCR and uniformly random task costs.
Both algorithms were shown to perform well compared to HEFT and ERT~\cite{scheduler:ert} despite their lower schedule generation times.

\textbf{GDL} (Generalized Dynamic Level), also called DLS (Dynamic Level Scheduling), is a variation on list scheduling where task priorities are updated each time a task is scheduled.
Due to this added computation in each iteration, the complexity of DLS is $O(|V|^3|T|)$ (a factor $|V|$ greater than HEFT and CPoP).
GDL was originally designed for the very general unrelated machines model and was shown to outperform HDLFET~\cite{scheduler:gdl} on randomly generated problem instances (though the method used for generating random task graphs is not well-described) and on four real digital signal processing applications (two sound synthesis algorithms, a telephone channel simulator, and a quadrature amplitude modulation transmitter).

\textbf{MCT} (Minimum Completion Time) and \textbf{MET} (Minimum Execution Time) are very simple algorithms originally designed for the unrelated machines model.
MET simply schedules tasks to the machine with the smallest execution time (regardless of task start/end time).
MCT assigns tasks in arbitrary order to the node with the smallest completion time given previously scheduled tasks (basically HEFT without insertion or its priority function).
MET and MCT have scheduling complexities of $O(|T||V|)$ and $O(|T|^2|V|)$, respectively.
The algorithms were evaluated on task graphs with 125 or 500 tasks, each task having one of five possible execution times.
They were shown to outperform a very naive baseline algorithm which does not use expected execution times for scheduling.

\textbf{OLB} (Opportunistic Load Balancing) has a runtime of just $O(|T|)$ and was designed for independent tasks under the unrelated machines model.
Probably useful only as a baseline for understanding the performance of other algorithms, OLB schedules tasks in arbitrary order on the earliest available compute node.
Its performance has been shown to be significantly worse than MET, MCT, and LBA~\cite{scheduler:mct}.

\textbf{WBA} (Workflow Based Application) is a scheduling algorithm developed for managing scientific workflows in cloud environments and was designed for the fully heterogeneous model discussed in this paper.
We observe that its scheduling complexity is at most $O(|T||D||V|)$ (the authors do not report this, however, and a more efficient implementation might be possible).
WBA operates by randomly assigning tasks to nodes, guided by a distribution that favors choices that least increase the schedule makespan in each iteration.

\textbf{FastestNode} is a simple baseline algorithm that schedules all tasks to execute in serial on the fastest compute node.
\textbf{BruteForce} is a naive algorithm that tries every possible schedule and returns that with the smallest makespan.
\textbf{SMT} uses an SMT (satisfiability modulo theory) solver and binary search to find a $(1+\epsilon)$-OPT schedule.

Because the BruteForce and SMT scheduling algorithms take much longer to run (exponential time) than the other algorithms, they are not included in the benchmarking or adversarial analysis results reported in this paper.

\subsection{Datasets}\label{sec:datasets}

\FrameworkName also includes a set of tools for generating, saving, and loading datasets.
Table~\ref{tab:datasets} lists the \NumDatasets dataset generators currently included in \FrameworkName.
\renewcommand{\arraystretch}{1.5}
\begin{table}[h]
    \centering
    \caption{Datasets available in \FrameworkName}\label{tab:datasets}
    \begin{tabular}{|r|lr|lr|}
        \hline
        \textbf{Name} & \textbf{Task Graph} & & \textbf{Network} & \\
        \hline
        in\_trees & in-trees & & \multirow{3}{*}{\shortstack[l]{randomly\\ weighted}} & \multirow{3}{*}{} \\
        \cline{1-3}
        out\_trees & out-trees & & & \\
        \cline{1-3}
        chains & parallel chains & & & \\
        \hline
        blast & Blast workflows & \cite{data:makeflow} & \multirow{9}{*}{\shortstack[l]{Chameleon \\ cloud \\ inspired}} & \multirow{9}{*}{\cite{data:chameleon}} \\
        \cline{1-3}
        bwa & BWA workflows & \cite{data:makeflow} & & \\
        \cline{1-3}
        cycles & Cycles workflows & \cite{data:cycles} & & \\
        \cline{1-3}
        epigenomics & Epigenomics workflows & \cite{data:epigenomics} & & \\
        \cline{1-3}
        genome & 1000genome workflows & \cite{data:1000genome} & & \\
        \cline{1-3}
        montage & Montage workflows & \cite{data:montage} & & \\
        \cline{1-3}
        seismology & Seismology workflows & \cite{data:seismology} & & \\
        \cline{1-3}
        soykb & SoyKB & \cite{data:soykb} & & \\
        \cline{1-3}
        srasearch & SRASearch workflows & \cite{data:srasearch} & & \\
        \hline
        etl & IoT ETL application & \cite{data:riotbench} & \multirow{4}{*}{\shortstack[l]{Edge, Fog, \\ \& Cloud \\ Networks}} & \multirow{4}{*}{\cite{data:fog_network}} \\
        \cline{1-3}
        predict & IoT PREDICT application & \cite{data:riotbench} & & \\
        \cline{1-3}
        stats & IoT STATS application & \cite{data:riotbench} & & \\
        \cline{1-3}
        train & IoT TRAIN application & \cite{data:riotbench} & & \\
        \hline
    \end{tabular}
\end{table}
The in\_trees, out\_trees, and chains datasets each consist of $1000$ randomly generated network/task graph pairs following a common methodology used in the literature~\cite{data:random_graphs}. In-trees and out-trees are generated with between $2$ and $4$ levels (chosen uniformly at random), a branching factor of either $2$ or $3$ (chosen uniformly at random), and node/edge-weights drawn from a clipped gaussian distribution (mean: $1$, standard deviation: $1/3$, min: $0$, max: $2$).
Parallel chains task graphs are generated with between $2$ and $5$ parallel chains (chosen uniformly at random) of length between $2$ and $5$ (chosen uniformly at random) and node/edge-weights drawn from the same clipped gaussian distribution.
Randomly weighted networks are complete graphs with between $3$ and $5$ nodes (chosen uniformly at random) and node/edge-weights drawn from the same clipped gaussian distribution.

The scientific workflow datasets blast, bwa, cycles, epigenomics, genome, montage, seismology, soykb, and srasearch each contain $100$ problem instances.
The task graphs are synthetically generated using the WfCommons Synthetic Workflow Generator~\cite{data:wfchef} and are based on real-world scientific workflows.
The Chameleon cloud inspired networks are generated by fitting a distribution to the machine speed data from the execution traces (detailed information from a real execution of the application including task start/end times, cpu usages/requirements, data I/O sizes, etc.) of real workflows on Chameleon that are available in WfCommons and then sampling from that distribution to generate random networks.
Because Chameleon uses a shared filesystem for data transfer, the communication cost can be absorbed into the computation cost and thus the communication strength between nodes is considered to be infinite.

The IoT/edge-inspired etl, predict, stats, and train datasets each contain $1000$ problem instances.
The task graphs and networks are generated using the approach described in~\cite{data:fog_network}.
The task graph structure is based on real-world IoT data streaming applications and the node weights are generated using a clipped gaussian distribution (mean: $35$, standard deviation: $25/3$, min: $10$, max: $60$).
The input size of the application is generated using a clipped gaussian distribution (mean: $1000$, standard deviation: $500/3$, min: $500$, max: $1500$) and the edge weights are determined by the known input/output ratios of the tasks.
The Edge/Fog/Cloud Networks are generated by constructing a complete graph with three types of nodes: edge nodes with CPU speed $1$, fog nodes with CPU speed $6$, and cloud nodes with CPU speed $50$.
The communication strength between edge and fog nodes is $60$ and between fog and cloud/fog nodes is $100$.
The number of edge, fog, and cloud nodes is between $75$ and $125$, $3$ and $7$, and $1$ and $10$, respectively (chosen uniformly at random).
In order to generate a complete graph (a requirement for many scheduling algorithms), edge-cloud and cloud-cloud communication strengths are set to $60$ and infinity (i.e., no communication delay), respectively.

Many other schedulers and datasets have been proposed and used in the literature and can be easily integrated into \FrameworkName in the future.
\FrameworkName is open-source~\cite{framework_repo} and designed to be modular and extensible.
We encourage the community to contribute new algorithms, datasets, and tools.

\section{Benchmarking Results}\label{sec:benchmarking}
\begin{figure*}[!ht]
    \centering
    \includegraphics[width=0.8\textwidth]{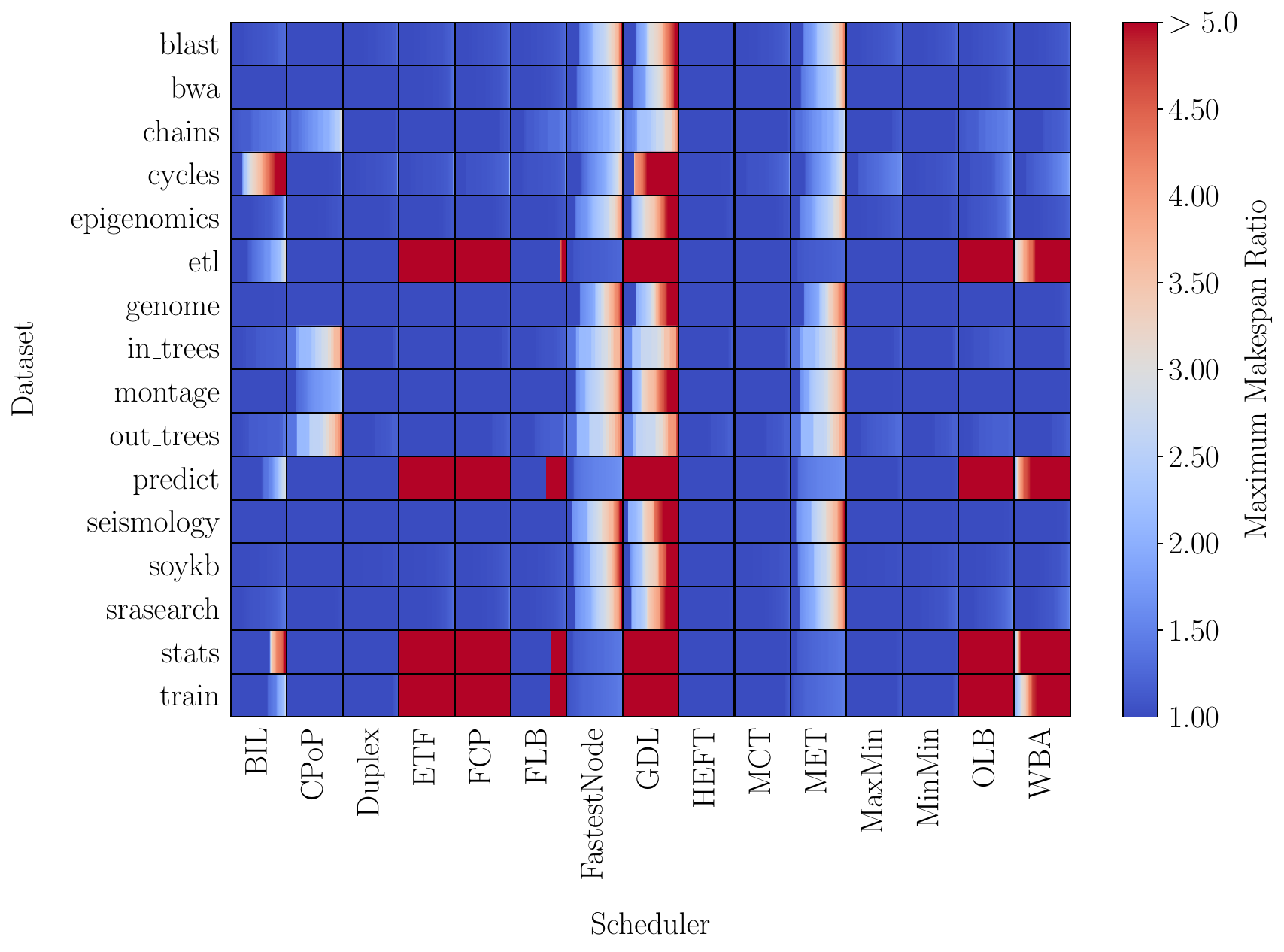}
    \caption{Makespan Ratios of \NumAlgorithms algorithms evaluated on \NumDatasets datasets. Gradients depict performance on different problem instances in each dataset.}
    \label{fig:benchmarking}
\end{figure*}

Figure~\ref{fig:benchmarking} shows the results of benchmarking \NumAlgorithms algorithms on \NumDatasets datasets.
Different scheduling algorithms perform better or worse depending on the dataset and algorithms that weren't designed for fully heterogeneous task graphs and networks (e.g., ETF, FastestNode) tend to perform poorly.
Many of the algorithms, though, perform similarly across the datasets.
While these experiments provide valuable information about the performance of each algorithm on each dataset, they provide much less information about the algorithms themselves.

\begin{figure*}[!htb]
    \centering
    
    \begin{subfigure}[b]{0.32\textwidth}
        \centering
        \includegraphics[trim={15pt 15pt 15pt 15pt},clip,width=\textwidth]{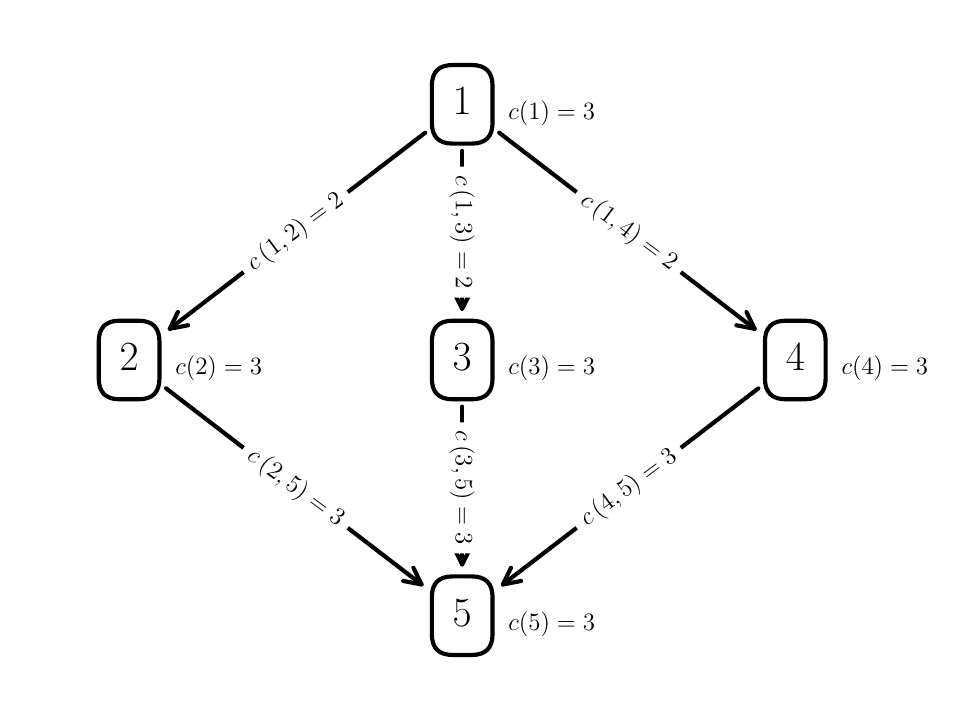}
        \caption{Task Graph}\label{fig:comp:task_graph}
    \end{subfigure}
    \hfill
    \begin{subfigure}[b]{0.32\textwidth}
        \centering
        \includegraphics[trim={15pt 15pt 15pt 15pt},clip,width=\textwidth]{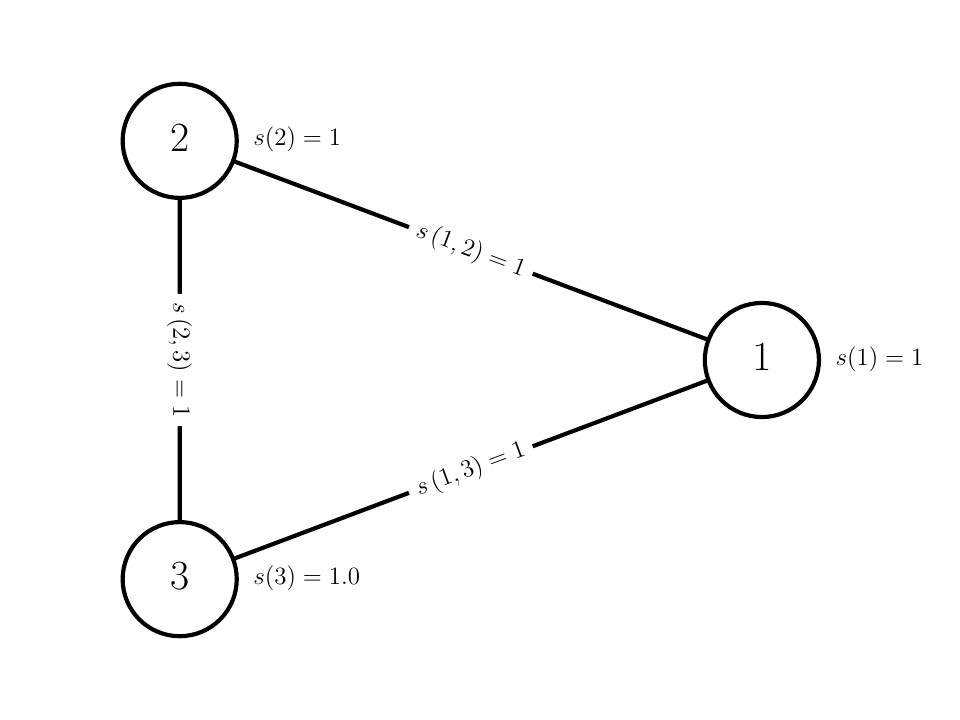}
        \caption{Original Network}\label{fig:comp:network}
    \end{subfigure}
    \hfill
    \begin{subfigure}[b]{0.32\textwidth}
        \centering
        \includegraphics[trim={15pt 15pt 15pt 15pt},clip,width=\textwidth]{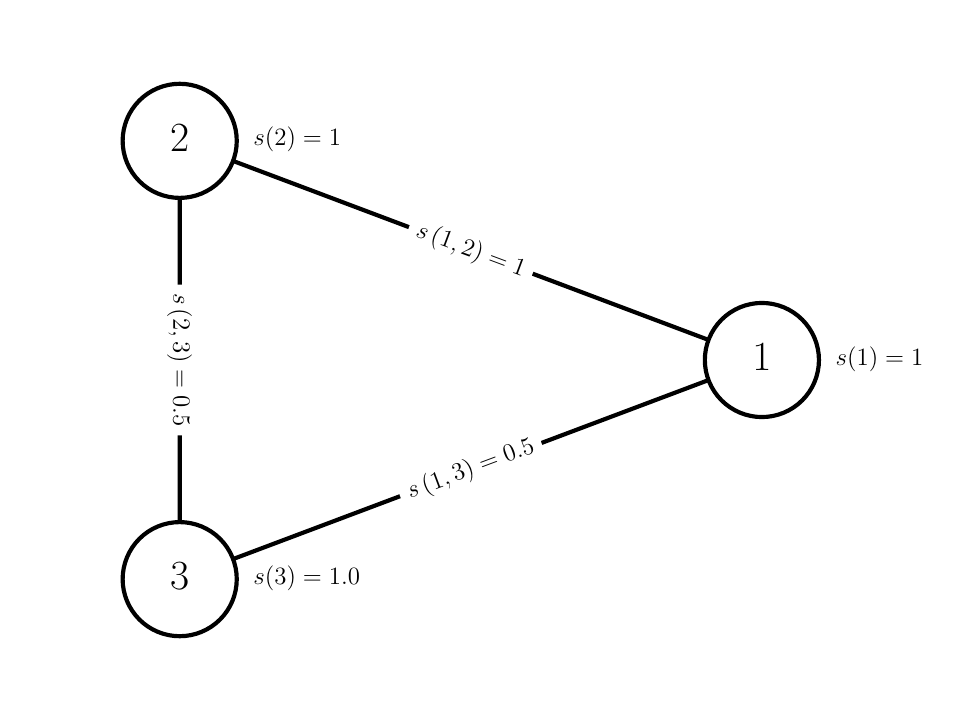}
        \caption{Modified Network}\label{fig:comp:modified_network}
    \end{subfigure}

    \vspace{0.25cm}

    \begin{subfigure}[b]{0.47\textwidth}
        \centering
        \includegraphics[width=\textwidth]{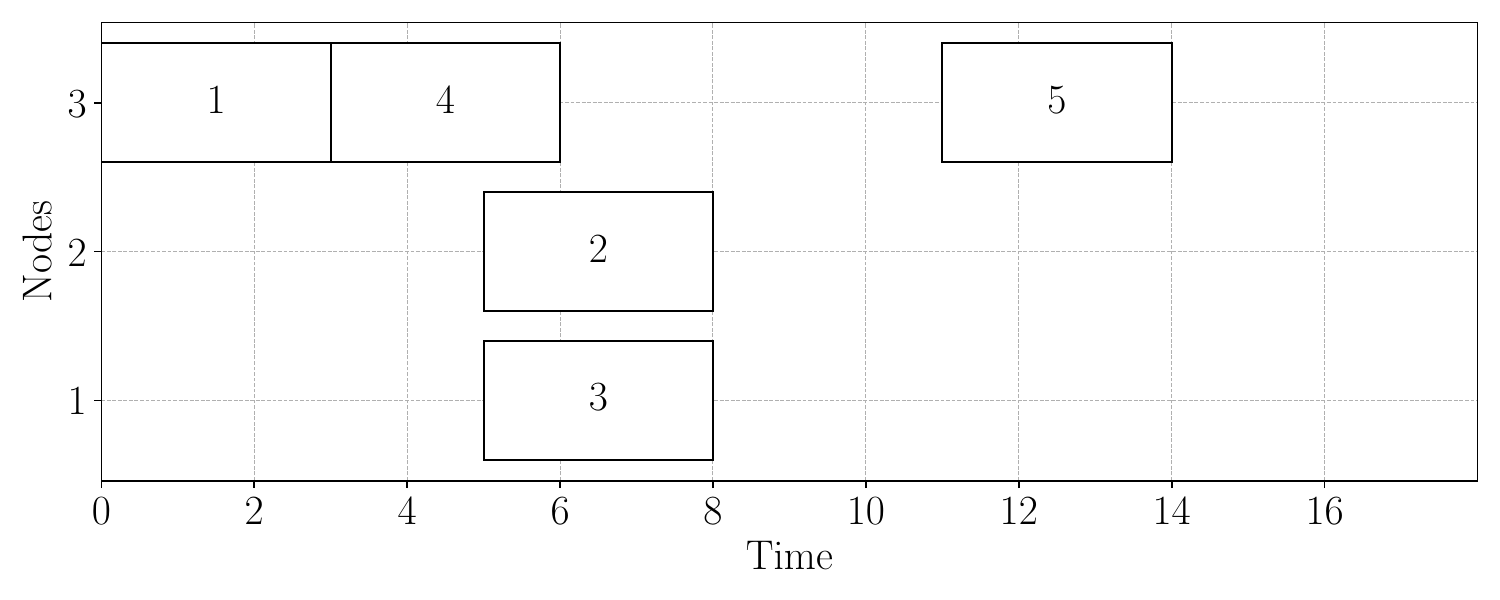}
        \caption{HEFT Schedule on Original Network}\label{fig:comp:heft_schedule}
    \end{subfigure}
    \hfill
    \begin{subfigure}[b]{0.47\textwidth}
        \centering
        \includegraphics[width=\textwidth]{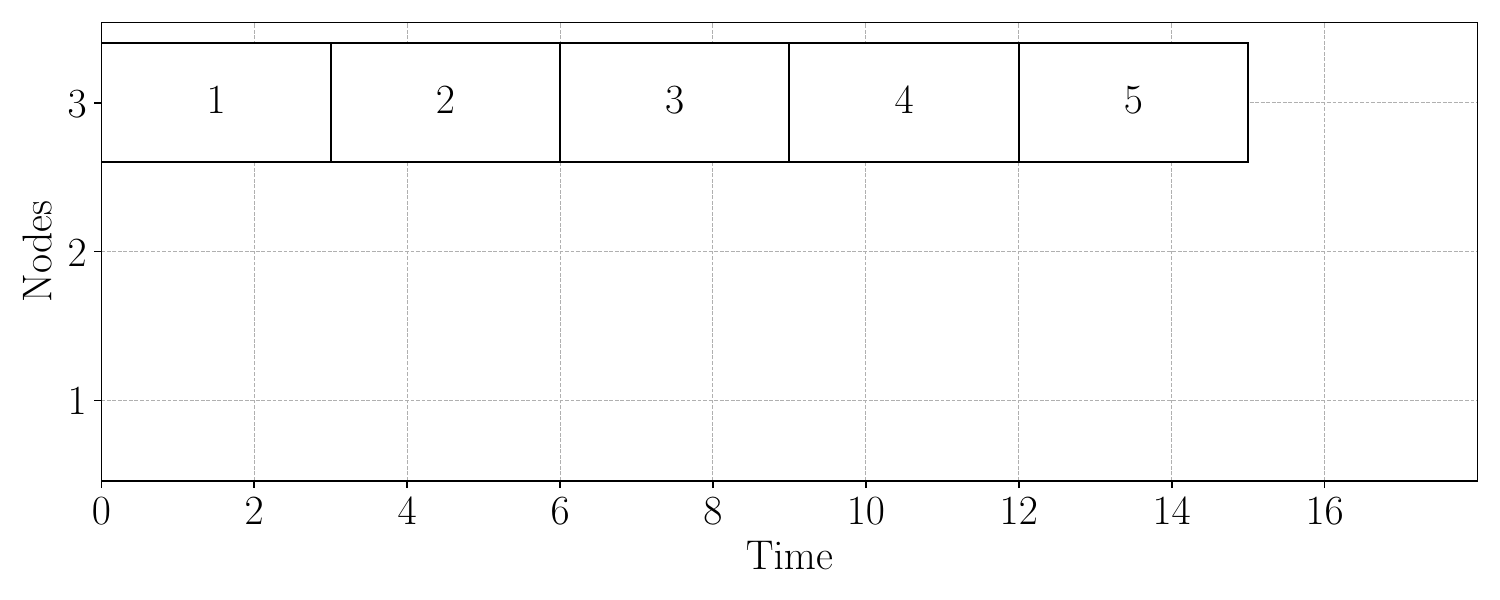}
        \caption{CPoP Schedule on Original Network}\label{fig:comp:cpop_schedule}
    \end{subfigure}

    \vspace{0.25cm}

    \begin{subfigure}[b]{0.47\textwidth}
        \centering
        \includegraphics[width=\textwidth]{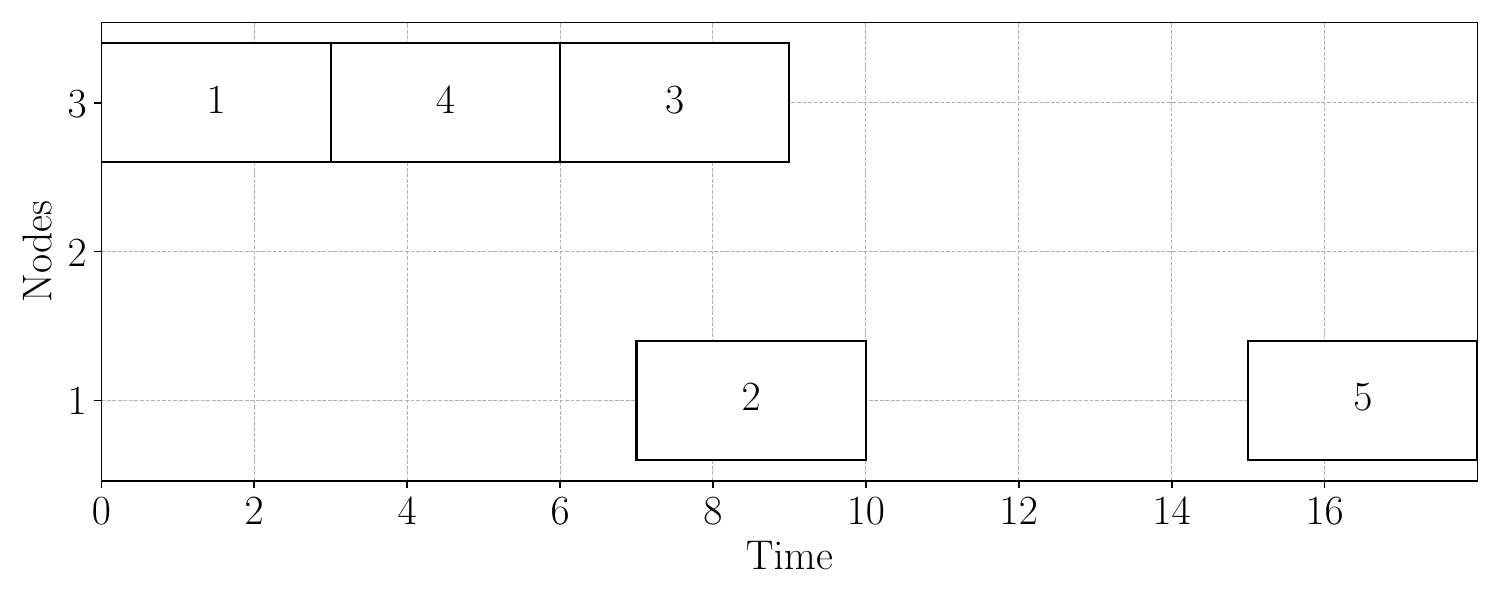}
        \caption{HEFT Schedule on Modified Network}\label{fig:comp:heft_schedule_modified_network}
    \end{subfigure}
    \hfill
    \begin{subfigure}[b]{0.47\textwidth}
        \centering
        \includegraphics[width=\textwidth]{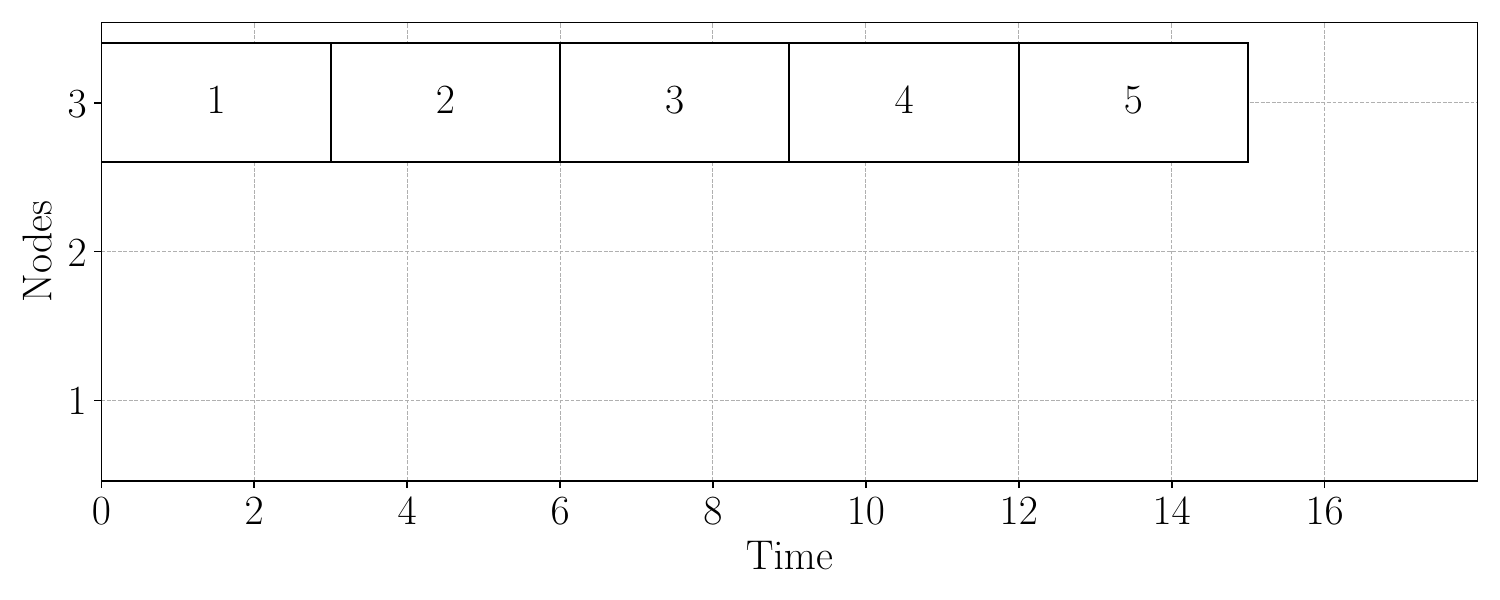}
        \caption{CPoP Schedule on Modified Network}\label{fig:comp:cpop_schedule_modified_network}
    \end{subfigure}
    
    \caption{Comparison of Scheduling Algorithms on Slightly Modified Networks}
    \label{fig:comp}
\end{figure*}

Consider the illustrative scenario in Figure~\ref{fig:comp}.

A simplistic task graph, as depicted in Figure~\ref{fig:comp:task_graph}, coupled with a minor alteration to the initial network (Figures~\ref{fig:comp:network} and~\ref{fig:comp:modified_network}) — a reduction in the strength of node $3$'s communication links — causes HEFT to perform worse than CPoP (Figures~\ref{fig:comp:heft_schedule}, \ref{fig:comp:cpop_schedule}, \ref{fig:comp:heft_schedule_modified_network}, and~\ref{fig:comp:cpop_schedule_modified_network}).
This example underscores the shortcoming of traditional benchmarking: it provides little insight into the conditions under which an algorithm performs well or poorly.
Observe that a structurally equivalent instance of this problem with all node/edge weights scaled so they are between $0$ and $1$ could have been generated by the Parallel Chains dataset generator in \FrameworkName.
So while the benchmarking results in Figure~\ref{fig:benchmarking} indicate that HEFT performs slightly better than CPoP on this dataset, there \textit{are}, in fact, Parallel Chains instances where CPoP performs \textit{significantly} better than HEFT.

\section{Adversarial Analysis}\label{sec:adversarial_analysis}
We propose a novel adversarial analysis method for comparing task scheduling algorithms called \AAName (\AADescription).
The goal of this method is to identify problem instances on which one algorithm maximally outperforms another.
More formally, the goal is to find a problem instance $(N,G)$ that maximizes the makespan ratio of algorithm $\mathcal{A}$ against algorithm $\mathcal{B}$:
$$
    \max_{N,G} \frac{M_{\mathcal{A}(N,G)}}{M_{\mathcal{B}(N,G)}}
$$
In doing so, we hope to fill in some of the gaps we know exist in the benchmarking results.
We propose a simulated annealing-based approach for finding such problem instances.
Simulated annealing is a meta-heuristic that is often used to find the global optimum of a function that has many local optima.
In the context of our problem, the optimizer starts with an initial problem instance $(N,G)$ and then randomly perturbs the problem instance by changing the network or task graph in some way.
If the perturbed problem instance has a higher makespan ratio than the current problem instance, the optimizer accepts the perturbation and continues the search from the new problem instance.
If the perturbed problem instance has a lower makespan ratio than the current problem instance, the optimizer \textit{still} accepts the perturbation with some probability.
This allows the optimizer to escape local optima and (potentially) find a global optimum.
Over time, the probability of accepting perturbations that decrease the makespan ratio decreases, allowing the optimizer to settle into a high-makespan ratio state before terminating.
The pseudocode for our proposed simulated annealing process is presented in Algorithm~\ref{alg:sa}.
\begin{algorithm}
    \caption{\AAName}\label{alg:sa}
    \begin{algorithmic}[1]
        \State Initialize solution $(N, G)$ and best solution $(N_{best}, G_{best})$
        \State Set initial temperature $T = T_{max}$
        \While{$T > T_{min}$ and iteration $< I_{max}$}
            \State Generate new candidate $(N', G')$ by perturbing $(N, G)$
            \State Evaluate new solution's quality $M'$
            \If{$M' > M_{best}$}
                \State Update best solution $(N_{best}, G_{best})$
            \Else
                \State Accept with probability $\exp\left(-\frac{M'/M_{best}}{T}\right)$
            \EndIf
            \State Reduce temperature: $T = T \cdot \alpha$
        \EndWhile
        \State \textbf{return} best solution $(N_{best}, G_{best})$
    \end{algorithmic}
\end{algorithm}%
In our implementation, we \textit{perturb} problem instances by randomly selecting (with equal probability) one of the following perturbations:
\begin{enumerate}
    \item \textbf{Change Network Node Weight:} Select a node $v \in V$ uniformly at random and change its weight by a uniformly random amount between $-1/10$ and $1/10$ with a minimum weight of $0$ and a maximum weight of $1$.
    \item \textbf{Change Network Edge Weight:} Same as \textbf{Change Network Node Weight}, but for (non-self) edges.
    \item \textbf{Change Task Weight:} Same as \textbf{Change Network Node Weight}, but for tasks.
    \item \textbf{Change Dependency Weight:} Same as \textbf{Change Network Edge Weight}, but for dependencies.
    \item \textbf{Add Dependency:} Select a task $t \in T$ uniformly at random and add a dependency from $t$ to a uniformly random task $t^\prime \in T$ such that $(t, t^\prime) \notin D$ and doing so does not create a cycle.
    \item \textbf{Remove Dependency:} Select a dependency $(t, t^\prime) \in D$ uniformly at random and remove it.
\end{enumerate}
Some of the algorithms we evaluate on were only designed for homogeneous compute nodes and/or communication links.
In these cases, we restrict the perturbations to only change the aspects of the network that are relevant to the algorithm.
For ETF, FCP, and FLB, we set all node weights to be $1$ initially and do not allow them to be changed.
For BIL, GDL, FCP, and FLB we set all communication link weights to be $1$ initially and do not allow them to be changed.

For every pair of schedulers, we run the simulated annealing algorithm $5$ times with different randomly generated initial problem instances.
The initial problem instance $(N,G)$ is such that $N$ is a complete graph with between $3$ and $5$ nodes (chosen uniformly at random) and node/edge-weights between $0$ and $1$ (generated uniformly at random, self-edges have weight $\infty$) and $G$ is a simple chain task graph with between $3$ and $5$ tasks (chosen uniformly at random) and task/dependency-weights between $0$ and $1$ (generated uniformly at random).
In our implementation, we set $T_{max} = 10$, $T_{min} = 0.1$, $I_{max} = 1000$, and $\alpha = 0.99$.

\subsection{Results}\label{sec:sa_results}
\begin{figure*}[!ht]
    \centering
    \includegraphics[width=0.8\textwidth]{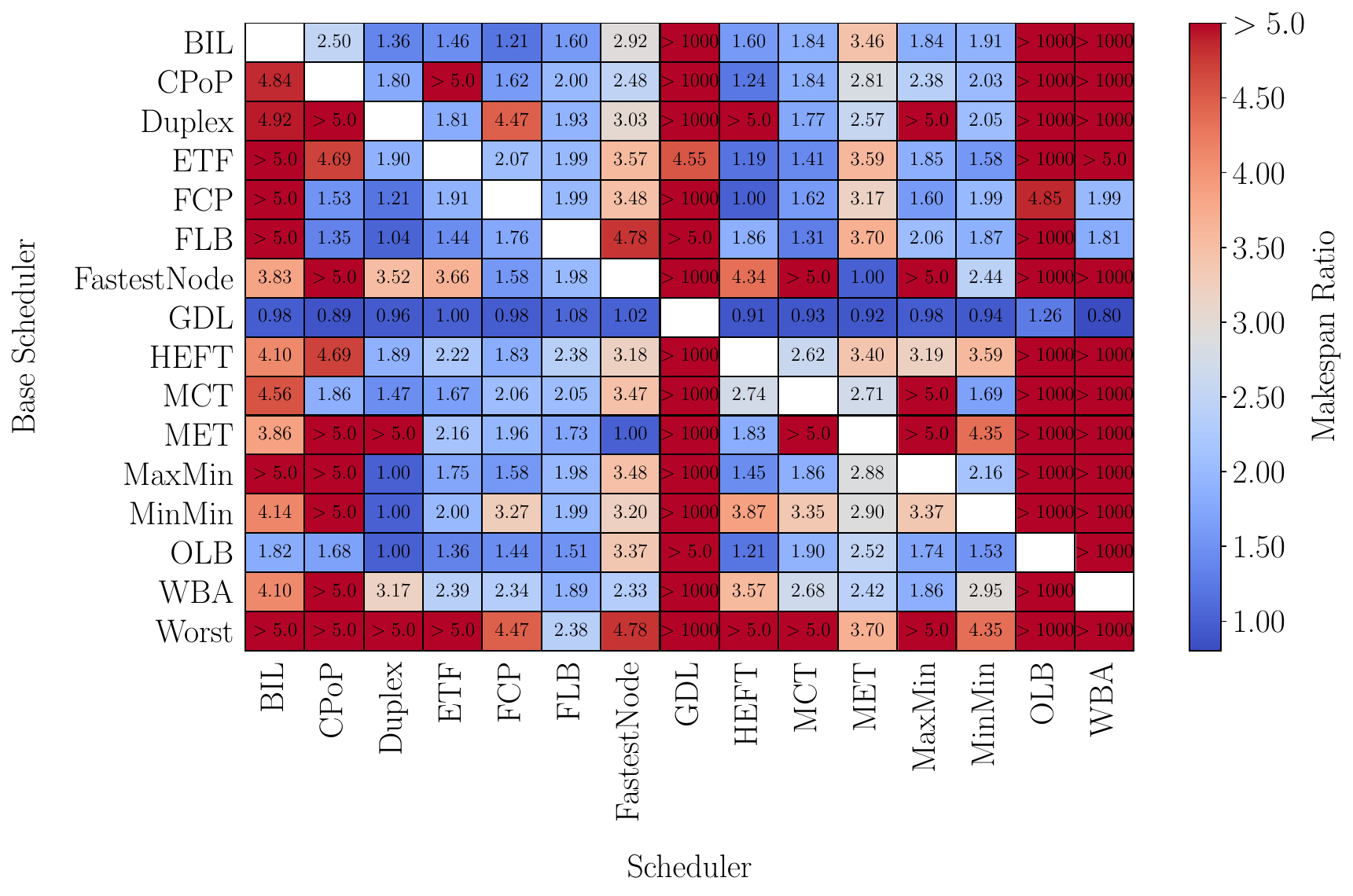}
    \caption{Heatmap of makespan ratios for all \NumAlgorithms Algorithms compared to each other. The cell value (and color) for row $i$ and column $j$ indicates the makespan ratio for the worst-case instance found by \AAName for scheduler $j$ against scheduler $i$.}
    \label{fig:sa_results}
\end{figure*}

Figure~\ref{fig:sa_results} shows the \AAName results for each pair of schedulers.
The benefit of our approach is clear from just a quick glance at Figure~\ref{fig:sa_results} (there's a lot more red!).
A closer look reveals even more interesting results.
First, \AAName finds, for \textit{every} scheduling algorithm, problem instances on which it performs at least twice as bad as another of the algorithms.
In fact, for most of the algorithms ($10$ of \NumAlgorithms), it finds a problem instance such that the algorithm performs at least  \textit{five times} worse than another algorithm.
For HEFT, one of the most popular scheduling algorithms, \AAName finds a problem instance where it performs $4.34$ times worse than FastestNode, a (generally poor-performing) simple baseline algorithm that schedules all tasks to execute in serial on the fastest compute node.
In fact, for many of the algorithms, \AAName finds problem instances where they perform quite poorly compared to FastestNode, perhaps suggesting that a common weakness of many scheduling algorithms is over-parallelization on some problem instances.

Furthermore, we note that for nearly every pair of algorithms $\mathcal{A}, \mathcal{B}$, \AAName identifies a problem instance where $\mathcal{A}$ outperforms $\mathcal{B}$ \textit{and} where $\mathcal{B}$ outperforms $\mathcal{A}$.
In other words, we don't see many algorithms that are strictly better or worse than others.
Each algorithm has its own strengths and weaknesses that are revealed by \AAName.
Finally, some of the cells in Figure~\ref{fig:sa_results} have values of $>1000$.
In this case, \AAName identified a problem instance where one algorithm \textit{drastically} outperforms the other.
For these cases, it's likely that there exist problem instances where the scheduler performs \textit{arbitrarily} poorly compared to the baseline scheduler.
Insights like these (of which Figure~\ref{fig:sa_results} contains many more), provided by \AAName, are invaluable for understanding the strengths and weaknesses of task scheduling algorithms and can be used to guide the development of new algorithms and distributed computing systems.

\subsection{Case Study: HEFT vs. CPoP}\label{sec:case_study}
\begin{figure*}[h!]
    \begin{subfigure}{0.4\textwidth}
        \centering
        \includegraphics[width=0.9\linewidth]{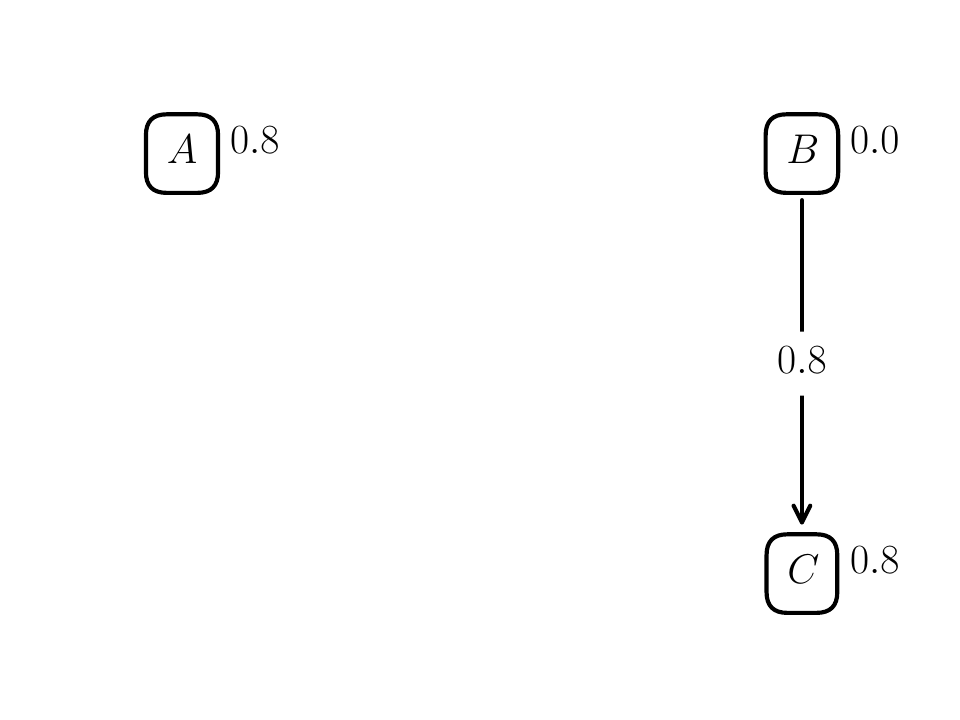}
        \caption{Task Graph}
        \label{fig:heft_vs_cpop:task_graph}
    \end{subfigure}%
    \begin{subfigure}{0.6\textwidth}
        \centering
        \includegraphics[width=0.9\linewidth]{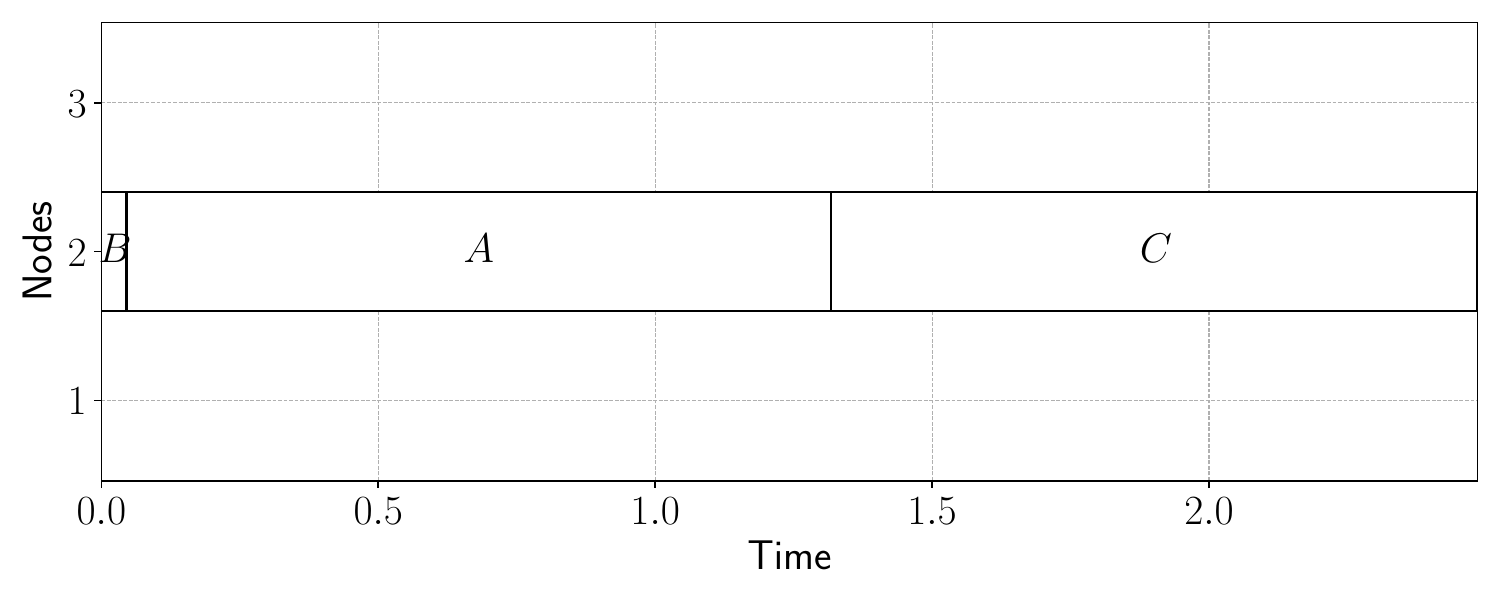}
        \caption{HEFT Schedule}
        \label{fig:heft_vs_cpop:heft}
    \end{subfigure}%
    
    \begin{subfigure}{0.4\textwidth}
        \centering
        \includegraphics[width=0.9\linewidth]{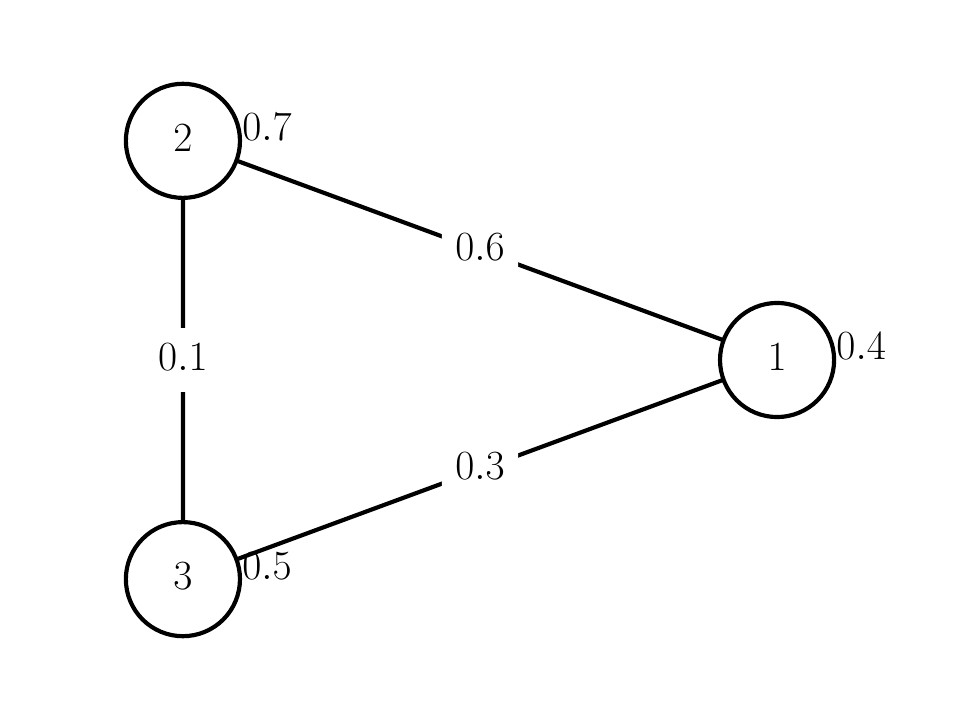}
        \caption{Network}
        \label{fig:heft_vs_cpop:network}
    \end{subfigure}%
    \begin{subfigure}{0.6\textwidth}
        \centering
        \includegraphics[width=0.9\linewidth]{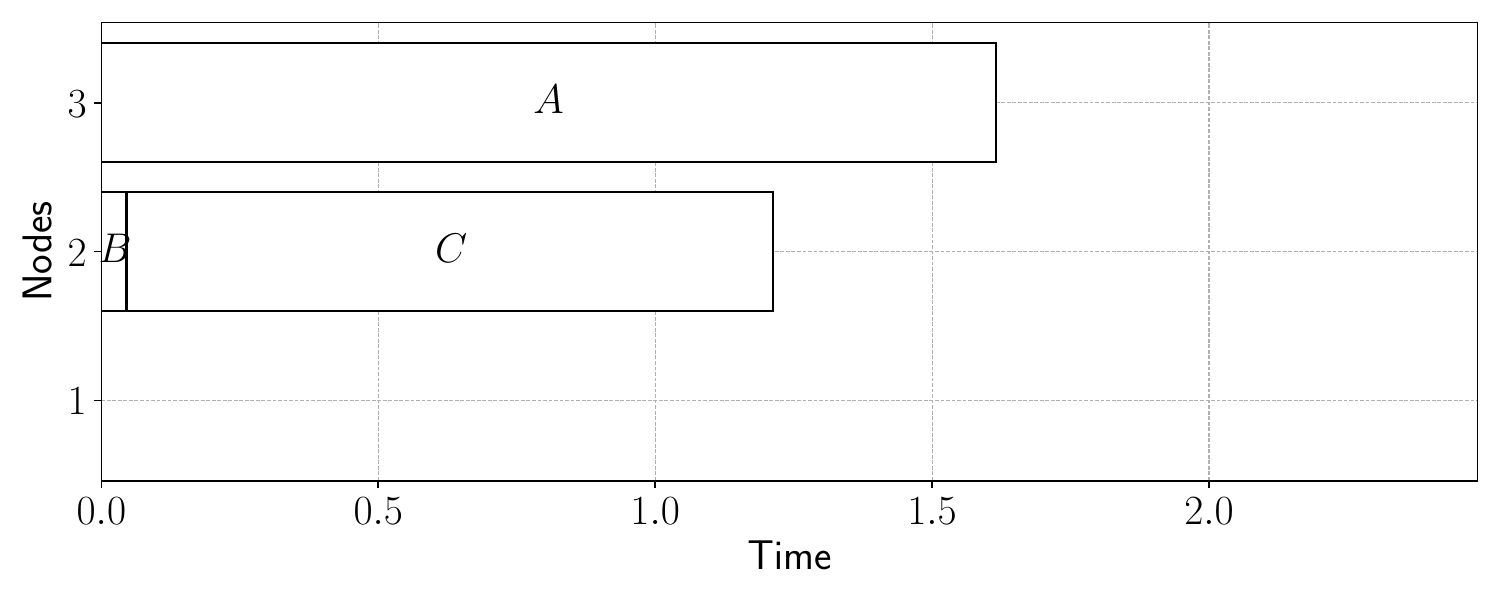}
        \caption{CPoP Schedule}
        \label{fig:heft_vs_cpop:cpop}
    \end{subfigure}%
    \caption{Problem Instance where HEFT performs $\approx 1.55$ times worse than CPoP.}
    \label{fig:heft_vs_cpop}
\end{figure*}

The HEFT and CPoP algorithms were proposed in the same paper by Topcuoglu et al. in 1999 and have remained two of the most popular task scheduling algorithms in decades since.
Both are list-scheduling algorithms that use a priority list to determine the order in which tasks are greedily scheduled.
In HEFT, each task is scheduled on the node that minimizes its finish time, given previous scheduling decisions.
CPoP is similar, but commits to scheduling all \textit{critical-path} tasks (those on the longest path in the task graph) on the fastest node\footnote{This statement is true for the related machines model. In general, CPoP schedules critical-path tasks to the node that minimizes the sum of execution times of critical-path tasks on that node.}.
The priority functions for these algorithms are slightly different from each other, but are both based on task distance (in terms of average compute and communication time) from the start and/or end of the task graph.
In HEFT, the priority of a sink task (one with no dependencies) is its average execution time over all nodes in the network.
A non-sink task's priority is the sum of its average execution time over all nodes in the network \textit{plus} the maximum communication and compute time of its successors.
In other words, it's the average amount of time it takes to execute the task on any node in the network plus the maximum amount of time it takes (in an ideal scenario) to execute the rest of the task graph.
In CPoP, the task priorities are computed based on the distance to the end \textit{and} from the start of the task graph.
This difference is a large part of what causes the two algorithms to perform differently on certain problem instances.
Observe in Figure~\ref{fig:heft_vs_cpop} a problem instance identified by \AAName where HEFT performs approximately $1.55$ times worse than CPoP.
In this instance, the algorithms differ primarily in whether task $A$ or task $C$ has higher priority.
In both algorithms, task $B$ has the highest priority and is thus scheduled first on node $2$ (the fastest node).
For CPoP, task $C$ \textit{must} have the highest priority because it is on the critical path $B \rightarrow C$.
For HEFT, though, task $A$ has the higher priority than $C$ because it is further from the end of the task graph.
As a result, CPoP schedules task $C$ on node $2$, which allows task $A$ to be scheduled and executed in parallel on node $3$ (the second fastest node).
HEFT, on the other hand, schedules all tasks on node $2$ and thus does not benefit from parallel execution.
CPoP succeeds in this instance because it prioritizes tasks that are on the critical path, keeping high-cost tasks/communications on the same node and allowing low-cost tasks/communications to be executed in parallel on other nodes.
HEFT greedily schedules high-cost tasks on fast nodes without taking into as much consideration how doing so might affect the rest of the task graph (especially the tasks on the critical path).

\begin{figure*}[h!]
    \begin{subfigure}{0.4\textwidth}
        \centering
        \includegraphics[width=0.9\linewidth]{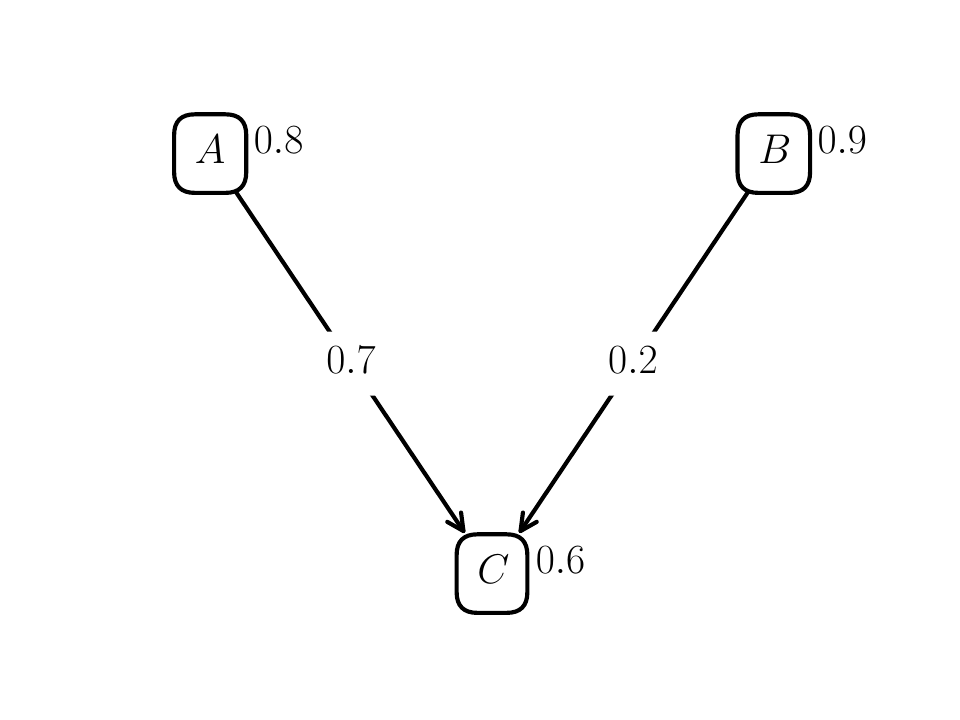}
        \caption{Task Graph}
        \label{fig:cpop_vs_heft:task_graph}
    \end{subfigure}%
    \begin{subfigure}{0.6\textwidth}
        \centering
        \includegraphics[width=0.9\linewidth]{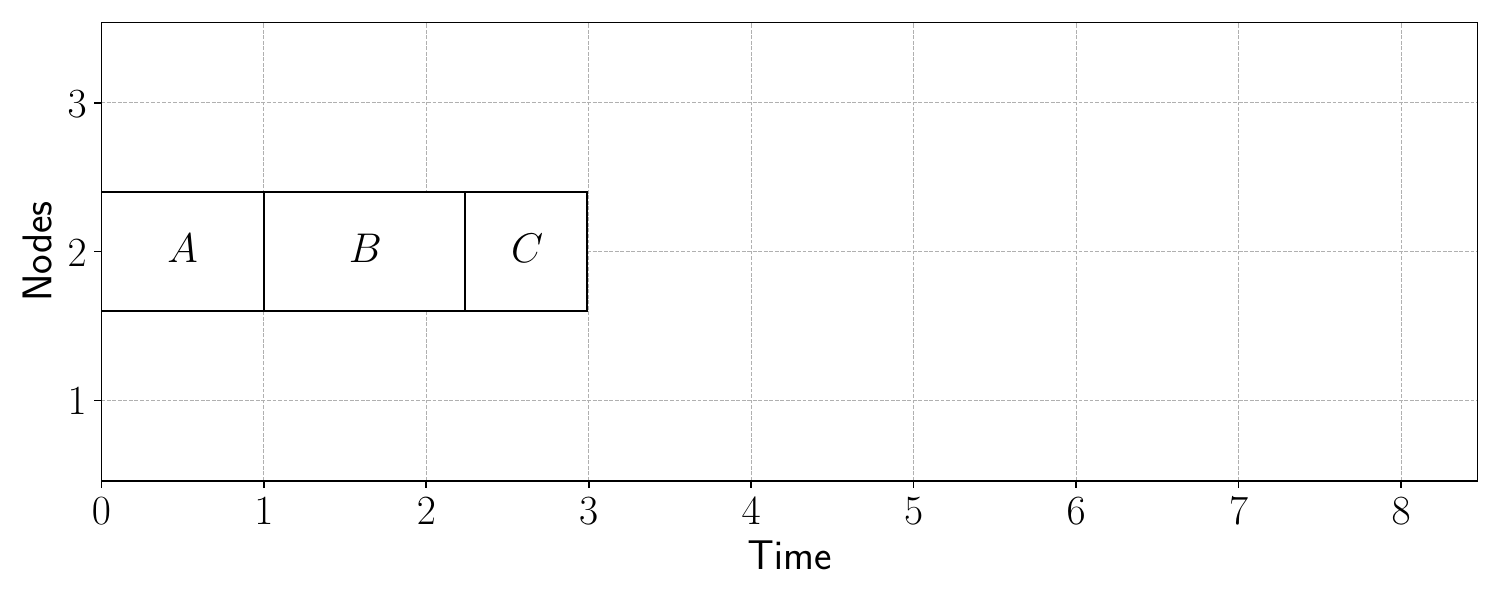}
        \caption{HEFT Schedule}
        \label{fig:cpop_vs_heft:heft}
    \end{subfigure}%

    \begin{subfigure}{0.4\textwidth}
        \centering
        \includegraphics[width=0.9\linewidth]{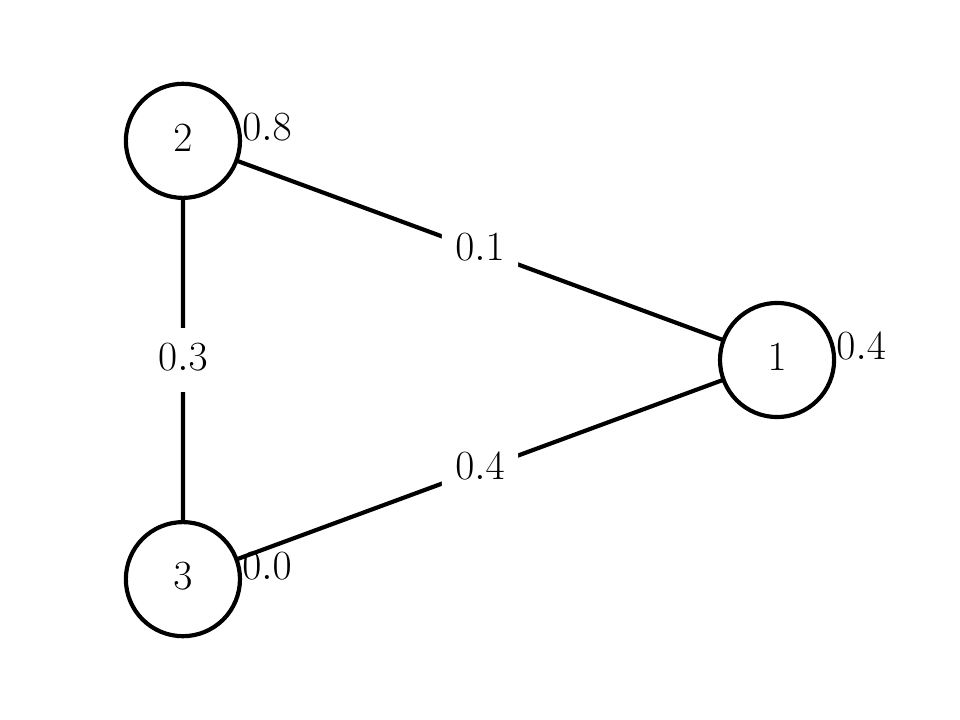}
        \caption{Network}
        \label{fig:cpop_vs_heft:network}
    \end{subfigure}%
    \begin{subfigure}{0.6\textwidth}
        \centering
        \includegraphics[width=0.9\linewidth]{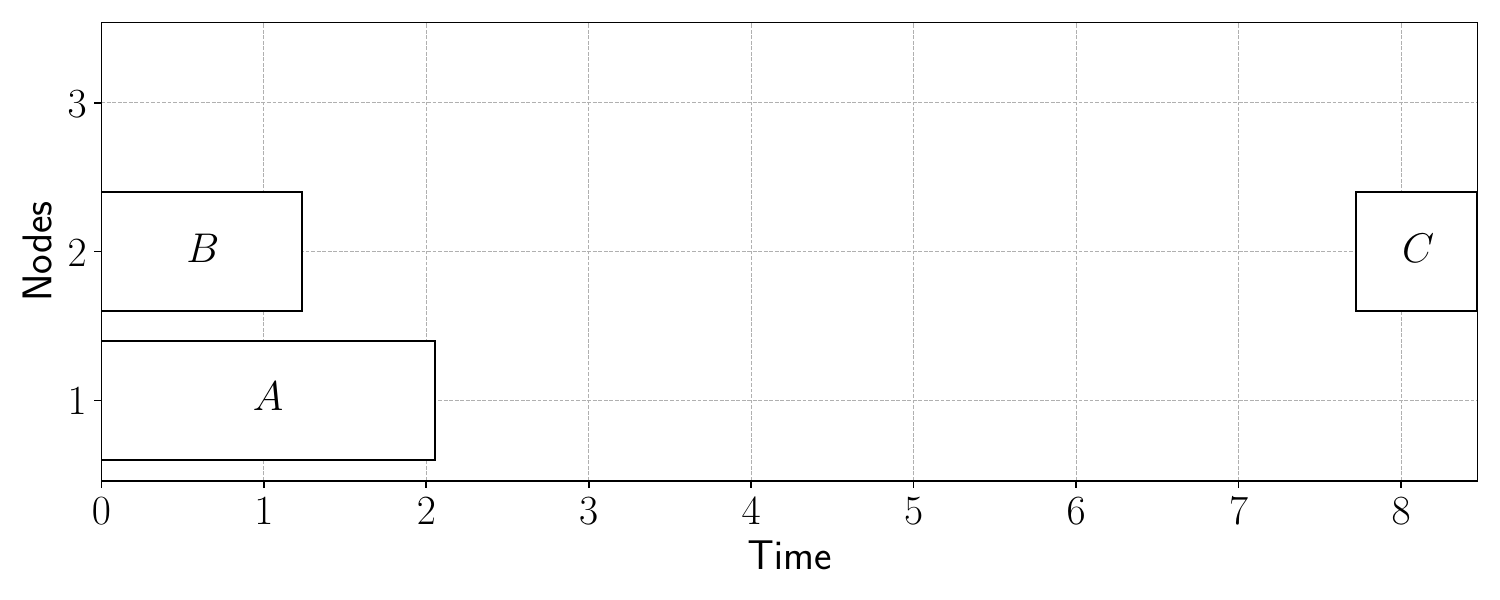}
        \caption{CPoP Schedule}
        \label{fig:cpop_vs_heft:cpop}
    \end{subfigure}%
    \caption{Problem Instance where CPoP performs $\approx 2.83$ times worse than HEFT.}
    \label{fig:cpop_vs_heft}
\end{figure*}

This does not mean that CPoP is better than HEFT, though.
Observe in Figure~\ref{fig:cpop_vs_heft} a problem instance identified by \AAName where CPoP performs approximately $2.83$ times worse than HEFT.
The problem that CPoP faces for this instance is exactly what allows it to succeed in the previous instance: committing to scheduling all critical-path tasks on the fastest node.
In this instance, the critical path is $B \rightarrow C$.
Thus, CPoP schedules task $C$ on node $2$ (the fastest node) even though it would have finished much faster (due to the communication cost incurred by its dependency on task $A$) on node $1$.
HEFT does not have this problem because it does not commit to scheduling all critical-path tasks on the fastest node.

\begin{figure*}[h!]
        \begin{subfigure}{0.5\textwidth}
            \centering
            \includegraphics[width=0.9\linewidth, trim=1cm 1cm 0cm 1cm, clip]{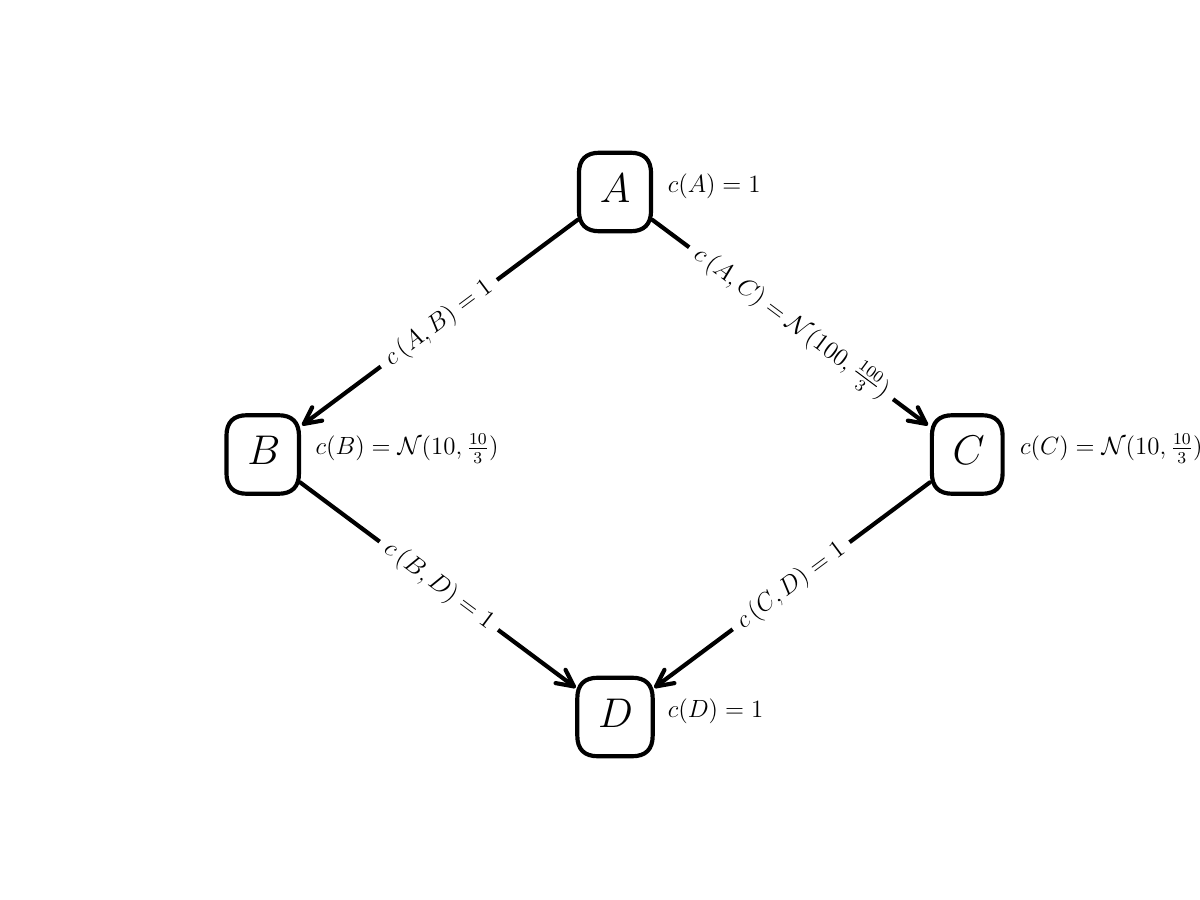}
            \caption{Task Graph}
            \label{fig:use_case:bad_heft:task_graph}
        \end{subfigure}%
        \begin{subfigure}{0.5\textwidth}
            \centering
            \includegraphics[width=0.9\linewidth, trim=0.5cm 1cm 1cm 1cm, clip]{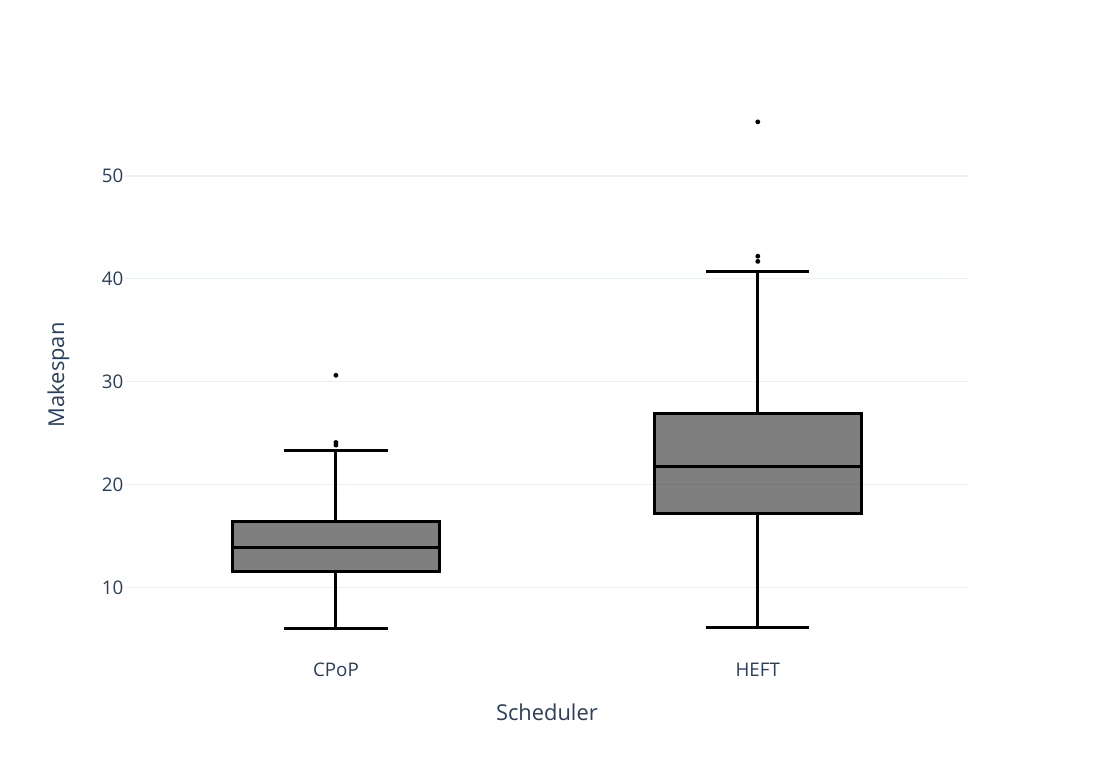}
            \caption{Makespans}
            \label{fig:use_case:bad_heft:makespans}
        \end{subfigure}%
        \caption{Task Graph and Makespans for HEFT and CPoP on a set of task graphs where HEFT performs poorly.}
        \label{fig:use_case:bad_heft}
    \end{figure*}

    One of the most promising aspects of \AAName is that it allows us to identify patterns where certain algorithms outperform others.
    We can generalize the patterns we observe in the above examples to formalize a set of conditions under which CPoP outperforms HEFT and vice versa.
    For example, we might hypothesize that HEFT will perform poorly against CPoP for fork-join task graphs where one chain of tasks has a much higher initial communication cost than the other.
    Formally, let us define a set of task graphs with tasks $A, B, C, D$ such that tasks $A$ and $D$ have cost $1$, tasks $B$ and $C$ have random cost (drawn from a clipped gaussian distribution with mean $10$, standard deviation $10/3$, and min $0$), and the dependencies $A \rightarrow B$, $A \rightarrow C$, and $B \rightarrow D$ have cost $1$ and $C \rightarrow D$ has high cost (drawn from a clipped gaussian distribution with mean $100$, standard deviation $100/3$, and min $0$).
    Observe in Figure~\ref{fig:use_case:bad_heft} an example of a task graph from this set and the makespans of both HEFT and CPoP on $1000$ randomly generated instances from this set (on a completely homogeneous network, for simplicity).
    Observe that HEFT performs significantly worse than CPoP on this family of task graphs.

\begin{figure*}[h!]
        \begin{subfigure}{0.5\textwidth}
            \centering
            \includegraphics[width=0.9\linewidth, trim=1cm 1cm 0cm 1cm, clip]{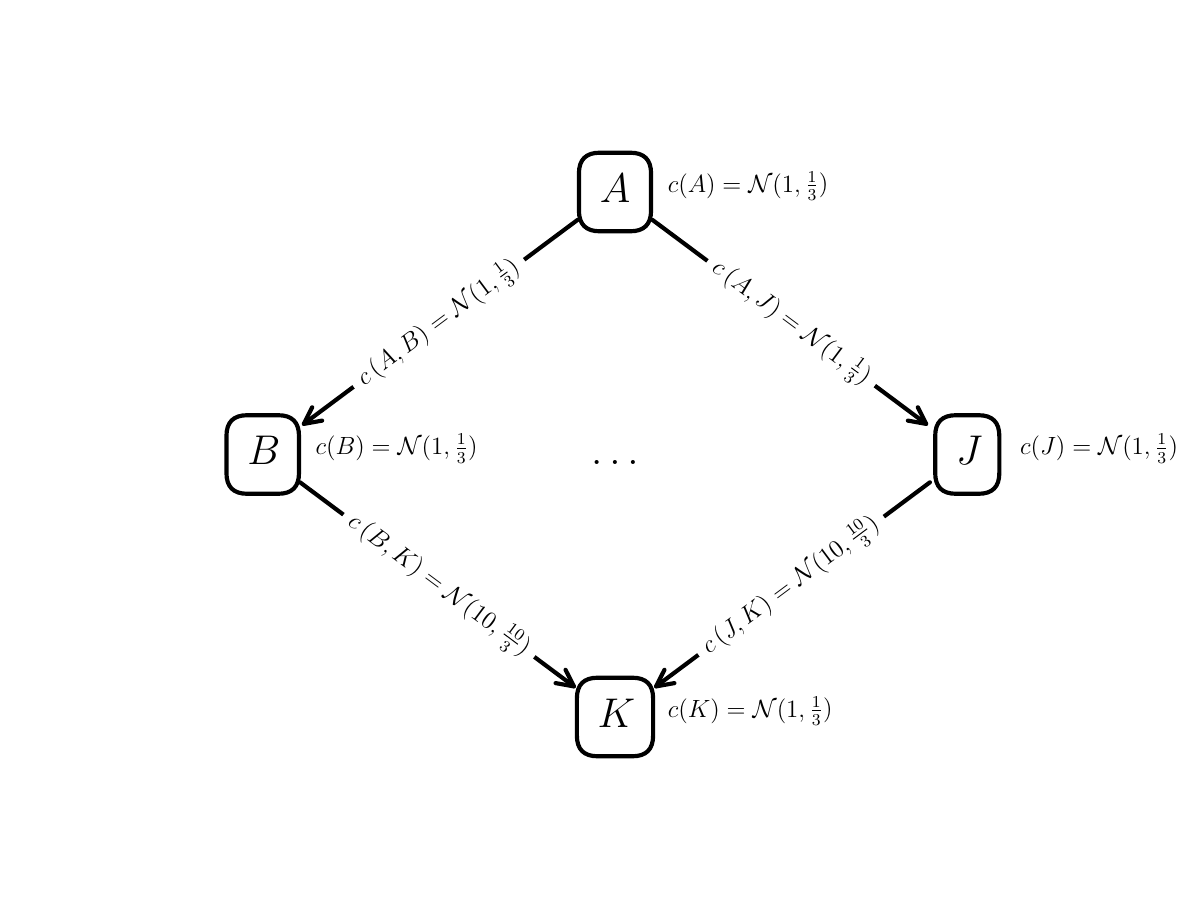}
            \caption{Task Graph}
            \label{fig:use_case:bad_cpop:task_graph}
        \end{subfigure}%
        \begin{subfigure}{0.5\textwidth}
            \centering
            \includegraphics[width=0.9\linewidth, trim=0.5cm 1cm 1cm 1cm, clip]{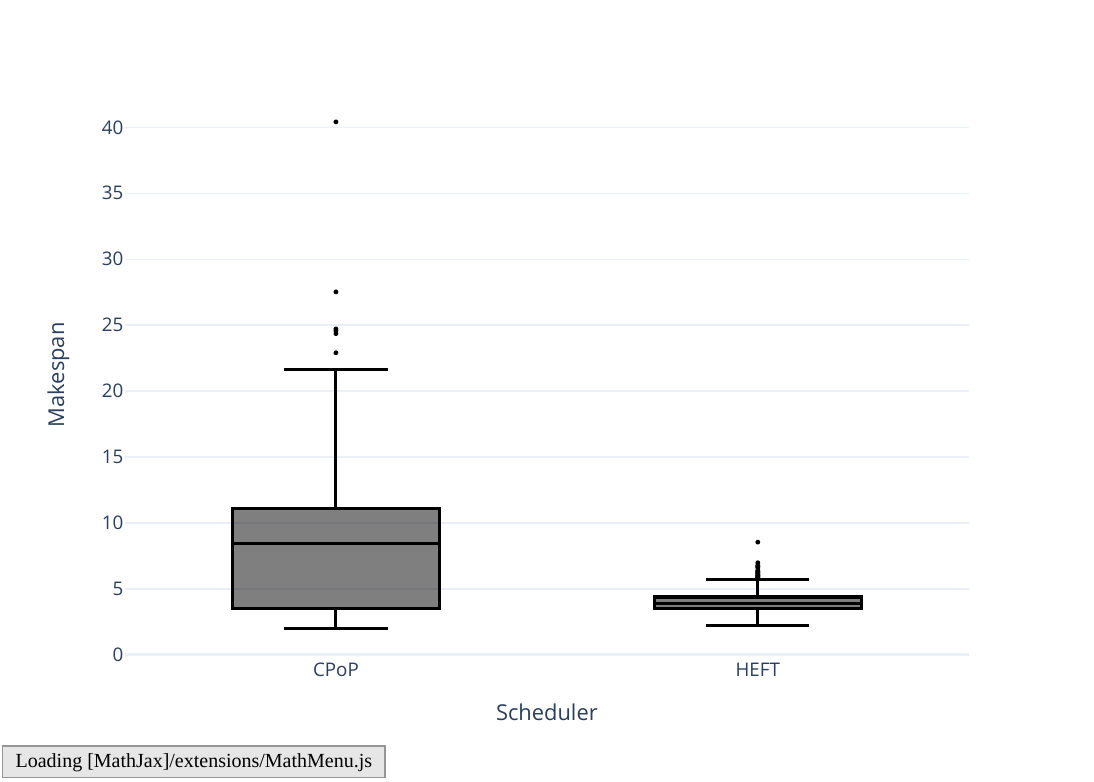}
            \caption{Makespans}
            \label{fig:use_case:bad_cpop:makespans}
        \end{subfigure}%
        \caption{Task Graph and Makespans for HEFT and CPoP on a set of task graphs where CPoP performs poorly.}
        \label{fig:use_case:bad_cpop}
    \end{figure*}

    On the other hand, we might hypothesize that CPoP will perform poorly against HEFT for task graphs where initially low communication costs guide CPoP into parallelizing tasks poorly and force it into a situation where it must perform a high-cost communication.
    Formally, let us define a set of task graphs with tasks $A$ through $K$ such that task $A$ is the start task, tasks $B$ through $J$ depend on task $A$, and task $K$ depends on tasks $B$ through $J$.
    All tasks have random costs drawn from a clipped gaussian distribution with mean $1$ and standard deviation $1/3$.
    The dependencies from task $A$ to the inner tasks $B$ through $J$ have random costs drawn from a clipped gaussian distribution with mean $1$ and standard deviation $1/3$ while the dependencies from the inner tasks to task $K$ have random costs drawn from a clipped gaussian distribution with mean $10$ and standard deviation $10/3$.
    In other words, the communication on the ``join'' part of the task graph is ten times more expensive than the communication on the ``fork'' part of the task graph.
    Let the network be a complete graph with four nodes where the fastest node has a weak link to the second fastest node.
    Let the fastest node have a speed of $3$ and the speeds of the other nodes be drawn from a clipped gaussian distribution with mean $1$ and standard deviation $1/3$ (they are, on average, three times slower than the fastest node).
    Let the communication costs between the fastest and second fastest node be drawn from a clipped gaussian distribution with mean $1$ and standard deviation $1/3$ and the communication costs between all other nodes be drawn from a clipped gaussian distribution with mean $10$ and standard deviation $5/3$ (the communication link between the two fastest nodes is on average ten times weaker than the communication links between all other nodes).
    Observe in Figure~\ref{fig:use_case:bad_cpop} an example of a task graph from this set and the makespans of both HEFT and CPoP on $1000$ randomly generated instances from this set.
    CPoP performs significantly worse than HEFT on this family of task graphs.

    By identifying problem instances where an algorithm performs significantly worse than another, \AAName allows us to gain insight into the strengths and weaknesses of different algorithms.
    Performing this kind of analysis on other pairs of algorithms remains an important task for future work.

\section{Application-Specific \AAName}\label{sec:app-specific}
In Section~\ref{sec:adversarial_analysis}, we introduced \AAName as an effective method for finding problem instances where an algorithm performs much worse than benchmarking results suggest.
The results, however, depend greatly on the initial problem instance and the implementation of the \textsc{Perturb} function.
These two things define the space of problem instances the algorithm searches over and also affect which problem instances are more or less likely to be explored.
We chose an implementation in Section~\ref{sec:adversarial_analysis} that kept problem instances relatively small (between three and five tasks and compute nodes) and allowed arbitrary task graph structure and CCRs (communication-to-computation ratios).
By keeping the problem instance size small but allowing for almost arbitrary task graph structure and CCR, this implementation allowed us to explore how the structure of problem instances affects schedules in order to find patterns where certain algorithms out-perform others.
In many more realistic scenarios, though, application developers have a better idea of what their task graph and/or compute network will look like.
\AAName can be easily restricted to searching over a space of \textit{realistic} problem instances by adjusting the \textsc{Perturb} implementation and initial problem instance.
In this section, we report results on experiments with application-specific \textsc{Perturb} implementations.
Again, these results support the main hypothesis of this paper that \AAName reveals performance boundaries that a traditional benchmarking approach does not.

\subsection{Experimental Setup}
One of the largest communities of researchers that depend on efficient task scheduling algorithms is that of scientists that use {scientific workflows} for simulation, experimentation, and more.
Scientific workflows are typically big-data scale task graphs that are scheduled to run on cloud or super computing platforms.
These researchers typically have little to no control over how their workflows are scheduled, instead relying on Workflow Management Systems (WFMS) like Pegasus~\cite{wfms:pegasus}, Makeflow~\cite{wfms:makeflow}, and Nextflow~\cite{wfms:nextflow} (to name a just few examples) to handle the technical execution details.
For this reason, it is especially important for WFMS developers/maintainers (who choose which scheduling algorithms their system uses) to understand the performance boundaries between different algorithms for the different types of scientific workflows and computing systems their clients use.

\FrameworkName currently has datasets based on nine real-world scientific workflows (blast, bwa, cycles, epigenomics, 1000genome, montage, seismology, soykb, and srasearch).
These applications come from a very wide range of scientific domains --- from astronomy (montage builds mosaics of astronomical imagery) to biology (1000genome performs human genome reconstruction) to agriculture (cycles is an agroecosystem model).
For each workflow, the runtime of each task, input/output sizes in bytes, and speedup factor (compute speed) for each machine are available from public execution trace information\footnote{\href{https://github.com/wfcommons/pegasus-instances}{https://github.com/wfcommons/pegasus-instances}, \href{https://github.com/wfcommons/makeflow-instances}{https://github.com/wfcommons/makeflow-instances}}.
The inter-node communication rate, however, is not available. 
We set communication rates to be homogeneous so that the average CCR, or $\frac{\text{average data size}}{\text{commmunication strength}}$, is $\frac{1}{5}, \frac{1}{2}, 1, 2, $ or $5$ (resulting in five experiments for each workflow).

Because these scientific workflows are much larger than the workflows used in the experiments from Section~\ref{sec:adversarial_analysis}, we evaluate a smaller subset of the schedulers available in \FrameworkName: FastestNode, HEFT, CPoP, MaxMin, MinMin, and WBA.
Performing these experiments for the rest of the schedulers remains a task for future work.
For generating a benchmarking dataset, we use the WfCommons Synthetic Workflow Generator~\cite{data:wfchef} to generate random in-family task graphs and create random networks using a best-fit distribution computed from the real execution trace data.
We also use this method to generate initial problem instances for \AAName.
Then, we adapt \AAName's \textsc{Perturb} implementation as follows:
\begin{itemize}
    \item \textbf{Change Network Node Weight}: Same as described in Section~\ref{sec:adversarial_analysis}, except the weight is scaled between to the range of speeds observed in the real execution trace data.
    \item \textbf{Change Network Edge Weight}: Removed since network edge weights are homogeneous and fixed to enforce a specific CCR.
    \item \textbf{Change Task Weight}: Same as described in Section~\ref{sec:adversarial_analysis}, except the weight is scaled to the range of task runtimes observed in the real execution trace data.
    \item \textbf{Change Dependency Weight}: Same as described in Section~\ref{sec:adversarial_analysis}, except the weight is scaled between the minimum and maximum task I/O size observed in the real execution trace data.
    \item \textbf{Add/Remove Dependency}: Removed so that the task graph structure is representative of the real application.
\end{itemize}
These adjustments to allow us to explore application-specific problem instances more realistically.

\subsection{Results}
\begin{figure*}[!htb]
    \centering
    
    \begin{subfigure}[b]{0.48\textwidth}
        \centering
        \includegraphics[width=\textwidth]{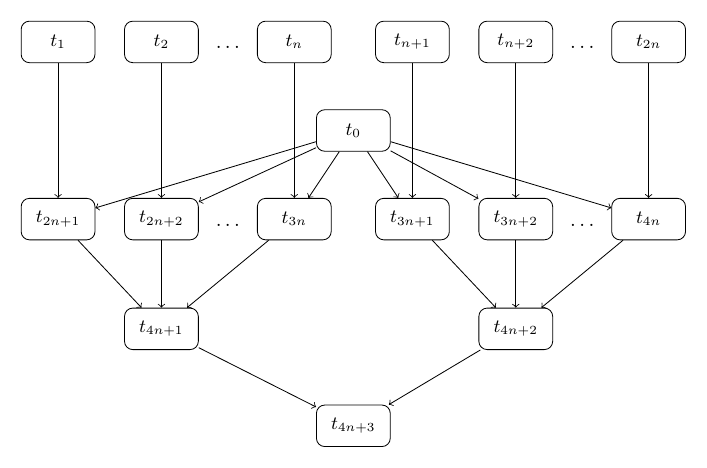}
        \caption{Srasearch workflow structure}
        \label{fig:workflow:srasearch}
    \end{subfigure}%
    \hfill
    \begin{subfigure}[b]{0.48\textwidth}
        \centering
        \includegraphics[width=\textwidth]{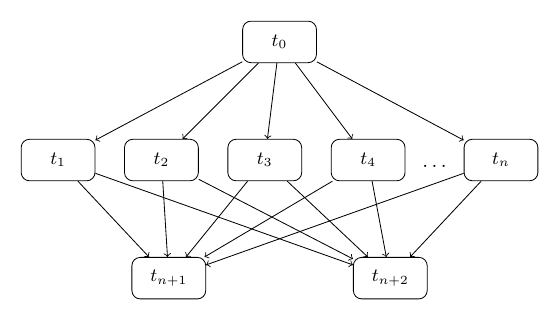}
        \caption{Blast workflow structure}
        \label{fig:workflow:blast}
    \end{subfigure}%
    \caption{Example workflow structures for blast and srasearch scientific workflows.}
    \label{fig:workflow}
\end{figure*}

In this section we present and discuss some results on two of the workflows analyzed: srasearch and blast.
Full results can be found in Appendix~\ref{sec:app-specific-full}.
Srasearch (workflow structure depicted in Figure~\ref{fig:workflow:srasearch}) is a toolkit for interacting with data in the INSDC Sequence Read Archives and blast (workflow structure depicted in Figure~\ref{fig:workflow:blast}) is a toolkit for finding regions of similarity between biological sequences.

\begin{figure*}[!htb]
    \centering
    
    \begin{subfigure}[b]{0.5\textwidth}
        \centering
        \includegraphics[width=\textwidth]{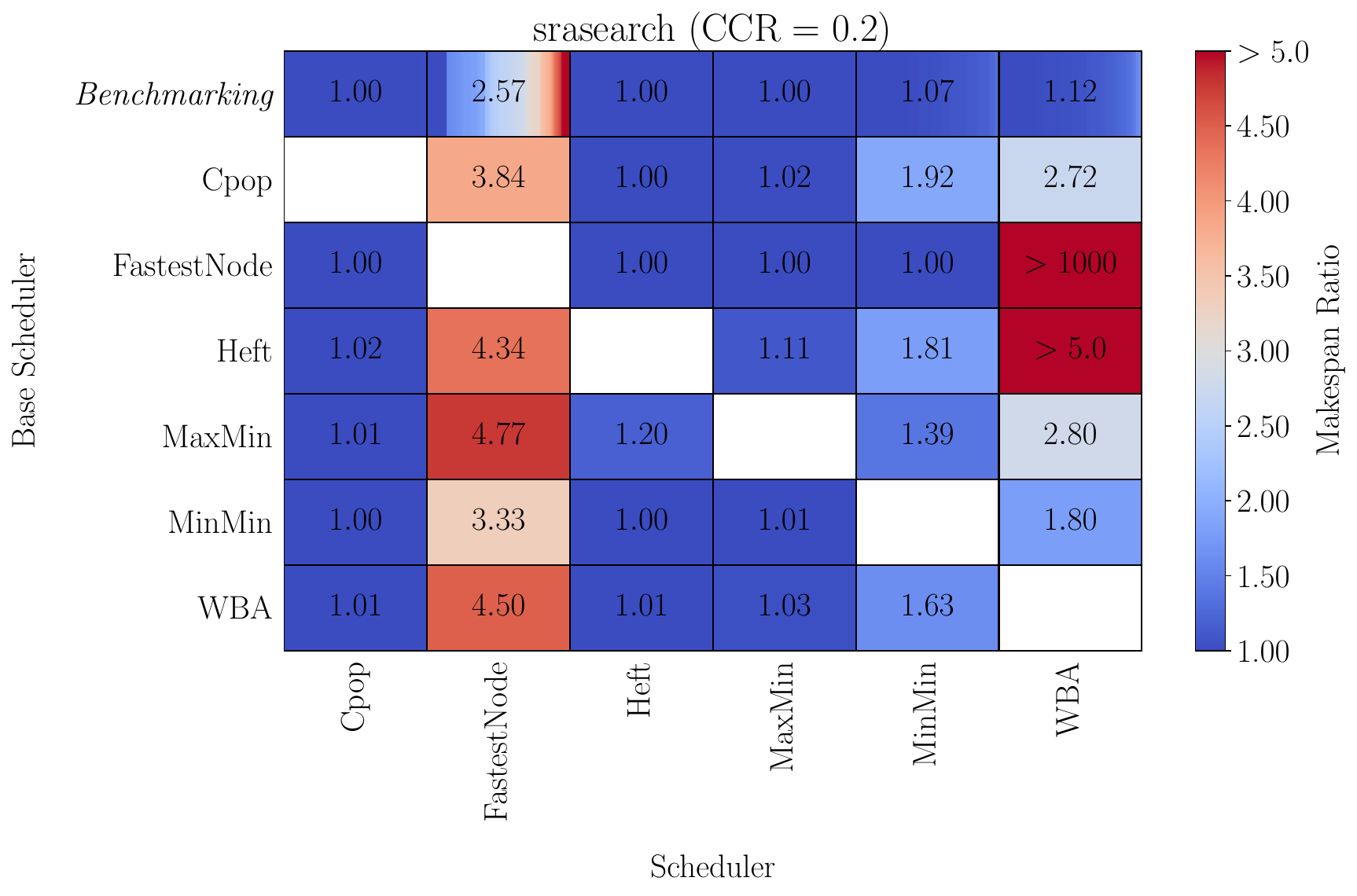}
    \end{subfigure}%
    \hfill
    \begin{subfigure}[b]{0.5\textwidth}
        \centering
        \includegraphics[width=\textwidth]{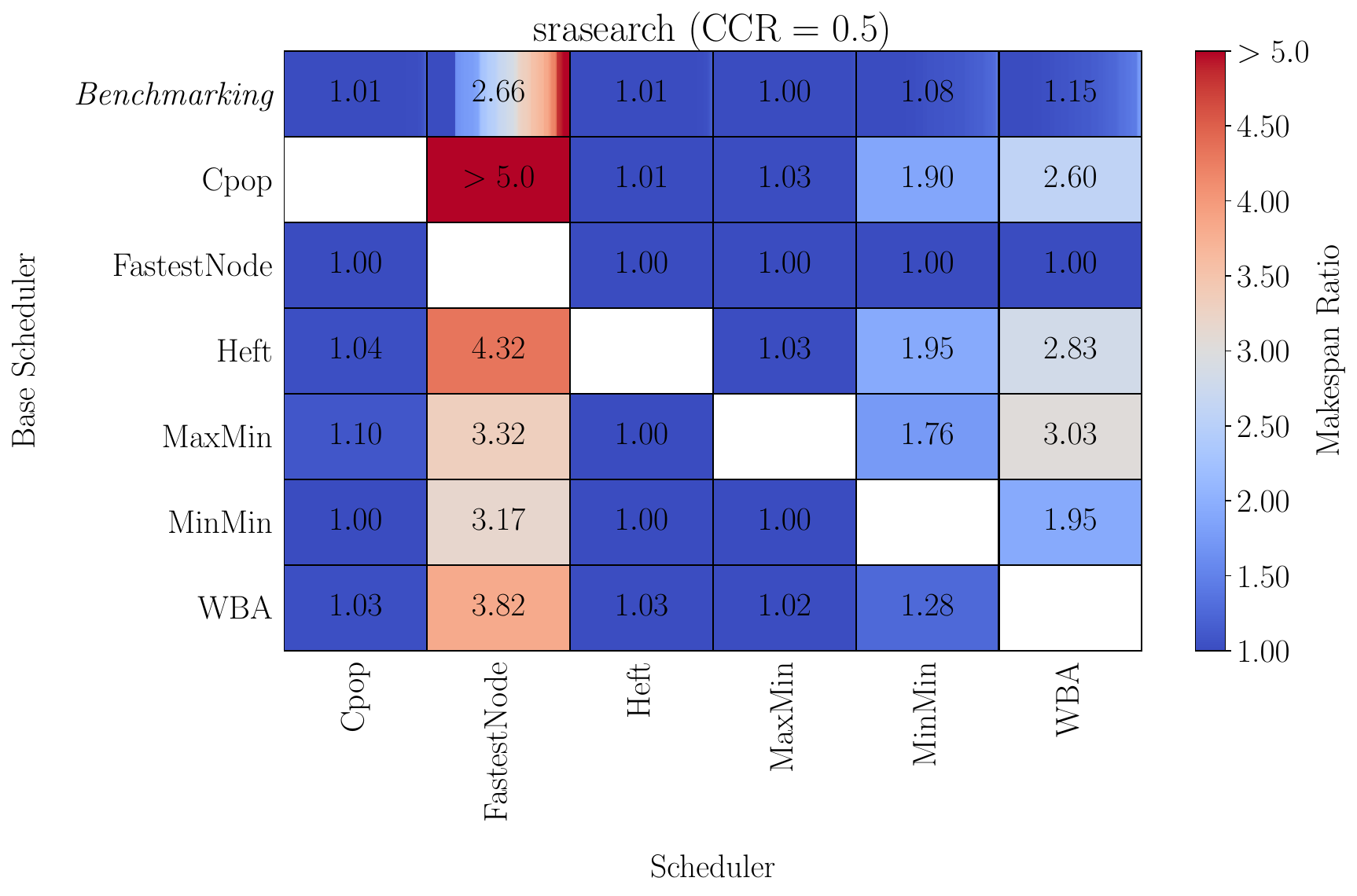}
    \end{subfigure}%
    
    \vspace{0.25cm}%
    
    \begin{subfigure}[b]{0.5\textwidth}
        \centering
        \includegraphics[width=\textwidth]{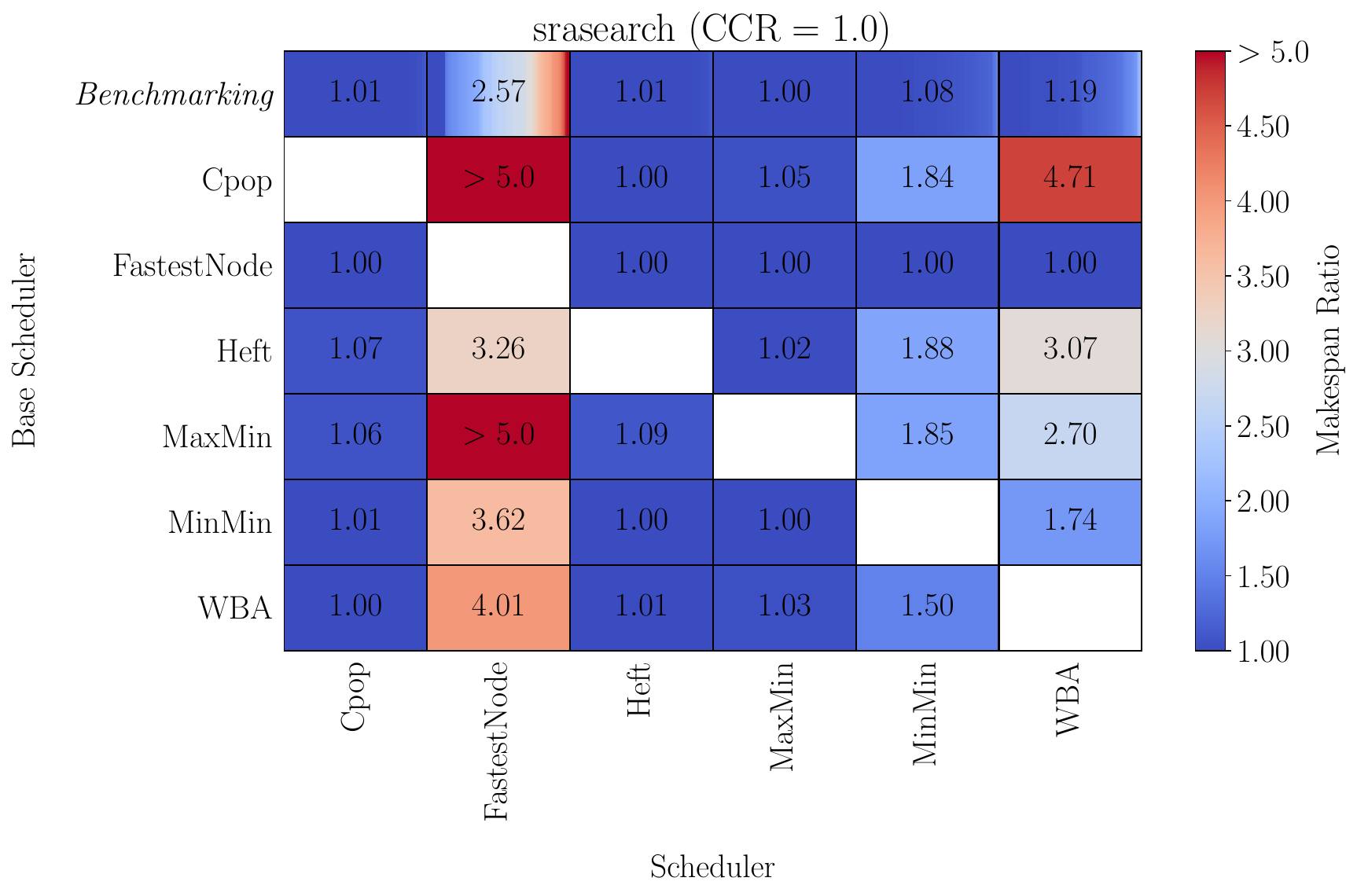}
    \end{subfigure}%
    \hfill
    \begin{subfigure}[b]{0.5\textwidth}
        \centering
        \includegraphics[width=\textwidth]{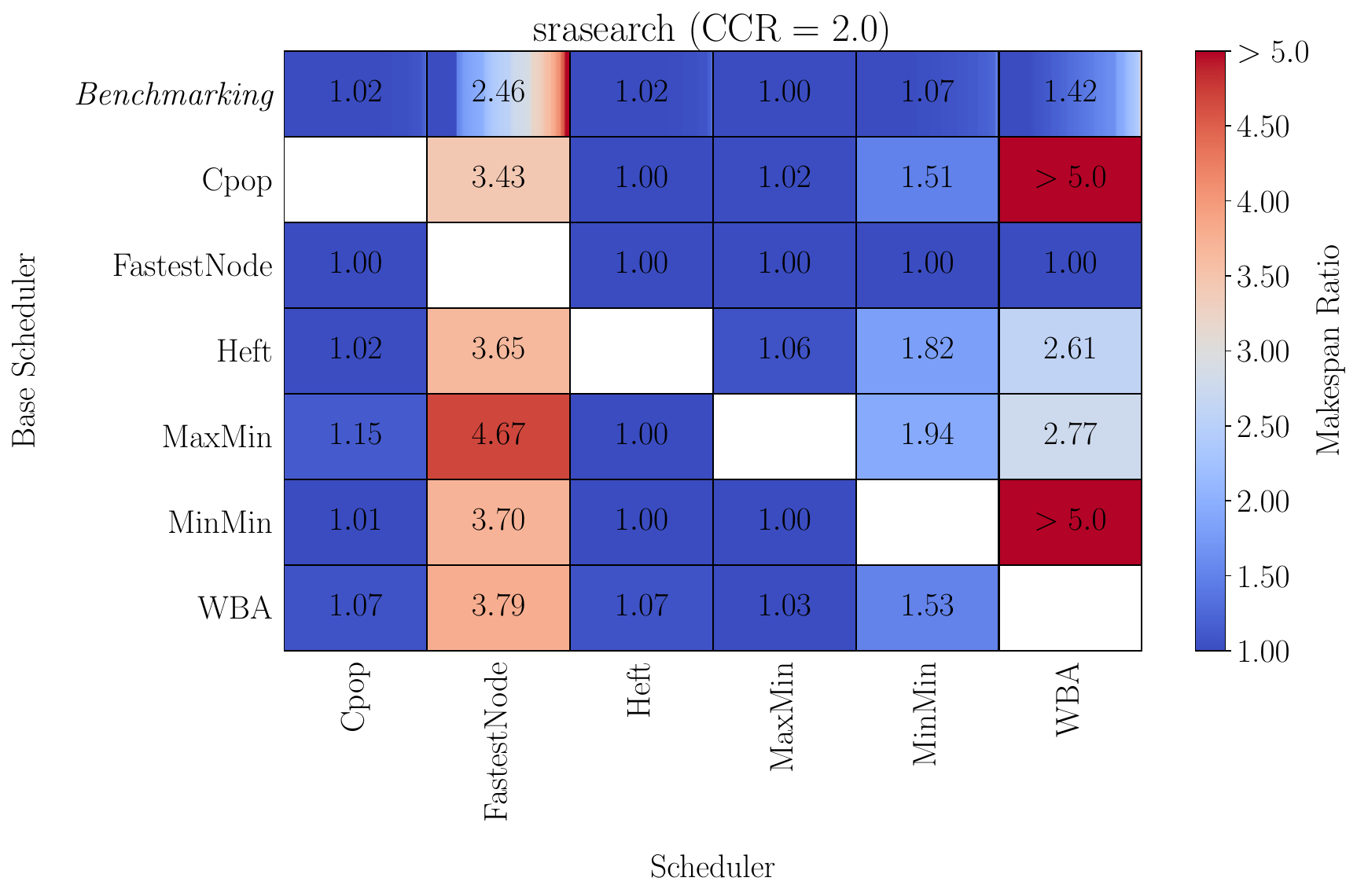}
    \end{subfigure}%
    
    \vspace{0.25cm}%
    
    \begin{subfigure}[b]{0.5\textwidth}
        \centering
        \includegraphics[width=\textwidth]{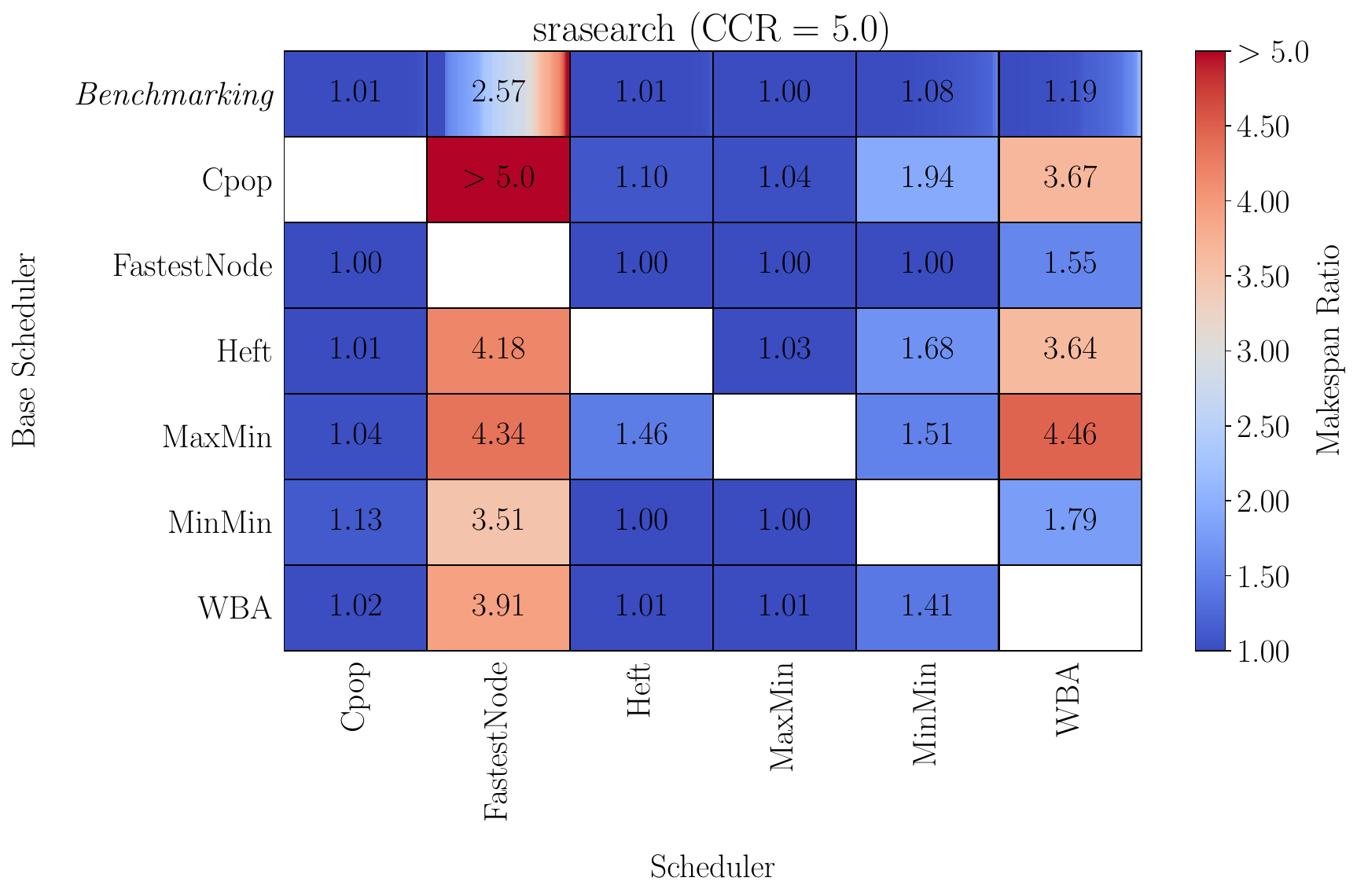}
    \end{subfigure}%
    \caption{Benchmarking and \AAName results for srasearch workflows with different average CCRs.
    For each CCR, the top row shows benchmarking results, where gradients indicate makespan ratios on different problem instances in the dataset. All other cells indicate the highest makespan ratio yielding problem instance discovered by \AAName.}
    \label{fig:app:srasearch}
\end{figure*}

\begin{figure*}[!htb]
    \centering
    
    \begin{subfigure}[b]{0.5\textwidth}
        \centering
        \includegraphics[width=\textwidth]{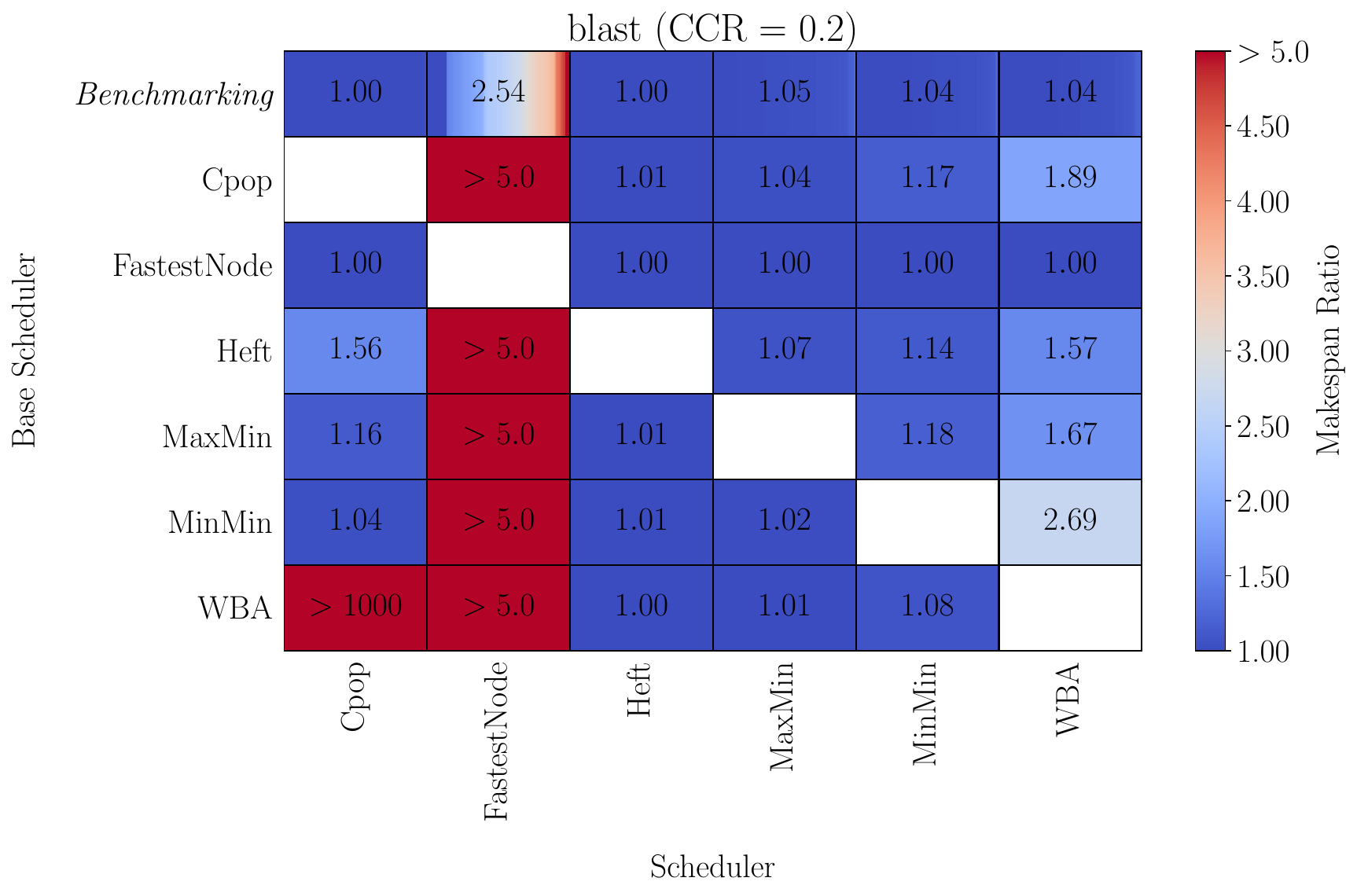}
    \end{subfigure}%
    \hfill
    \begin{subfigure}[b]{0.5\textwidth}
        \centering
        \includegraphics[width=\textwidth]{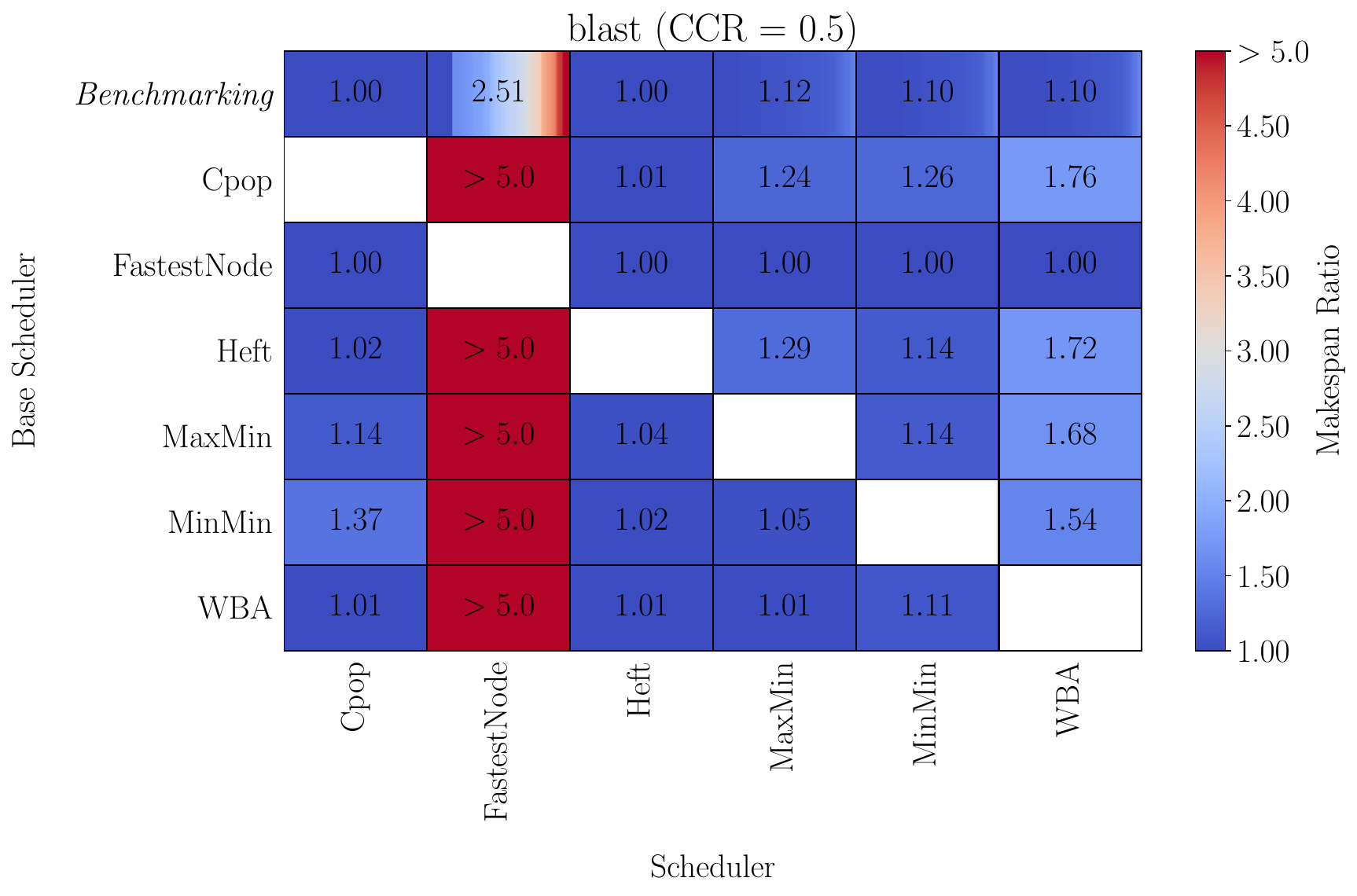}
    \end{subfigure}%

    \vspace{0.25cm}%

    \begin{subfigure}[b]{0.5\textwidth}
        \centering
        \includegraphics[width=\textwidth]{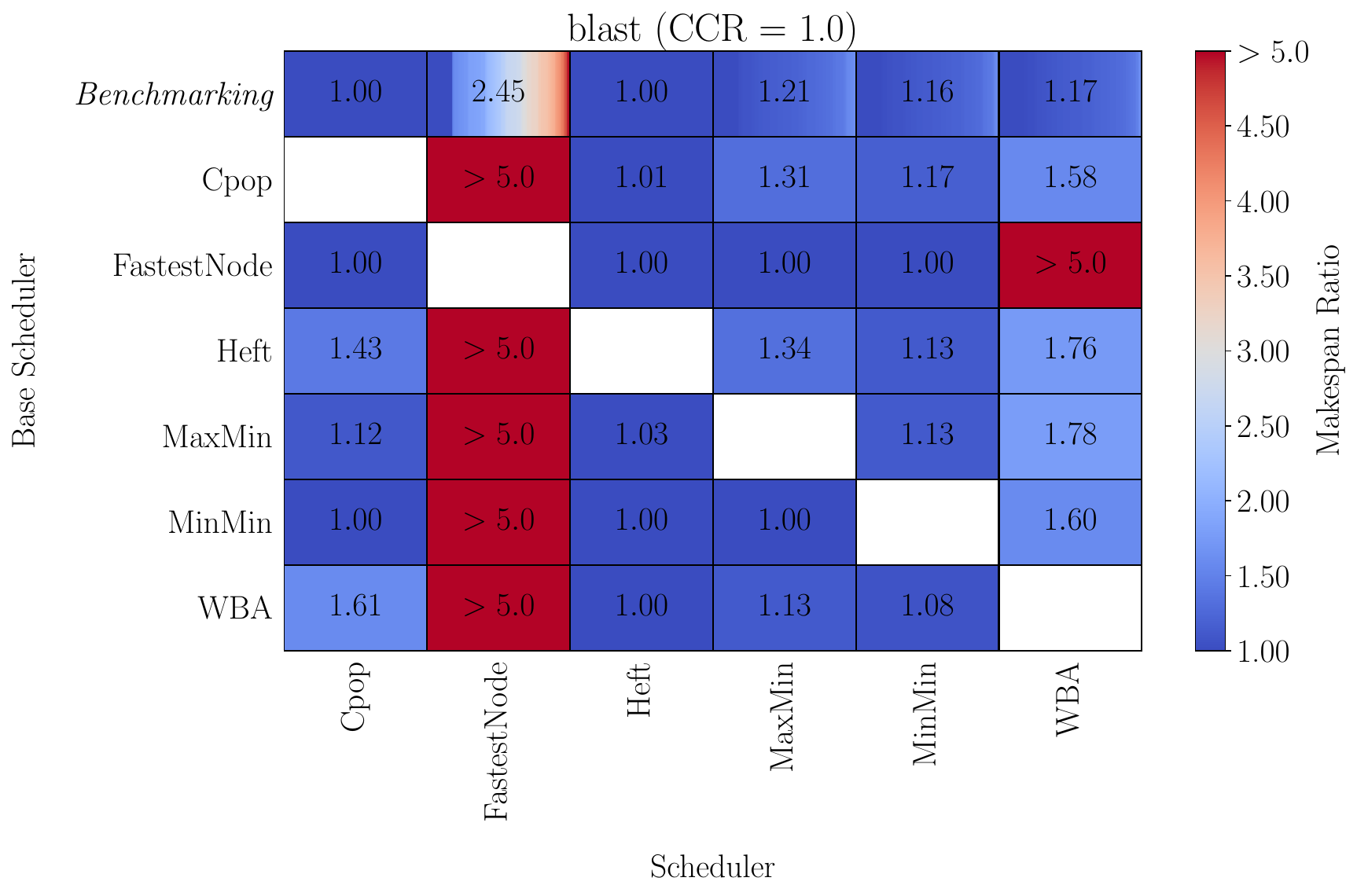}
    \end{subfigure}%
    \hfill
    \begin{subfigure}[b]{0.5\textwidth}
        \centering
        \includegraphics[width=\textwidth]{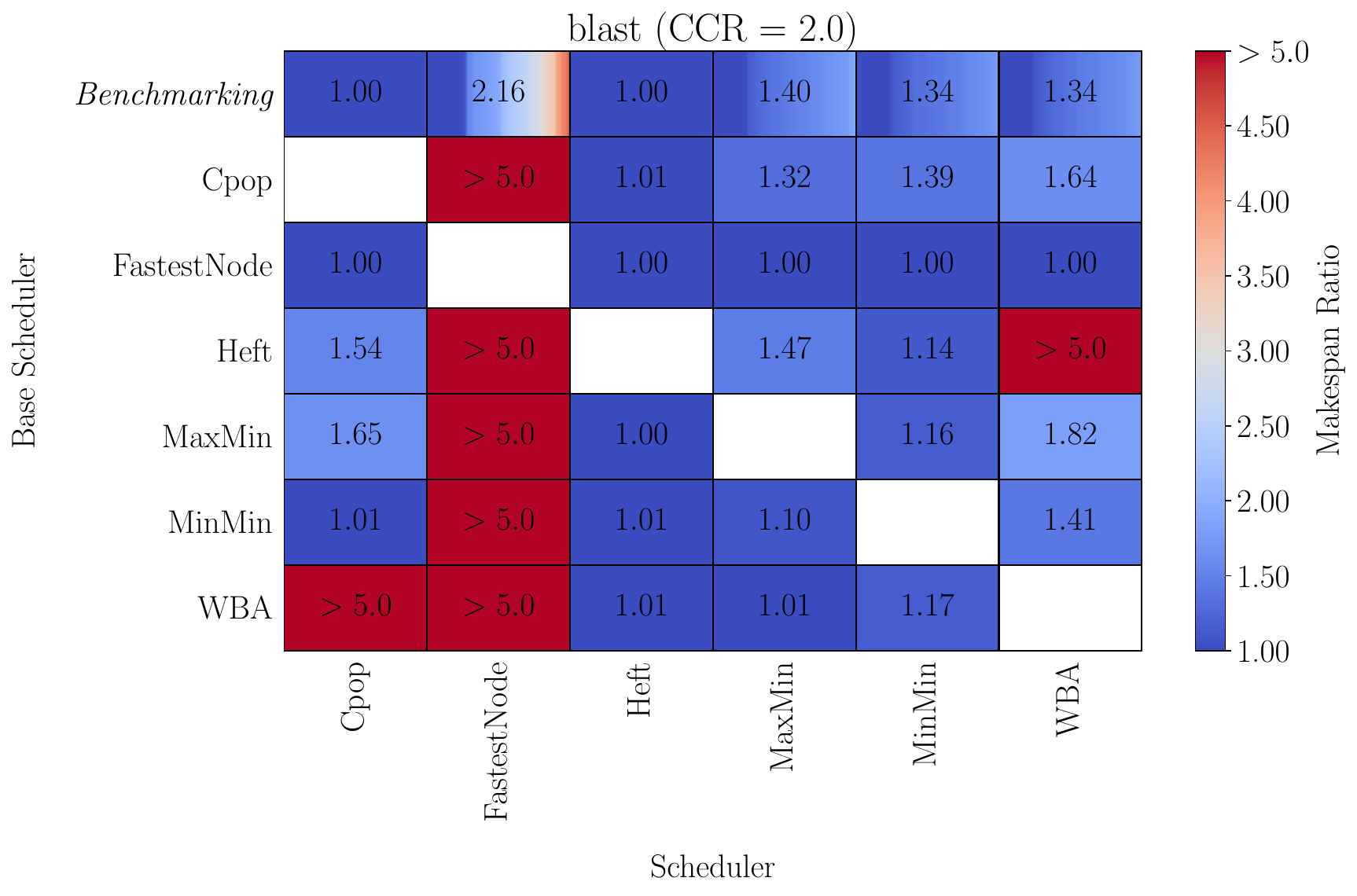}
    \end{subfigure}%
    
    \vspace{0.25cm}%
    
    \begin{subfigure}[b]{0.5\textwidth}
        \centering
        \includegraphics[width=\textwidth]{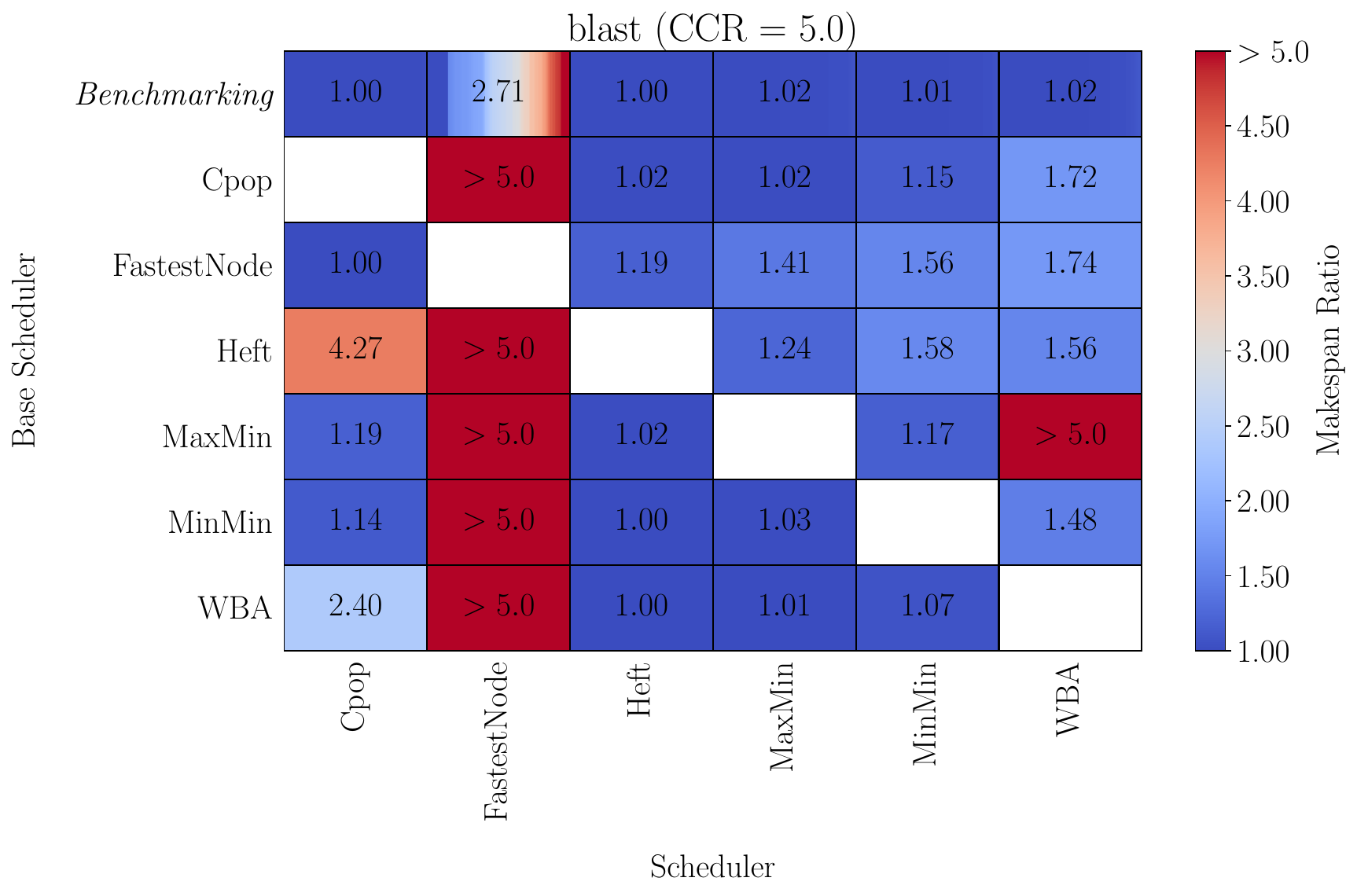}
    \end{subfigure}%
    \caption{Benchmarking and \AAName results for blast workflows with different average CCRs.
    For each CCR, the top row shows benchmarking results, where gradients indicate makespan ratios on different problem instances in the dataset. All other cells indicate the highest makespan ratio yielding problem instance discovered by \AAName.}
    \label{fig:app:blast}
\end{figure*}

Observe in Figure~\ref{fig:workflow} that while the number of tasks may vary, the structure of both workflows is very rigid.
Our new \textsc{Perturb} implementation, though, guarantees that the search space contains only task graphs with appropriate structure.
Figures~\ref{fig:app:srasearch} and~\ref{fig:app:blast} show the benchmarking and \AAName results for srasearch and blast, respectively (results for some CCRs are excluded due to space constraints and can be found in the full version of the paper~\cite{full_paper}).
First, observe that the benchmarking approach suggests all algorithms (except FastestNode) perform very well on the srasearch applications with a CCR of $1/5$.
Using \AAName, however, we are able to identify problem instances where WBA performs thousands of times worse than FastestNode!
Also, we're able to find instances where MinMin performs almost twice as bad CPoP.
Even among the ``good'' algorithms, though, we see interesting behavior.
Observe the results of HEFT and MaxMin.
We are able to find both a problem instance where HEFT performs approximately $20\%$ worse that MaxMin \textit{and} an instance where MaxMin performs approximately $11\%$ worse than HEFT.

The results for srasearch seem to imply that  CPoP, which appears to perform consistently well for all CCRs tested would be the best choice for a WFMS designer.
Workflow Management Systems do not support just one type of scientific workflow, though, and CPoP's effectiveness for srasearch workflows may not extend to others.
This is, in fact, true for blast (see results in Figure~\ref{fig:app:blast}) where CPoP performs generally poorly for all CCRs tested and in one case yields a schedule with 1000 times the makespan of the schedule WBA produces!

These results are evidence that the traditional benchmarking approach to comparing algorithms is insufficient even for highly regular, application-specific problem instances.
They also have implications for the design of Workflow Management Systems (and other task scheduling systems in IoT, edge computing environments, etc.).
It may be reasonable for a WFMS to run a \textit{set} of scheduling algorithms that best covers the different types of client scientific workflows and computing systems.
For example, a WFMS designer might run \AAName and choose the three algorithms with the combined minimum maximum makespan ratio.
Exploring different methods for constructing and comparing such hybrid algorithms is an interesting topic for future work.

\section{Conclusion}\label{sec:conclusion}
In this paper, we presented \FrameworkName, a Python framework for running, evaluating, and comparing task scheduling algorithms.
We evaluated \NumAlgorithms of the scheduling algorithms implemented in \FrameworkName on \NumDatasets datasets and demonstrated how our proposed adversarial analysis method, \AAName, provides useful information that traditional benchmarking does not.
We showed that many algorithms that appear to perform well on benchmarking datasets can perform poorly on adversarially chosen problem instances, effectively presenting new lower bounds on the worst-case performance of algorithms compared to popular baselines.
We showed that \AAName's effectiveness extends even to highly restricted, application-specific scenarios.
We observed that there exist problem instances where HEFT, a popular task scheduling algorithm, can perform significantly worse than FastestNode, a very simple, generally quite poor scheduling algorithm.
Applying this kind of analysis to other pairs of algorithms and exploring how it might be automated is an important and interesting direction for future work.

There are many other directions for future work as well.
First, we plan to extend \FrameworkName to include more algorithms and datasets.
Another logical next step is to extend \FrameworkName and \AAName to support other problem variants (e.g., throughput maximization, energy minimization, etc.).
We also plan to develop a framework for publishing the problem instances identified by \AAName so that other researchers can use them to evaluate their own algorithms.

Due to our particular interest in task scheduling for dynamic environments, we plan to add support for stochastic problem instances (with stochastic task costs, data sizes, computation speeds, and communication costs).
It would also be interesting to explore other meta-heuristics for adversarial analysis (e.g., genetic algorithms) and other performance metrics (e.g., throughput, energy consumption, cost, etc.).
Some other directions that could be interesting to explore include ensemble methods (e.g., running multiple algorithms and choosing the best schedule) and online scheduling (e.g., scheduling tasks as they arrive).
Finally, the application-specific scenario we explored in Section~\ref{sec:app-specific} suggests that an exploration into different methodologies for constructing and comparing hybrid scheduling algorithms using \AAName might be fruitful.

\FloatBarrier

\bibliographystyle{IEEEtran}
\bibliography{IEEEabrv,pisa}

\newpage
\appendix
\section{Appendix}

\subsection{Application-Specific PISA Results}\label{sec:app-specific-full}
Here we include the Section~\ref{sec:app-specific} benchmarking and \AAName results for all evaluated scientific workflows.
For each CCR, the top row shows benchmarking results, where gradients indicate makespan ratios on different problem instances in the dataset.
All other cells indicate the highest makespan ratio yielding problem instance discovered by \AAName.

\begin{figure*}[!htb]
    \centering
    
    \begin{subfigure}[b]{0.5\textwidth}
        \centering
        \includegraphics[width=\textwidth]{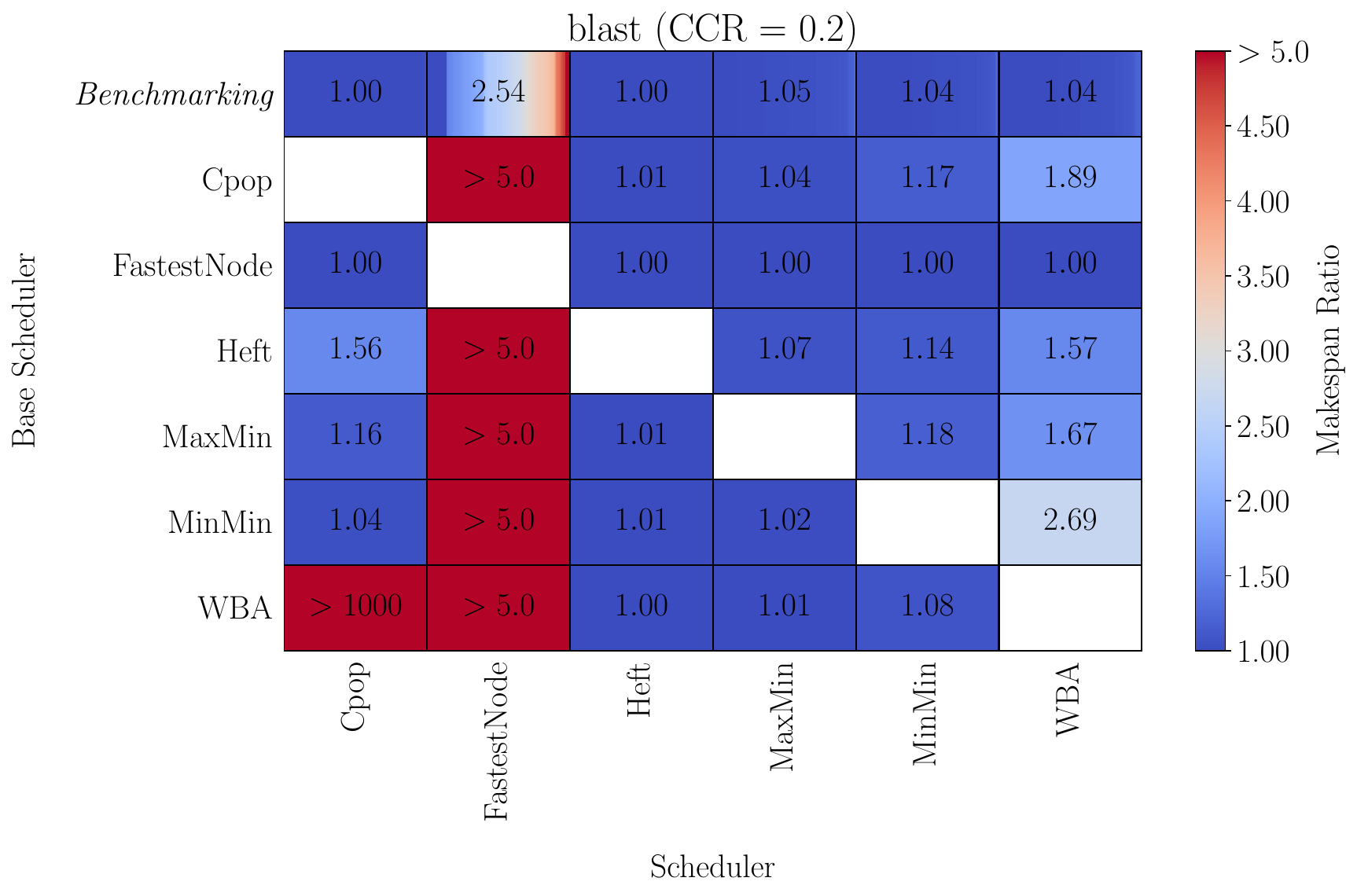}
    \end{subfigure}%
    \hfill
    \begin{subfigure}[b]{0.5\textwidth}
        \centering
        \includegraphics[width=\textwidth]{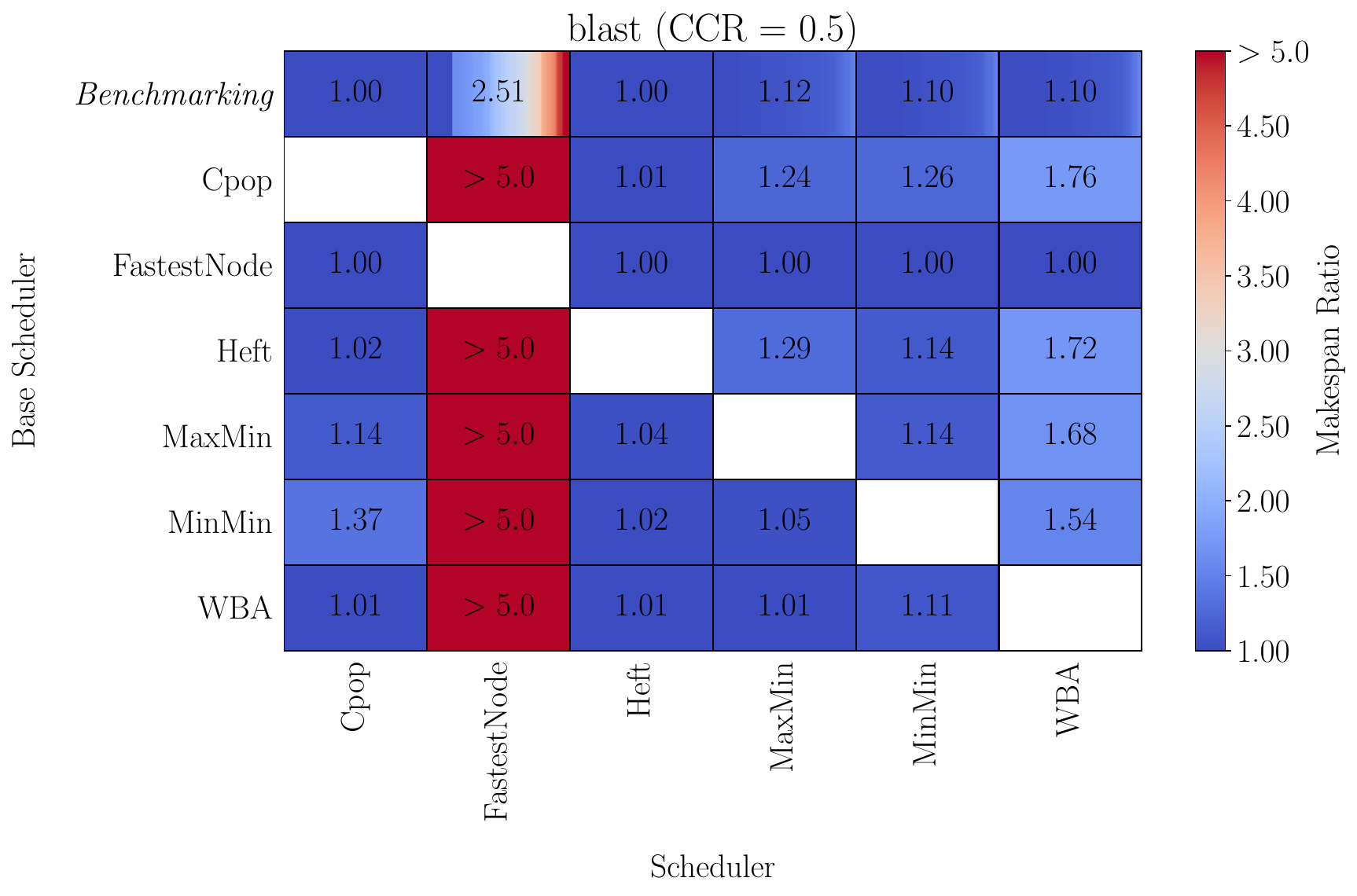}
    \end{subfigure}%
    
    \vspace{0.25cm}%
    
    \begin{subfigure}[b]{0.5\textwidth}
        \centering
        \includegraphics[width=\textwidth]{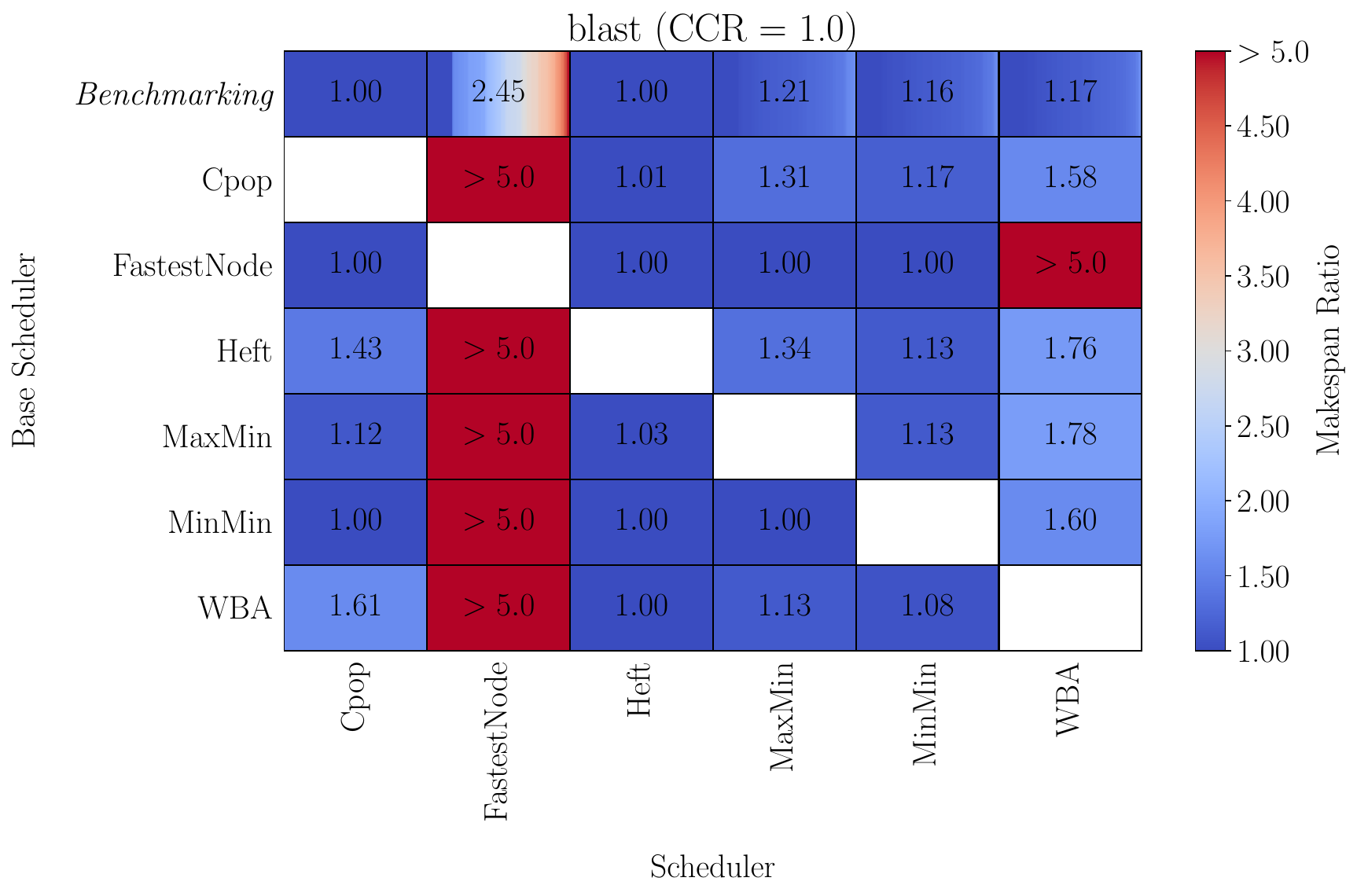}
    \end{subfigure}%
    \hfill
    \begin{subfigure}[b]{0.5\textwidth}
        \centering
        \includegraphics[width=\textwidth]{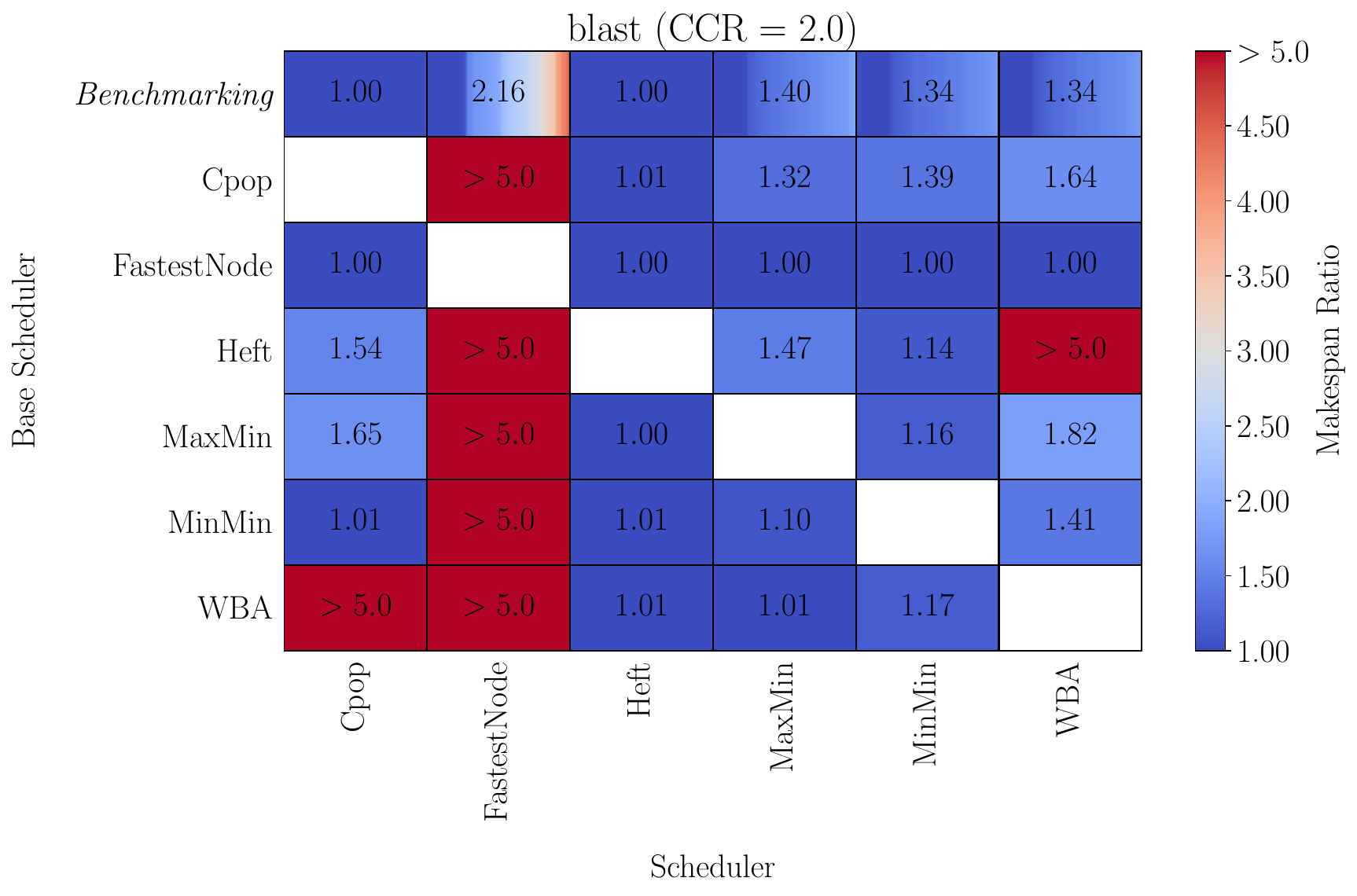}
    \end{subfigure}%
    
    \vspace{0.25cm}%
    
    \begin{subfigure}[b]{0.5\textwidth}
        \centering
        \includegraphics[width=\textwidth]{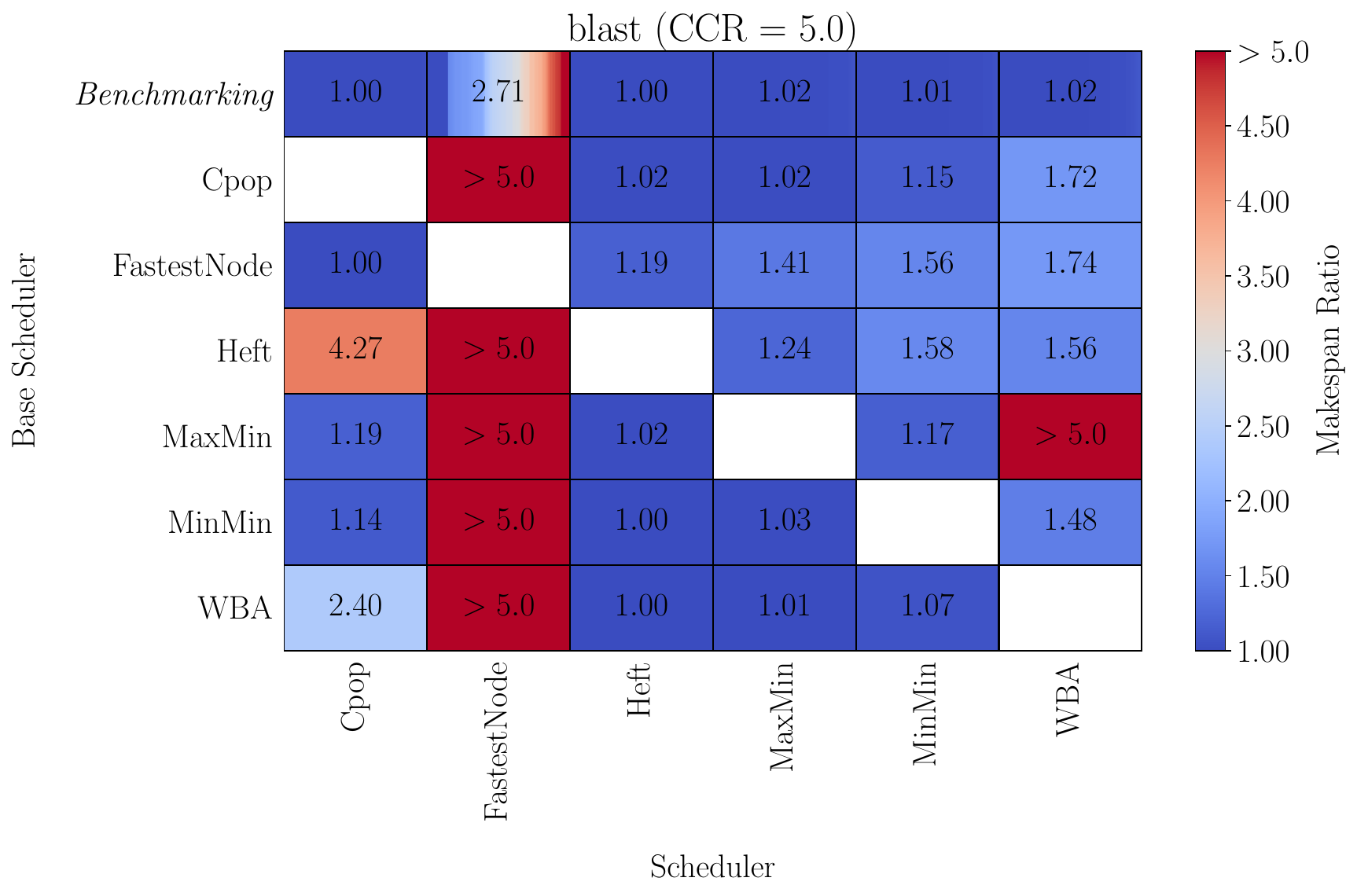}
    \end{subfigure}%
    \caption{Results for the blast scientific workflow.}
    \label{fig:apx-app:blast}
\end{figure*}

\begin{figure*}[!htb]
    \centering
    
    \begin{subfigure}[b]{0.5\textwidth}
        \centering
        \includegraphics[width=\textwidth]{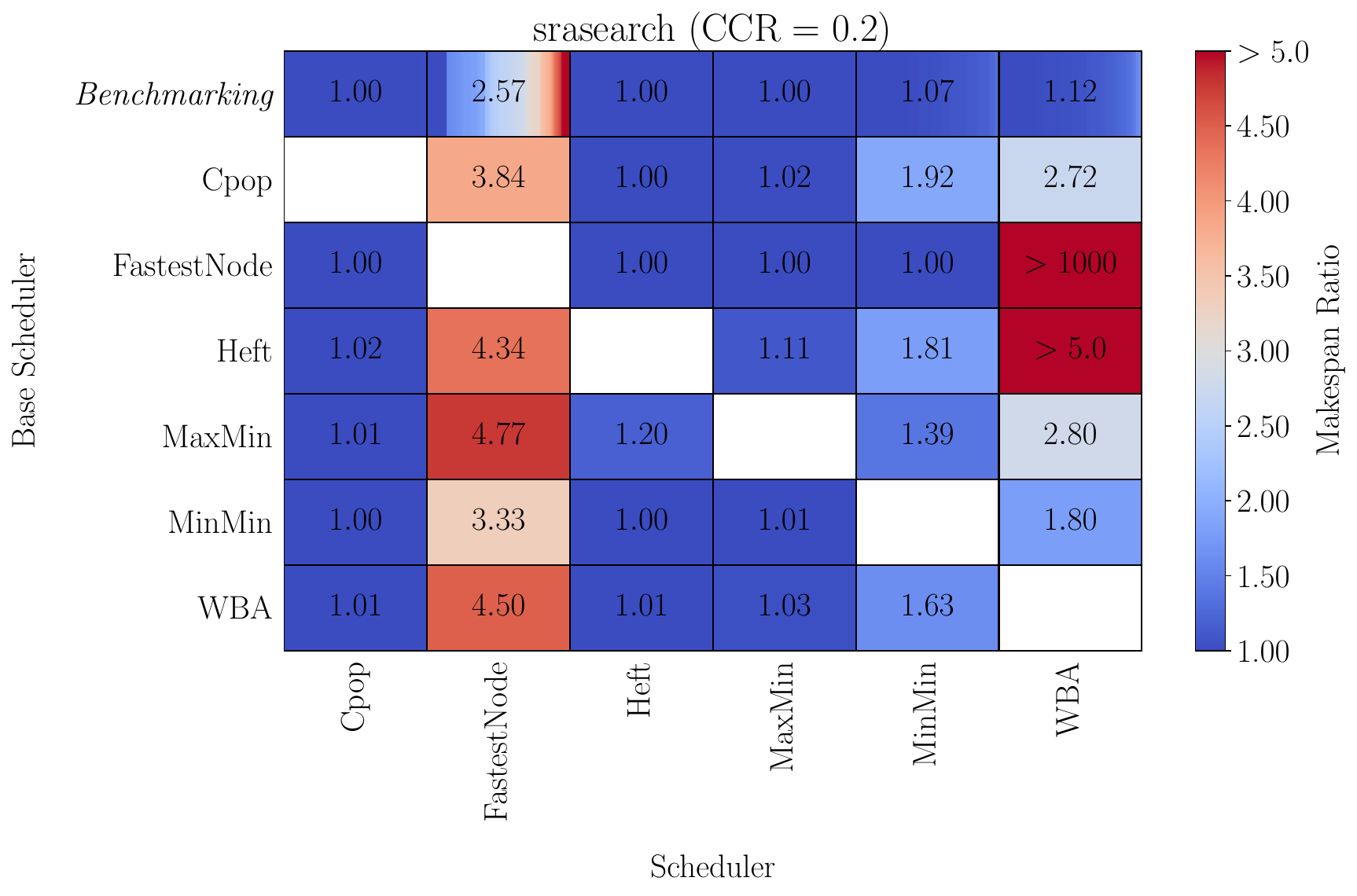}
    \end{subfigure}%
    \hfill
    \begin{subfigure}[b]{0.5\textwidth}
        \centering
        \includegraphics[width=\textwidth]{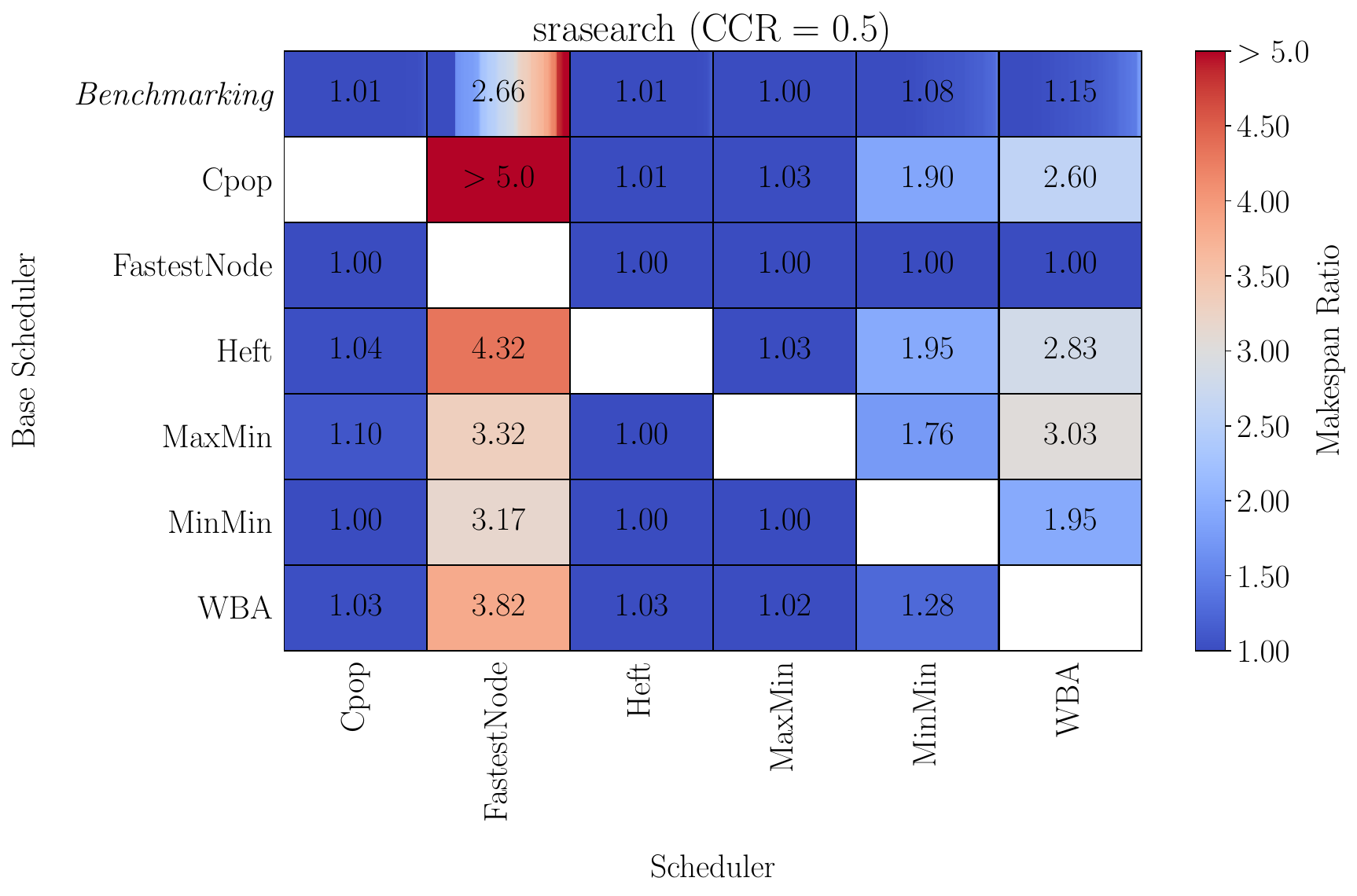}
    \end{subfigure}%
    
    \vspace{0.25cm}%
    
    \begin{subfigure}[b]{0.5\textwidth}
        \centering
        \includegraphics[width=\textwidth]{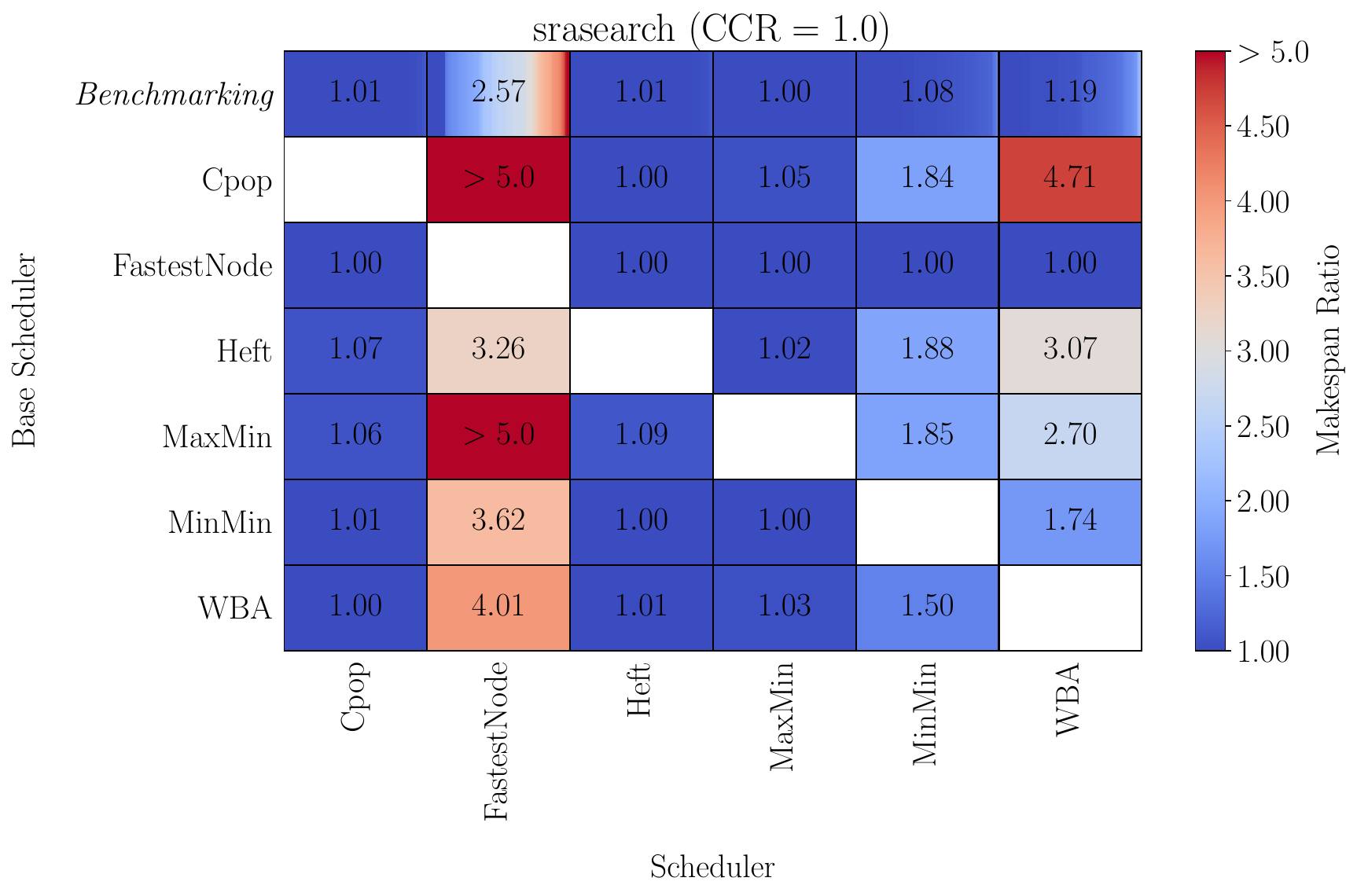}
    \end{subfigure}%
    \hfill
    \begin{subfigure}[b]{0.5\textwidth}
        \centering
        \includegraphics[width=\textwidth]{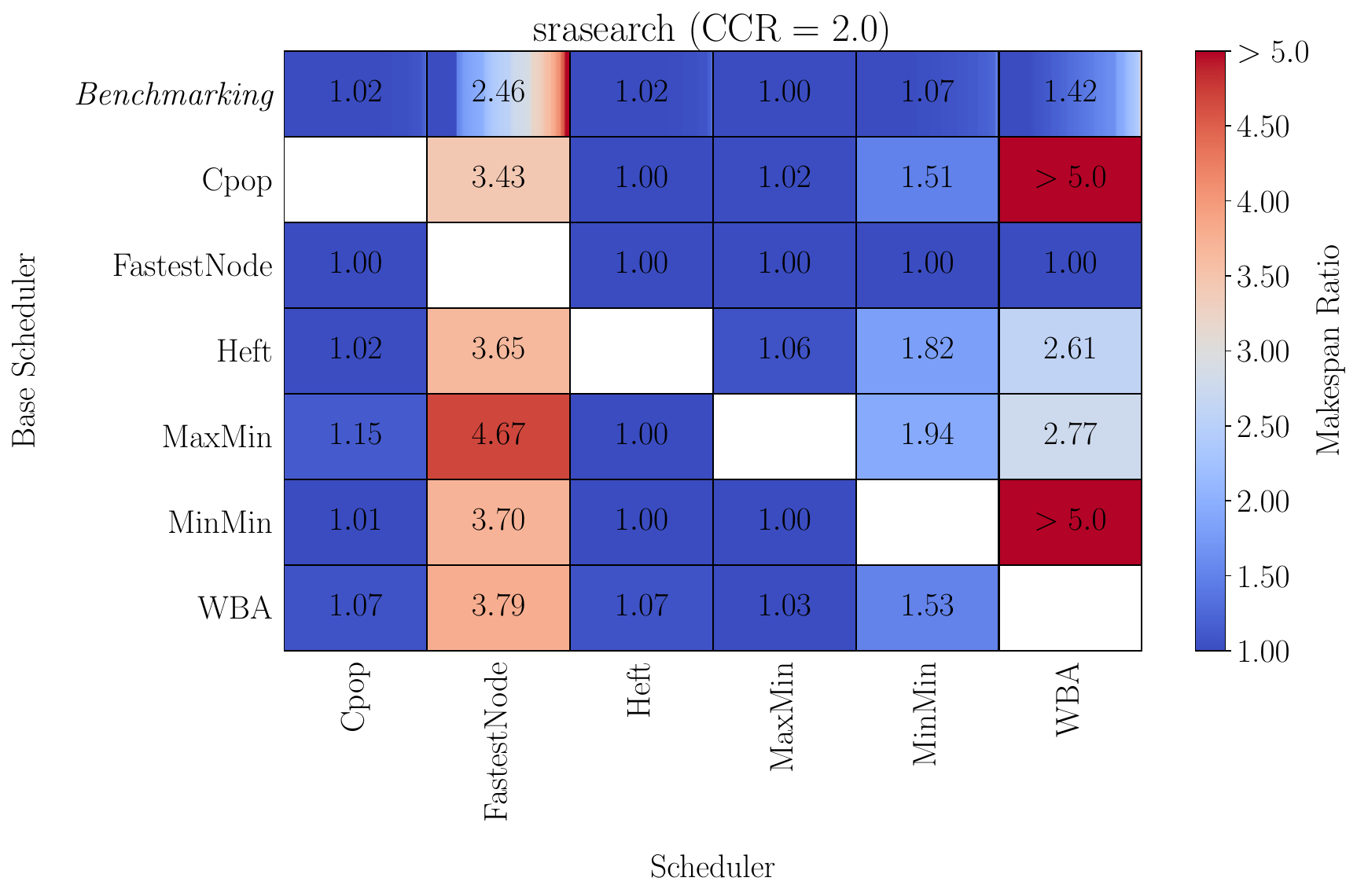}
    \end{subfigure}%
    
    \vspace{0.25cm}%
    
    \begin{subfigure}[b]{0.5\textwidth}
        \centering
        \includegraphics[width=\textwidth]{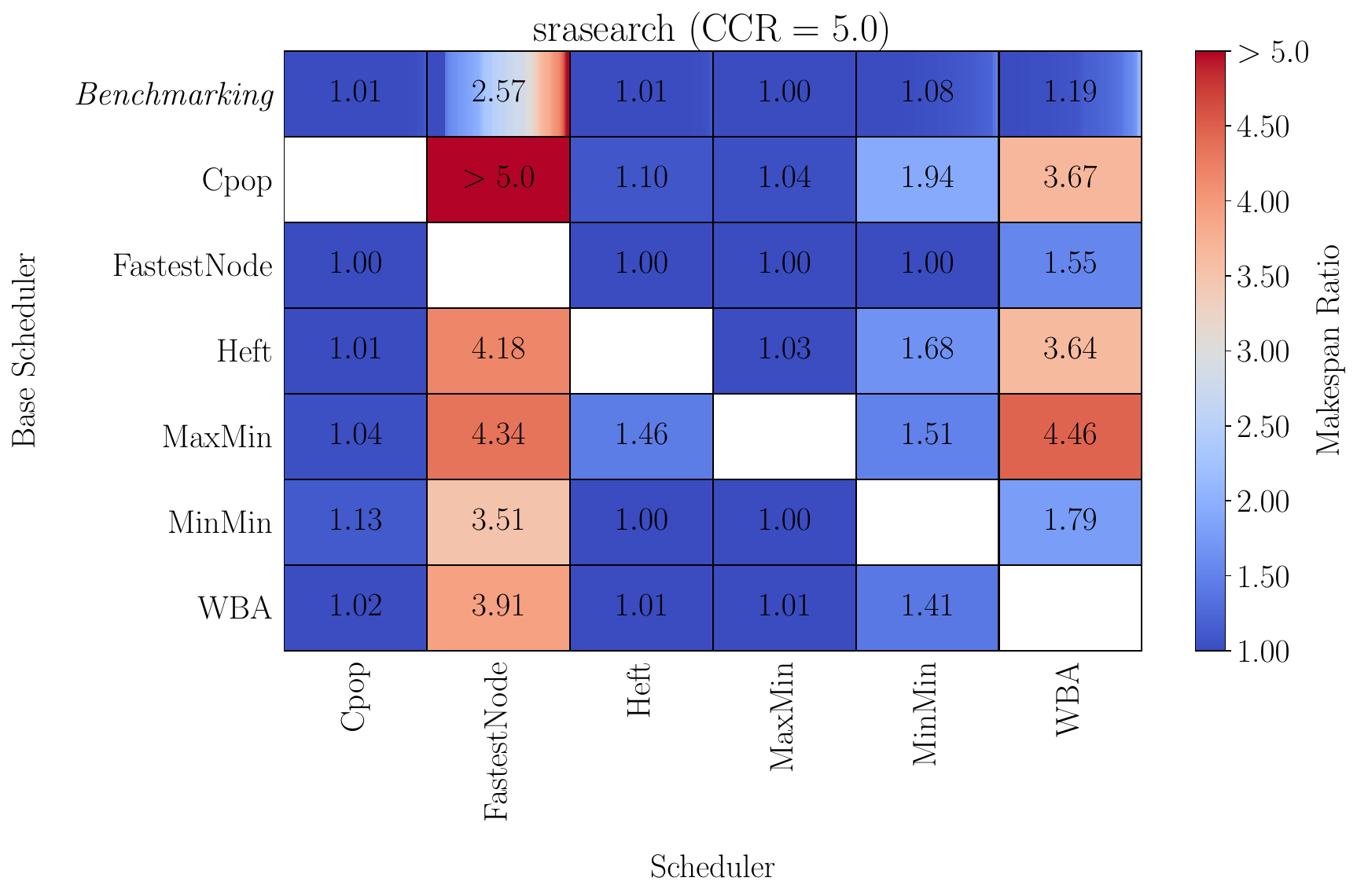}
    \end{subfigure}%
    \caption{Results for the srasearch scientific workflow.}
    \label{fig:apx-app:srasearch}
\end{figure*}

\begin{figure*}[!htb]
    \centering
    
    \begin{subfigure}[b]{0.5\textwidth}
        \centering
        \includegraphics[width=\textwidth]{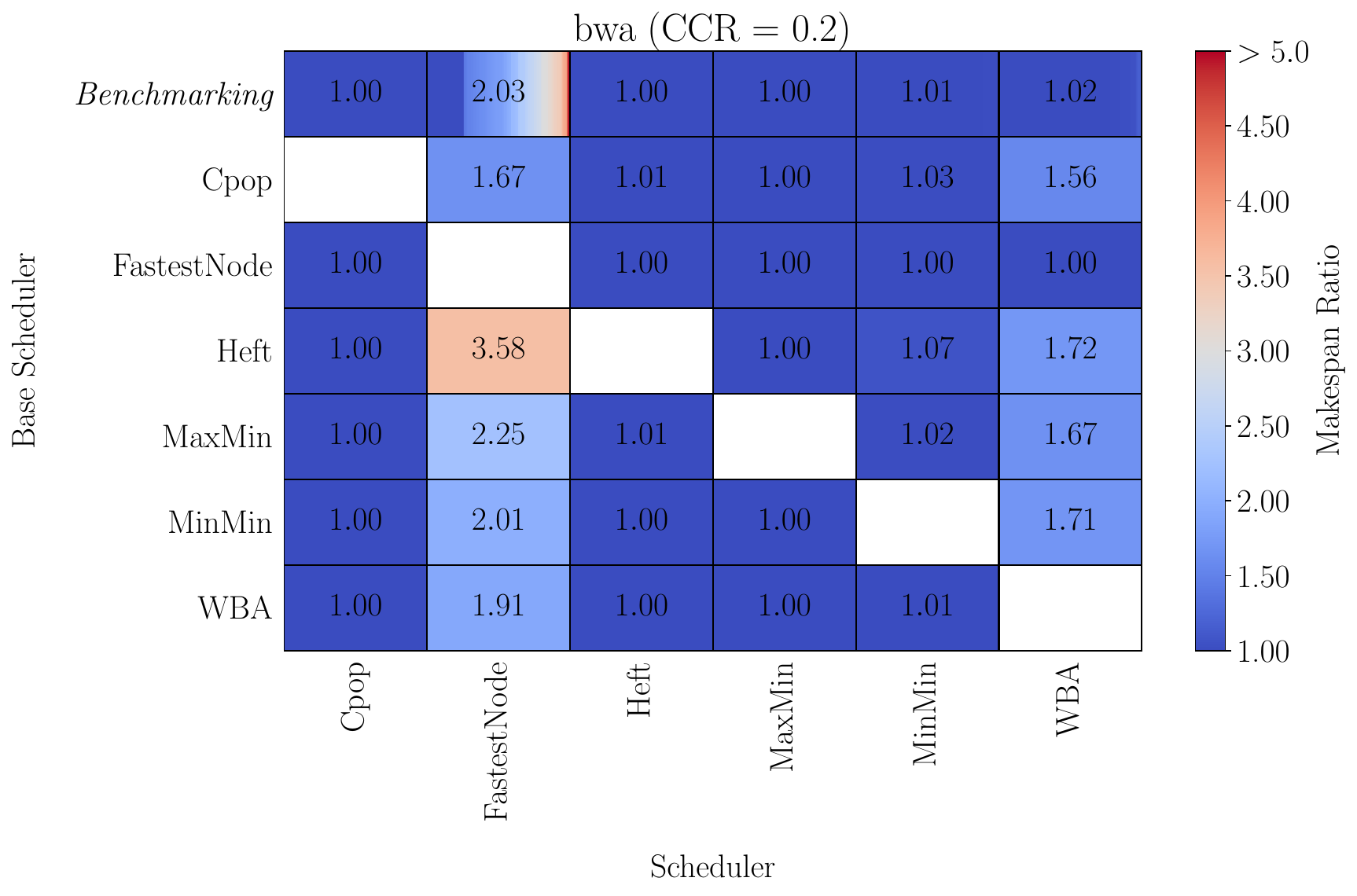}
    \end{subfigure}%
    \hfill
    \begin{subfigure}[b]{0.5\textwidth}
        \centering
        \includegraphics[width=\textwidth]{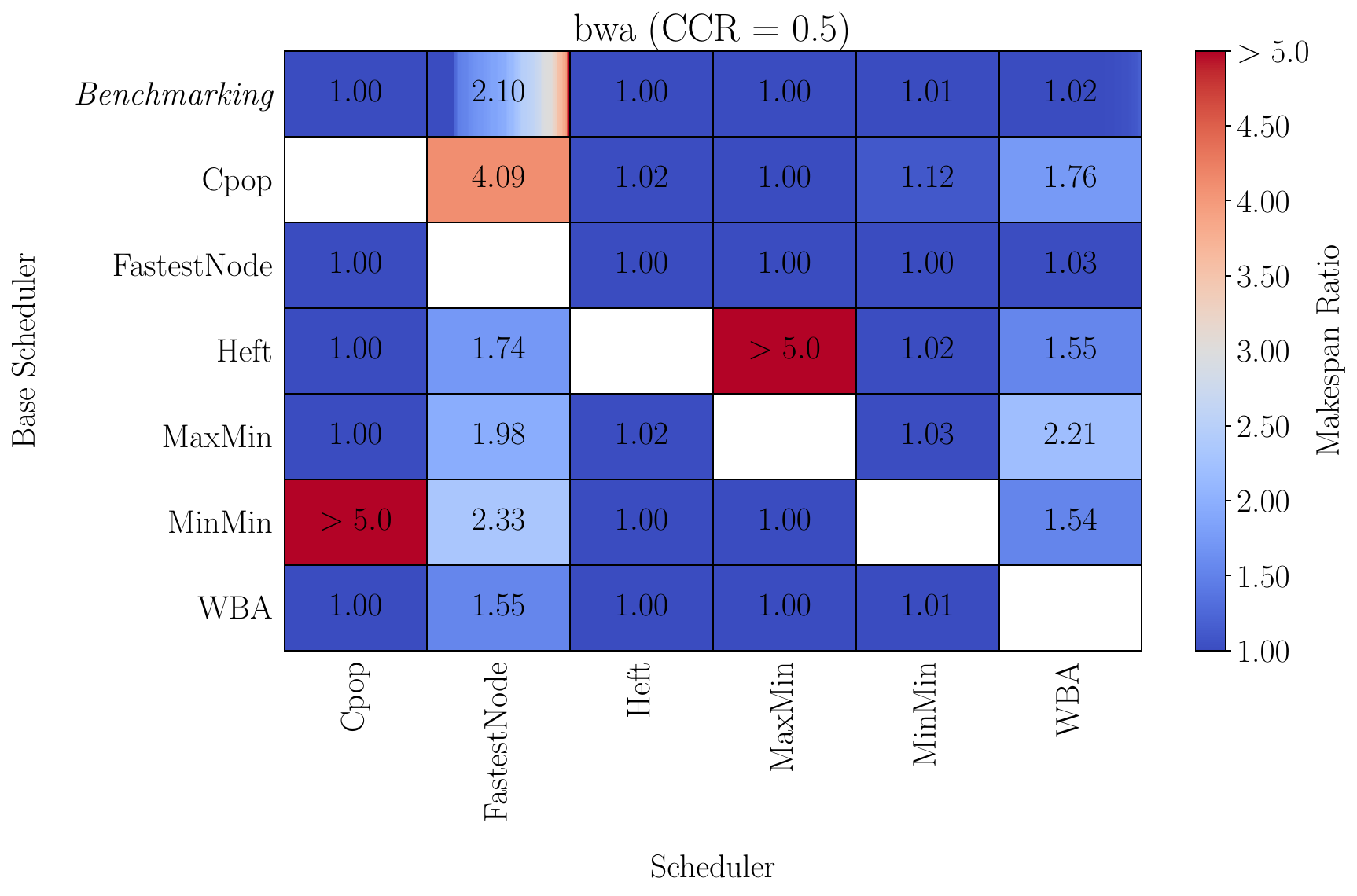}
    \end{subfigure}%
    
    \vspace{0.25cm}%
    
    \begin{subfigure}[b]{0.5\textwidth}
        \centering
        \includegraphics[width=\textwidth]{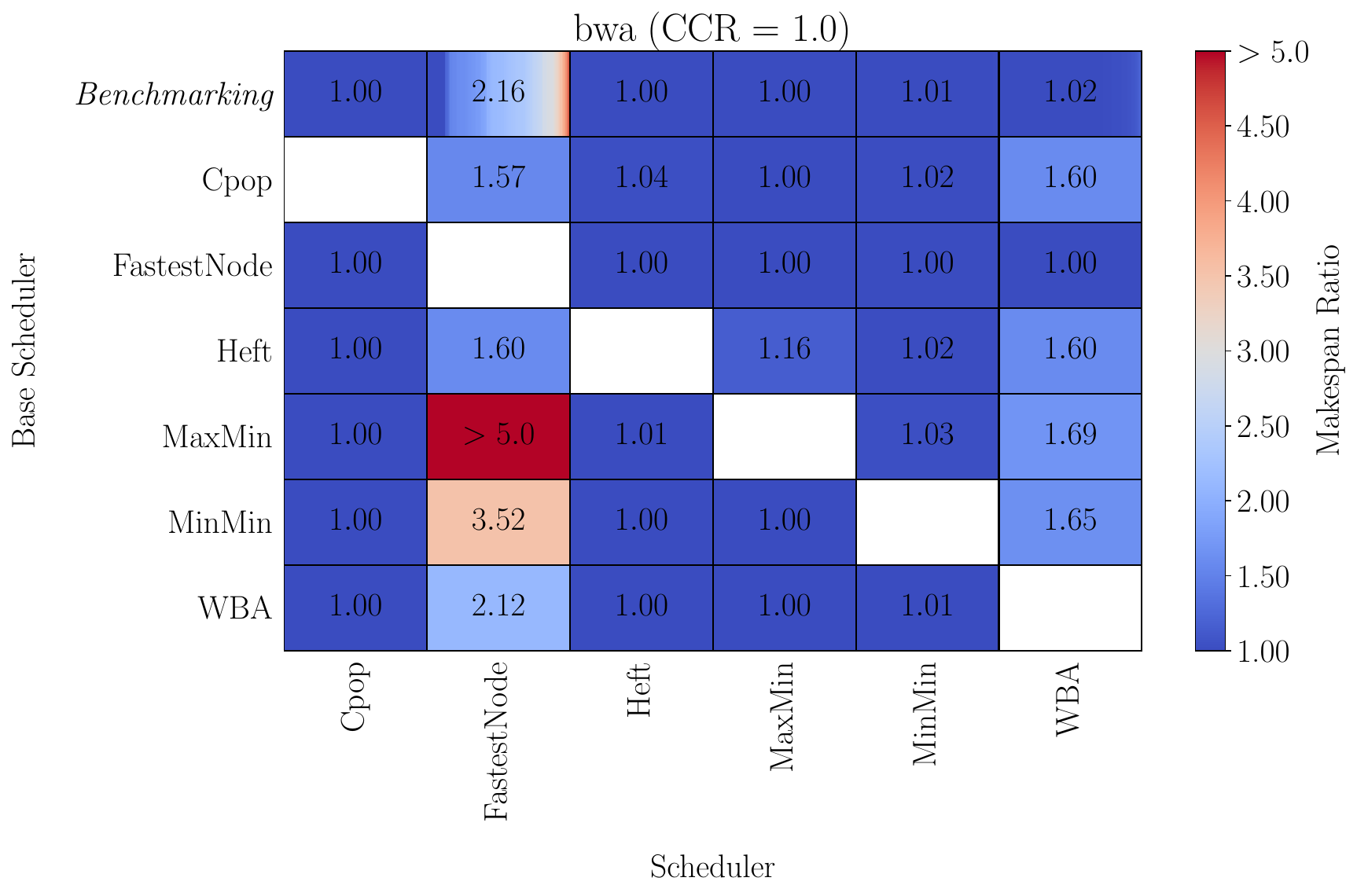}
    \end{subfigure}%
    \hfill
    \begin{subfigure}[b]{0.5\textwidth}
        \centering
        \includegraphics[width=\textwidth]{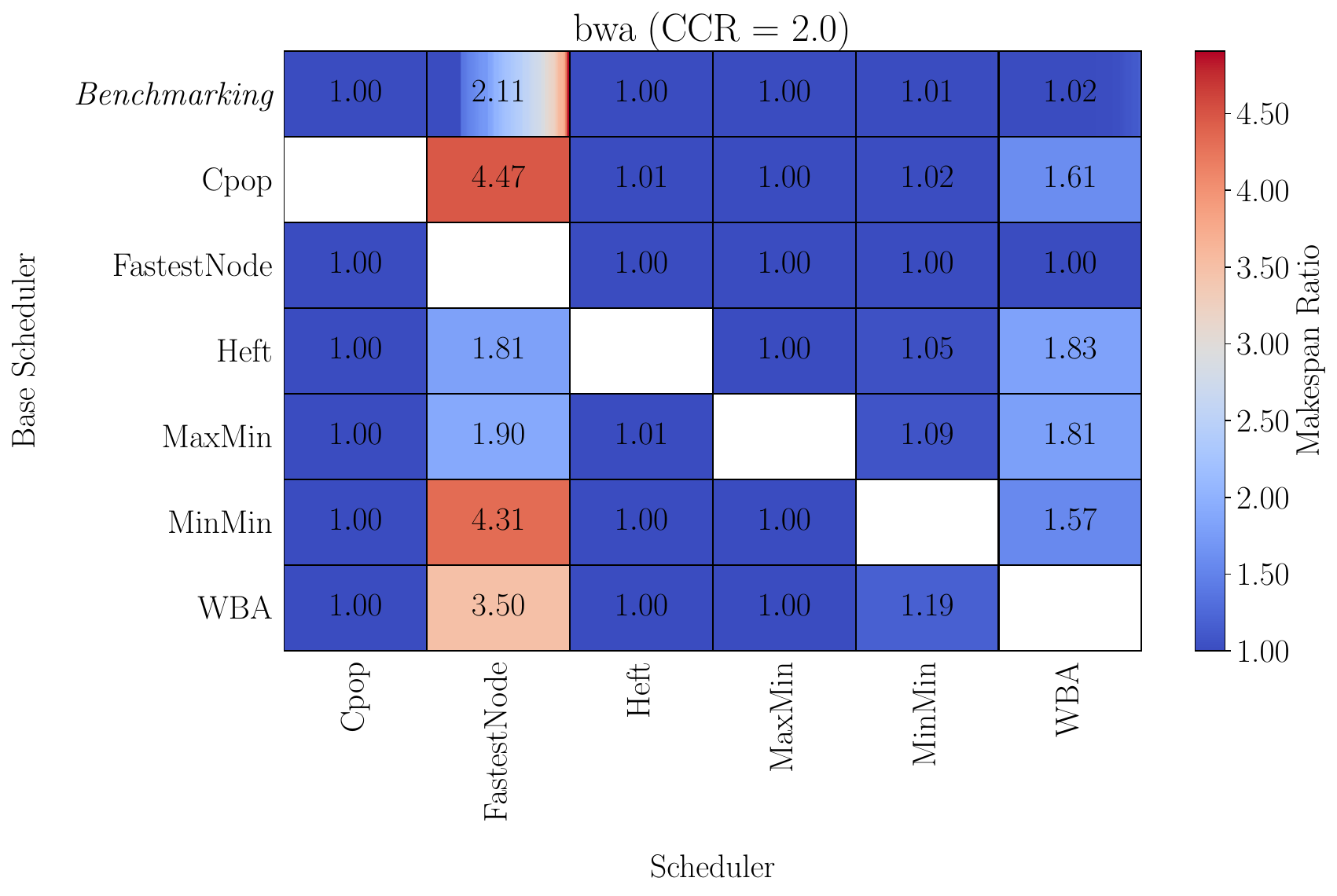}
    \end{subfigure}%
    
    \vspace{0.25cm}%
    
    \begin{subfigure}[b]{0.5\textwidth}
        \centering
        \includegraphics[width=\textwidth]{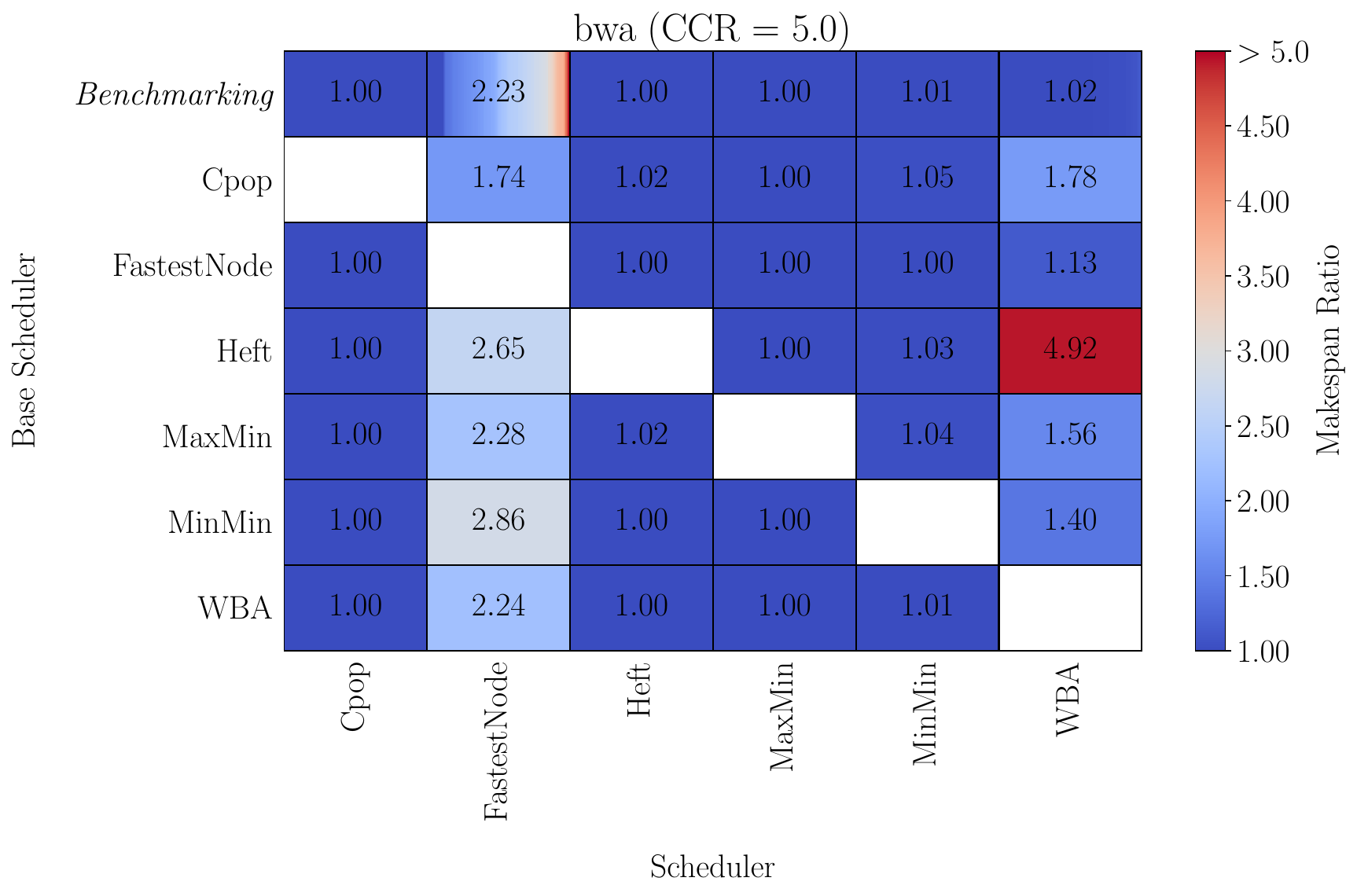}
    \end{subfigure}%
    \caption{Results for the bwa scientific workflow.}
    \label{fig:apx-app:bwa}
\end{figure*}

\begin{figure*}[!htb]
    \centering
    
    \begin{subfigure}[b]{0.5\textwidth}
        \centering
        \includegraphics[width=\textwidth]{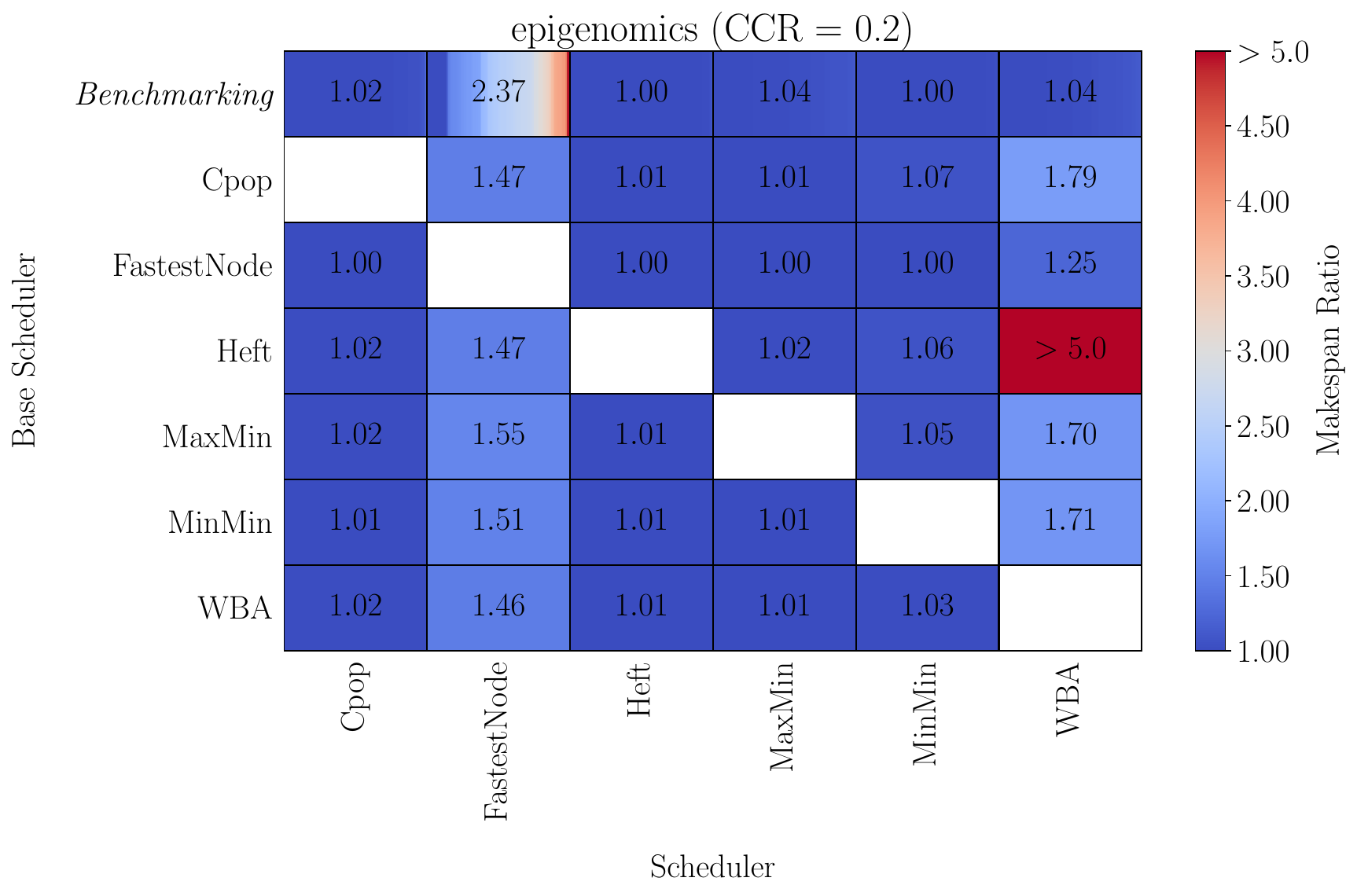}
    \end{subfigure}%
    \hfill
    \begin{subfigure}[b]{0.5\textwidth}
        \centering
        \includegraphics[width=\textwidth]{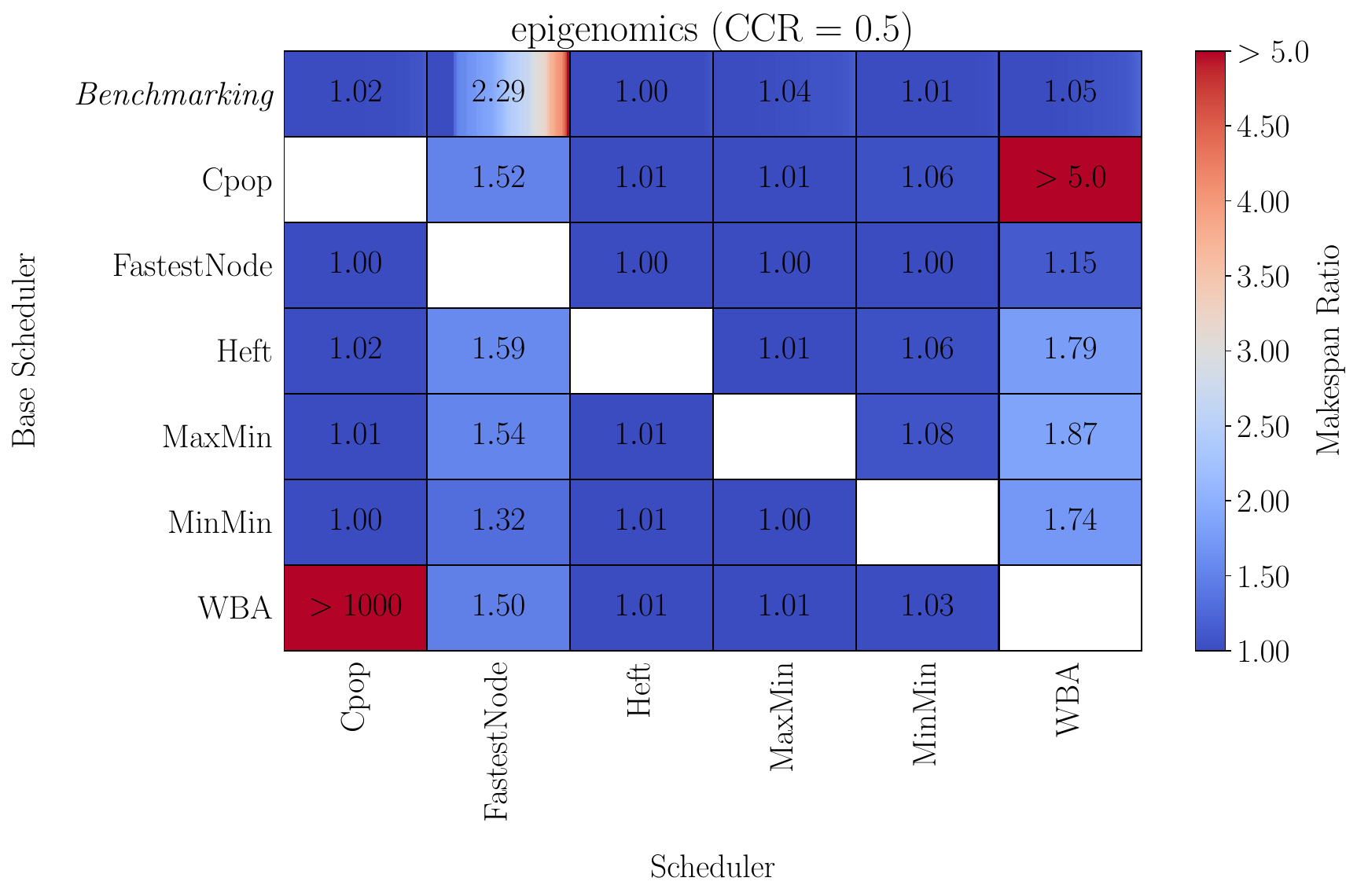}
    \end{subfigure}%
    
    \vspace{0.25cm}%
    
    \begin{subfigure}[b]{0.5\textwidth}
        \centering
        \includegraphics[width=\textwidth]{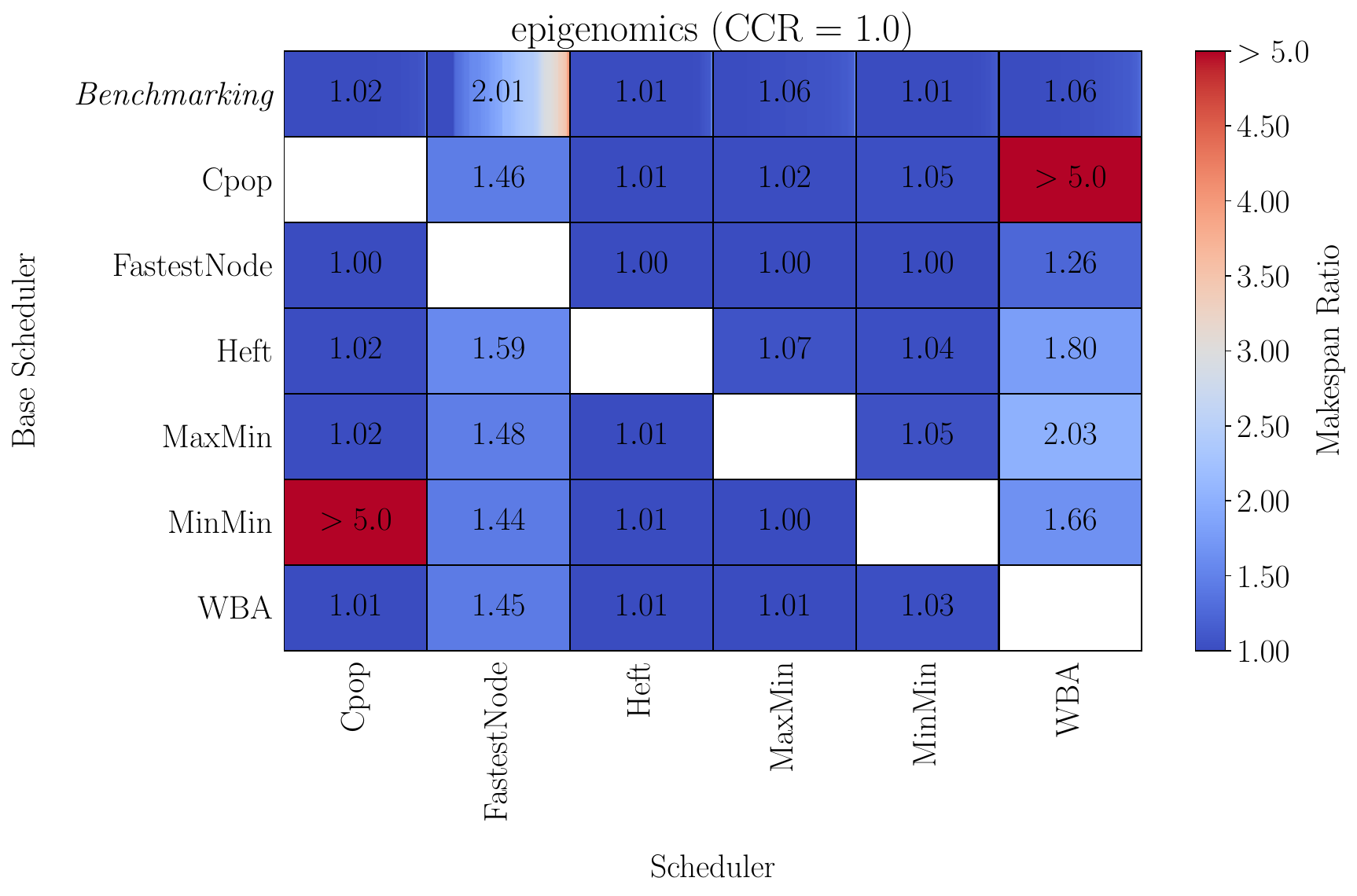}
    \end{subfigure}%
    \hfill
    \begin{subfigure}[b]{0.5\textwidth}
        \centering
        \includegraphics[width=\textwidth]{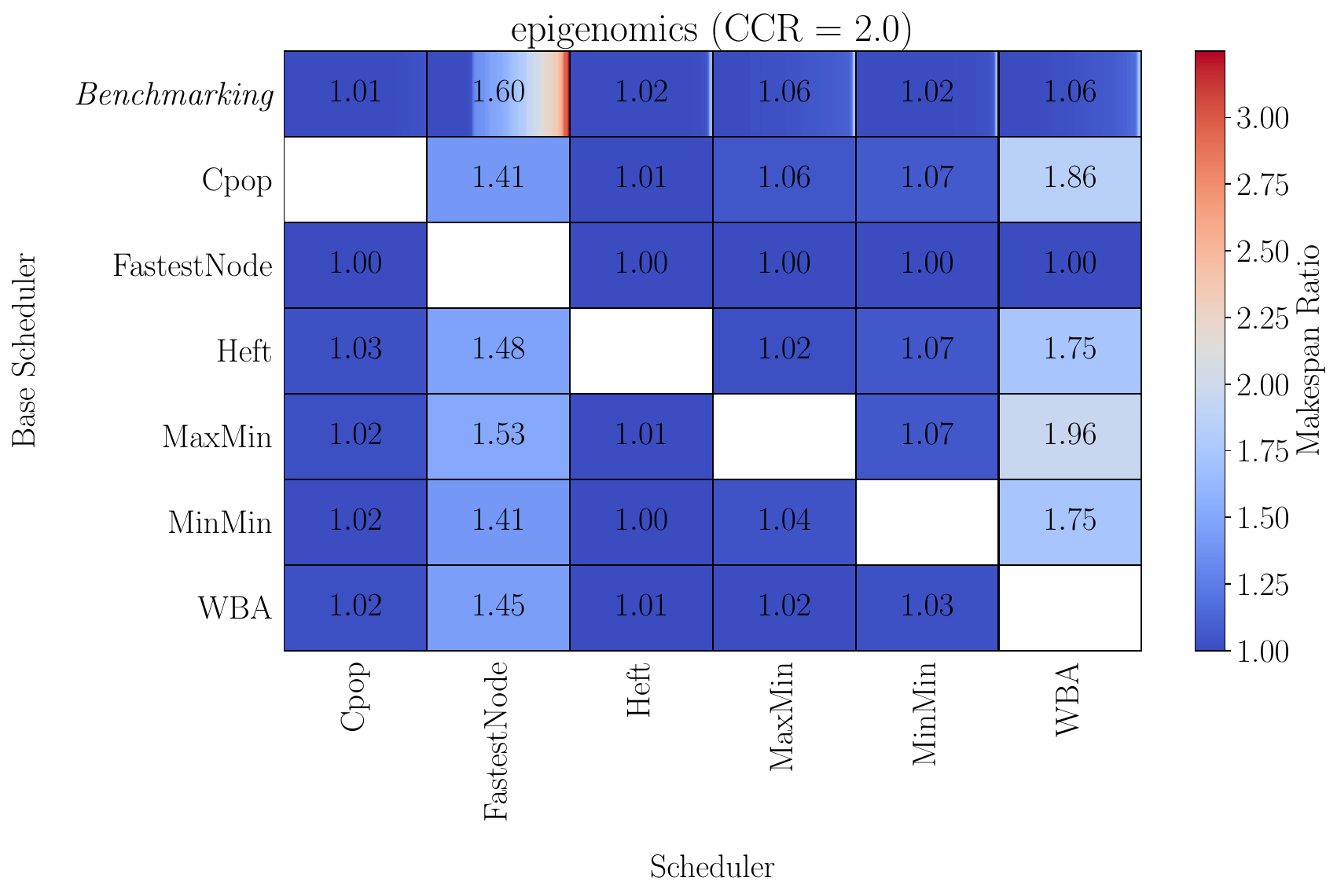}
    \end{subfigure}%
    
    \vspace{0.25cm}%
    
    \begin{subfigure}[b]{0.5\textwidth}
        \centering
        \includegraphics[width=\textwidth]{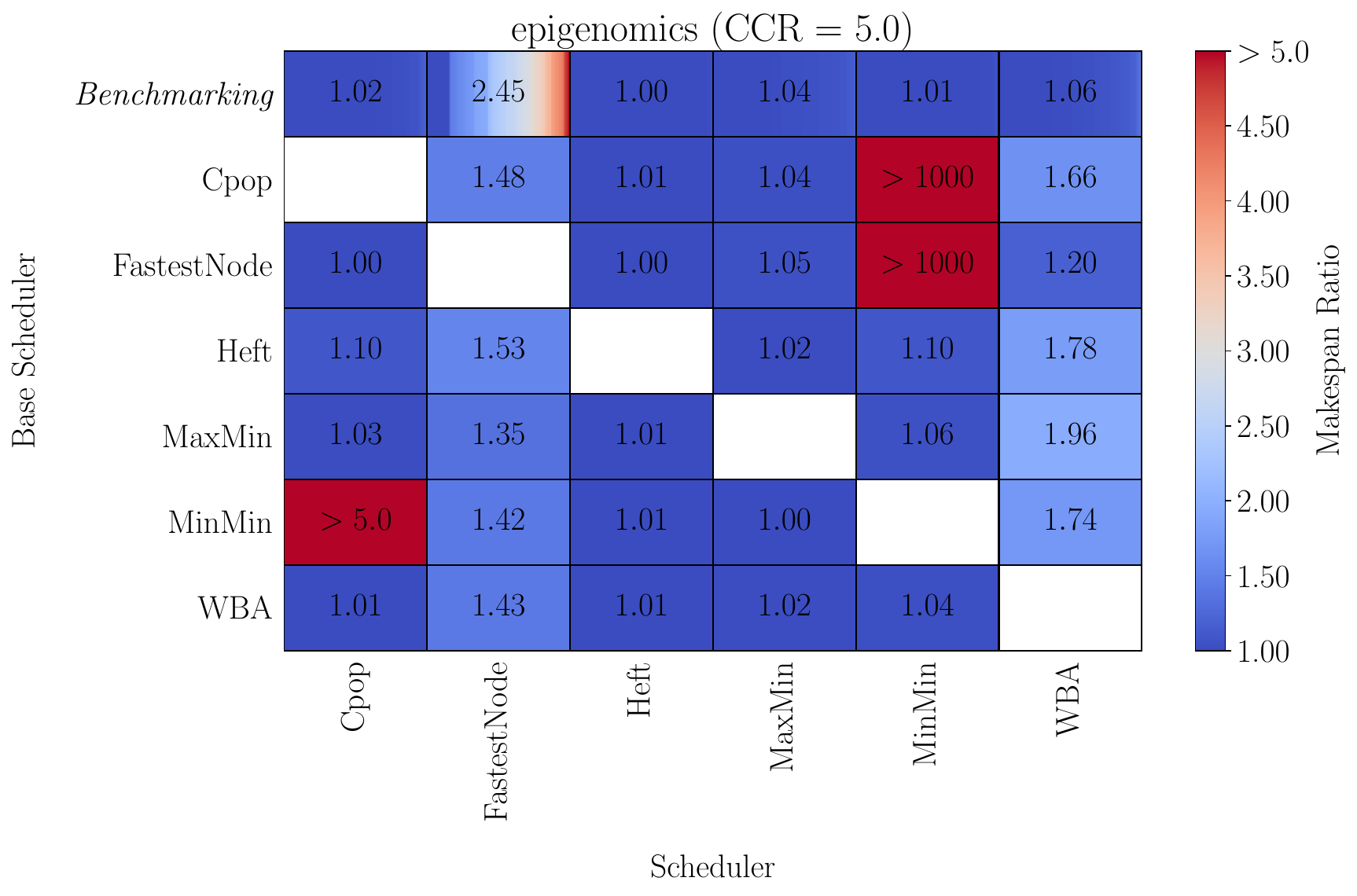}
    \end{subfigure}%
    \caption{Results for the epigenomics scientific workflow.}
    \label{fig:apx-app:epigenomics}
\end{figure*}

\begin{figure*}[!htb]
    \centering
    
    \begin{subfigure}[b]{0.5\textwidth}
        \centering
        \includegraphics[width=\textwidth]{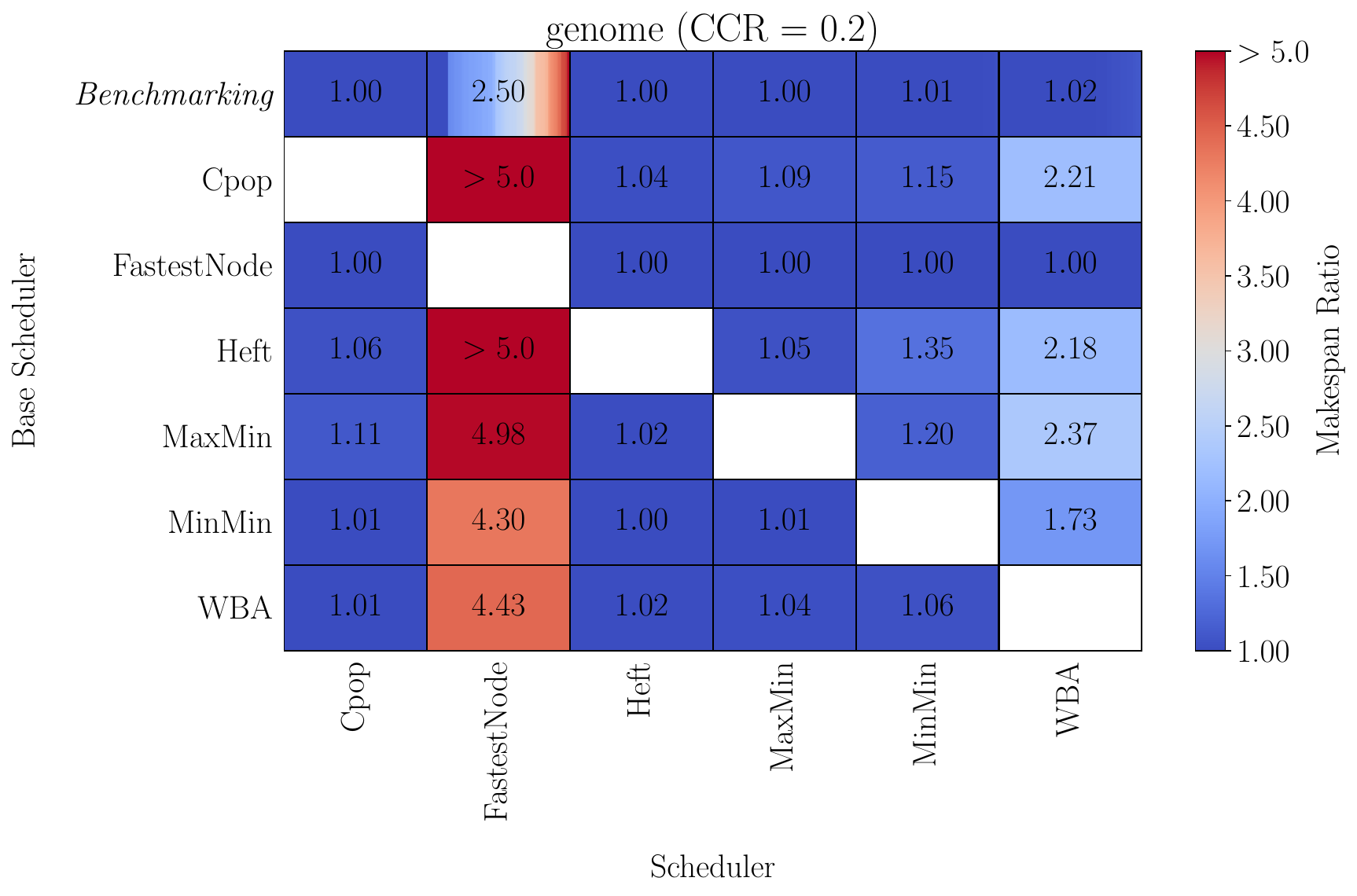}
    \end{subfigure}%
    \hfill
    \begin{subfigure}[b]{0.5\textwidth}
        \centering
        \includegraphics[width=\textwidth]{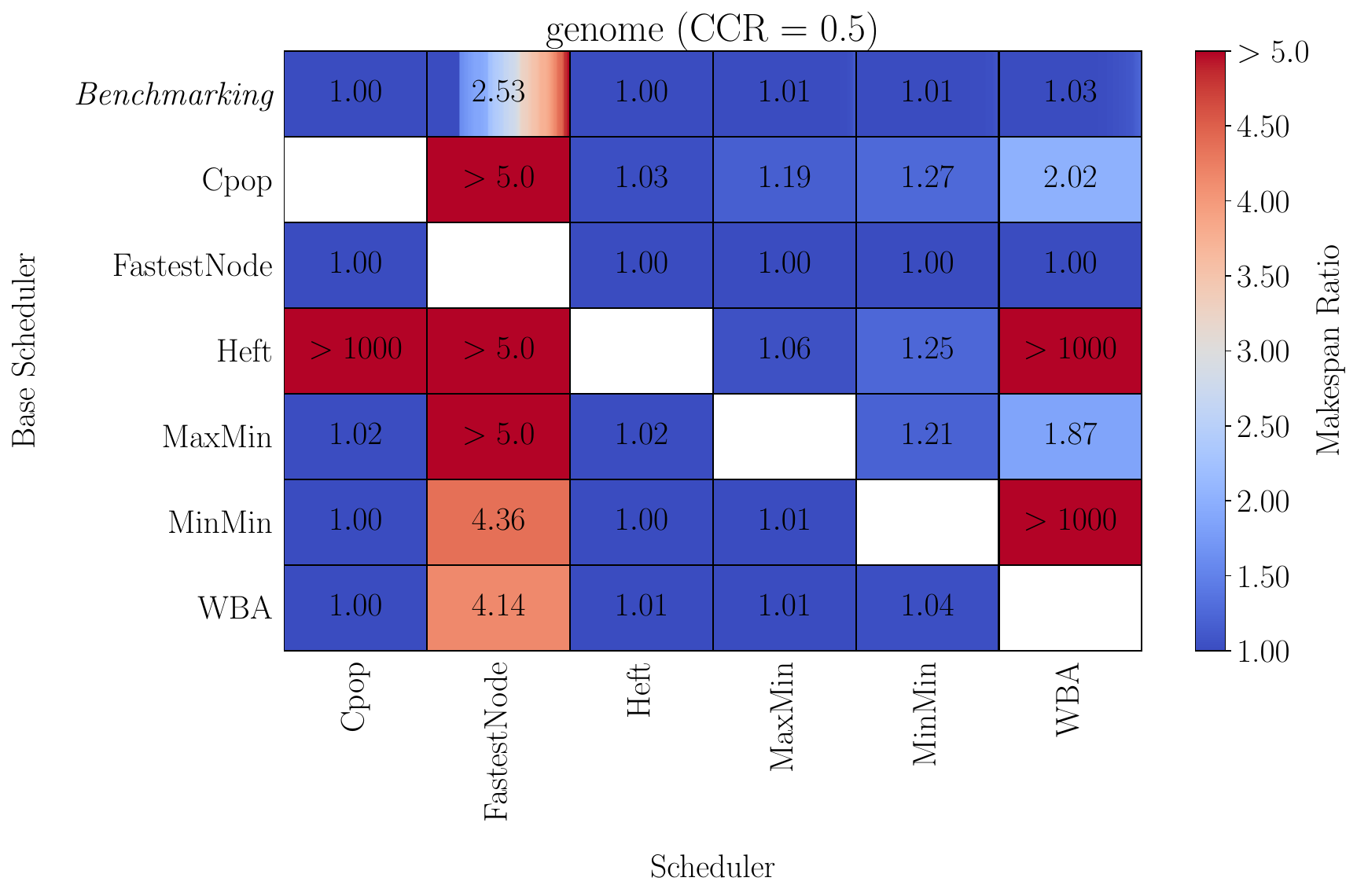}
    \end{subfigure}%
    
    \vspace{0.25cm}%
    
    \begin{subfigure}[b]{0.5\textwidth}
        \centering
        \includegraphics[width=\textwidth]{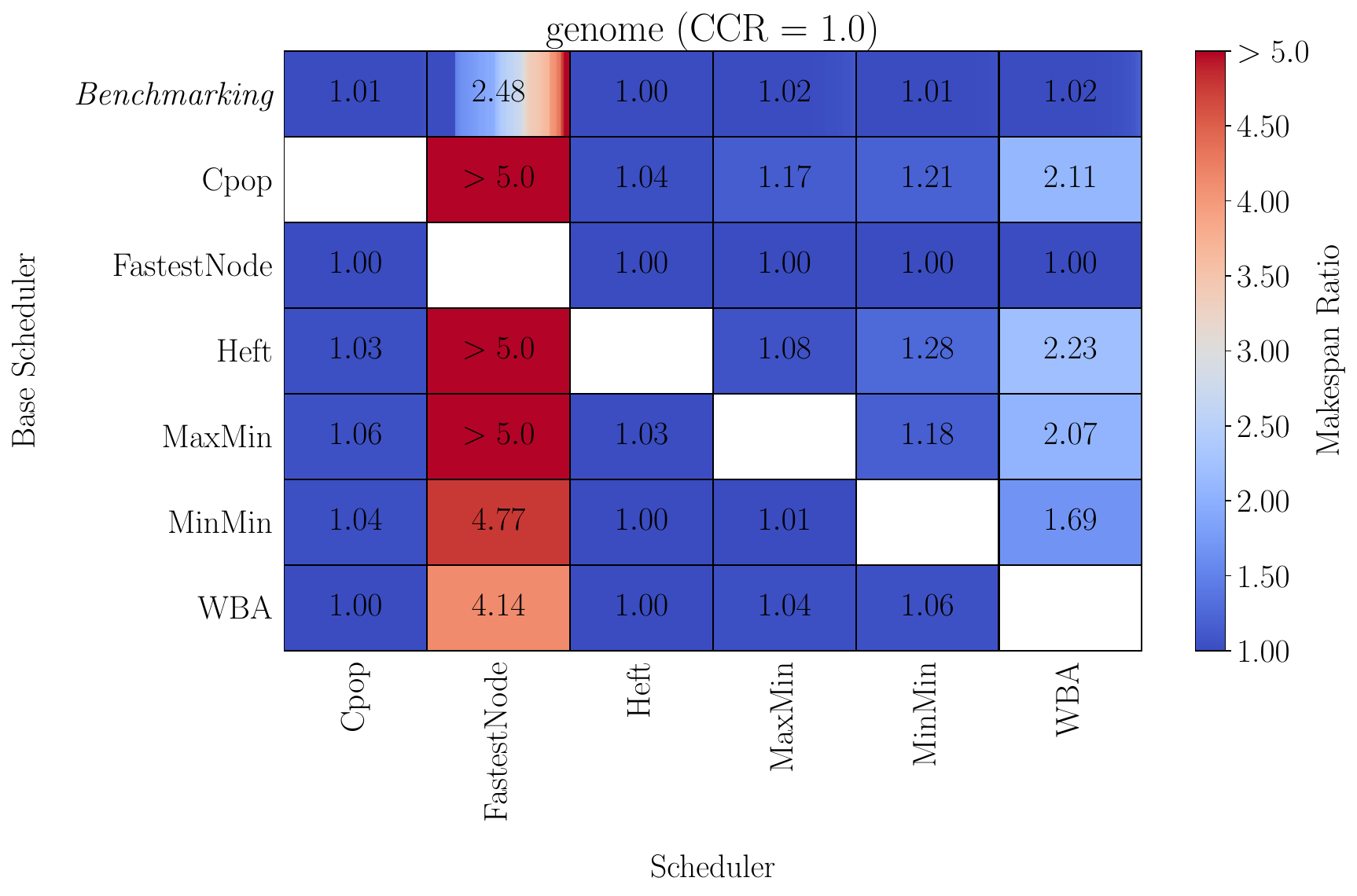}
    \end{subfigure}%
    \hfill
    \begin{subfigure}[b]{0.5\textwidth}
        \centering
        \includegraphics[width=\textwidth]{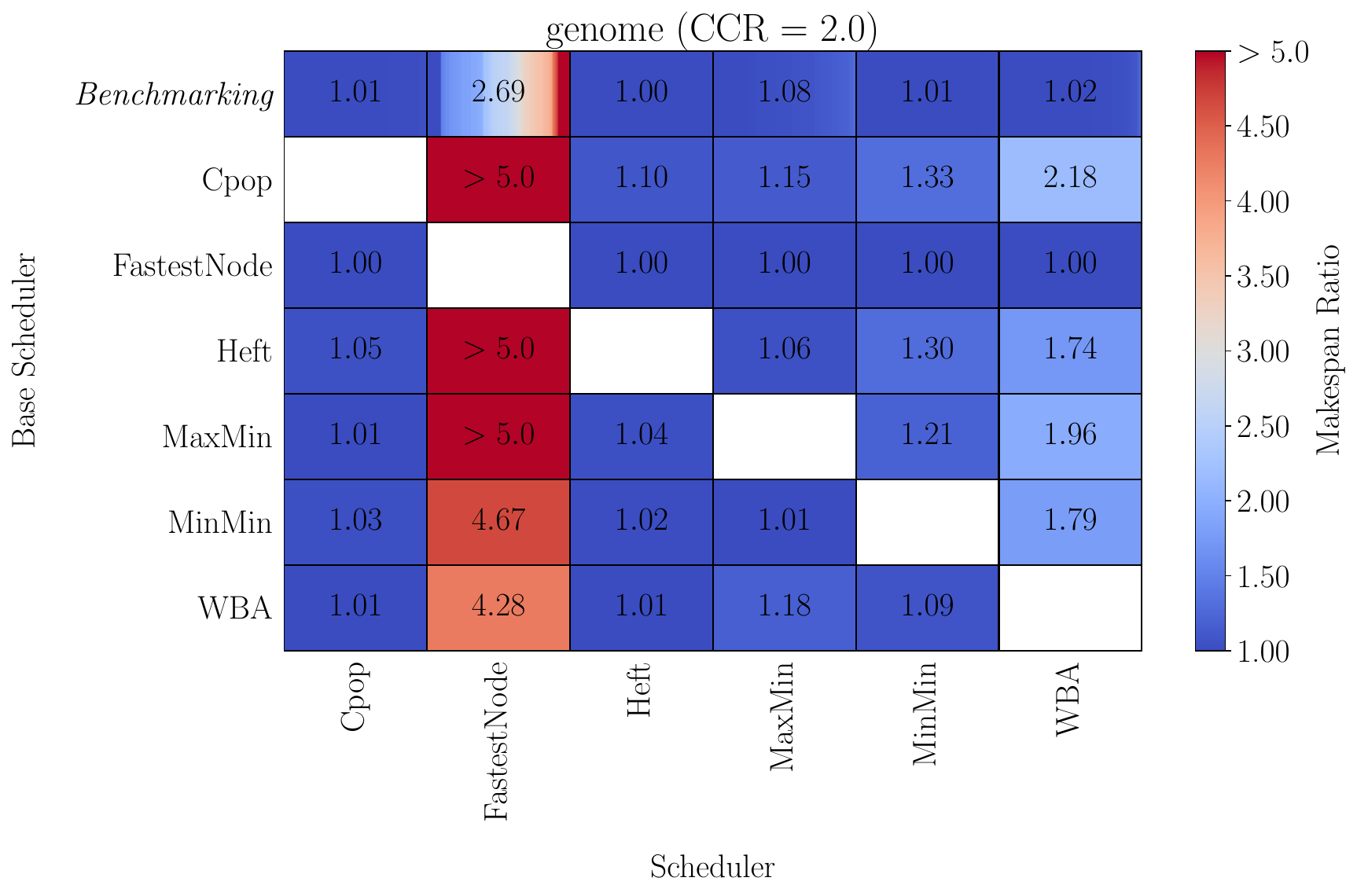}
    \end{subfigure}%
    
    \vspace{0.25cm}%
    
    \begin{subfigure}[b]{0.5\textwidth}
        \centering
        \includegraphics[width=\textwidth]{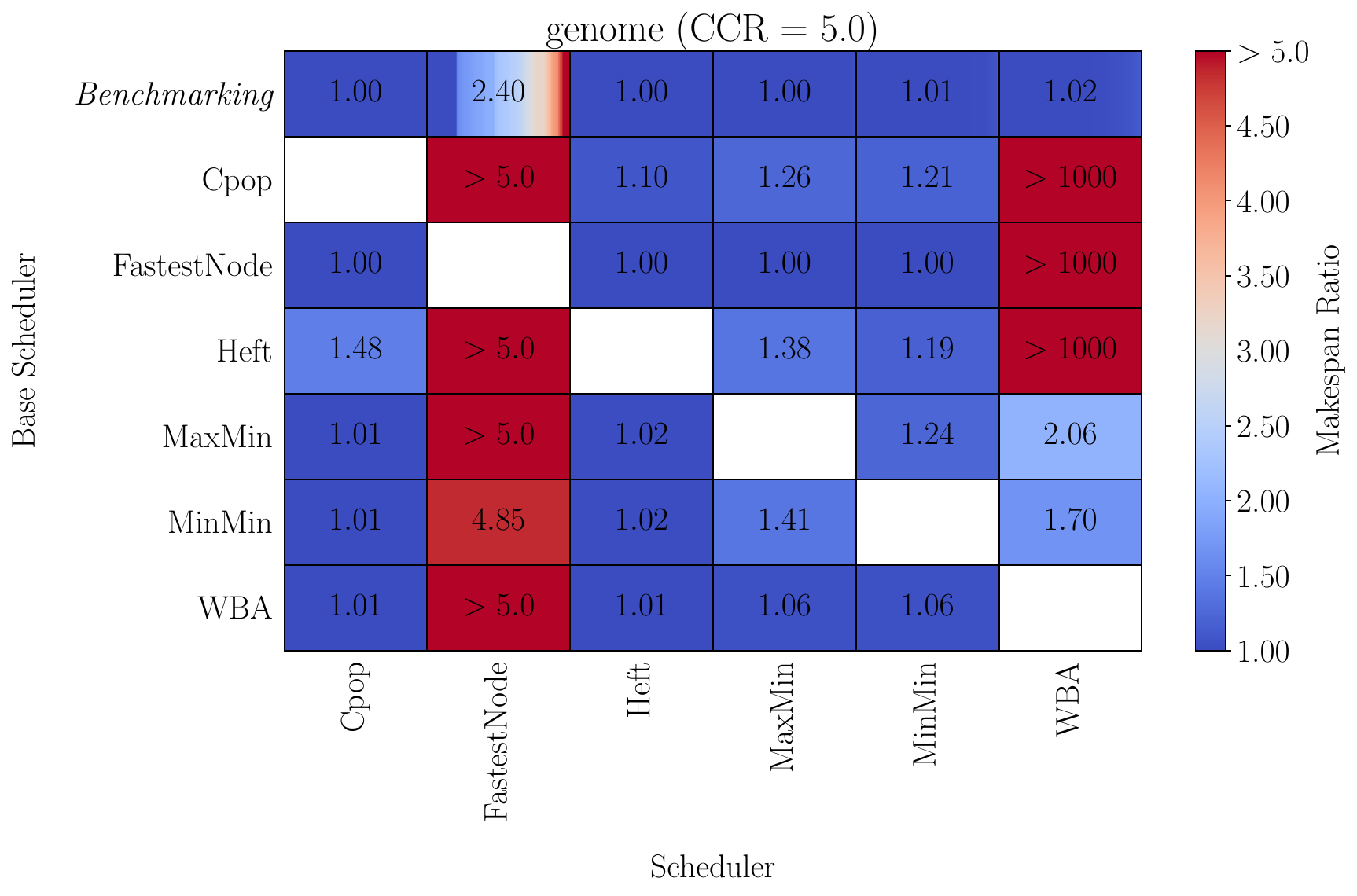}
    \end{subfigure}%
    \caption{Results for the 1000genome scientific workflow.}
    \label{fig:apx-app:genome}
\end{figure*}

\begin{figure*}[!htb]
    \centering
    
    \begin{subfigure}[b]{0.5\textwidth}
        \centering
        \includegraphics[width=\textwidth]{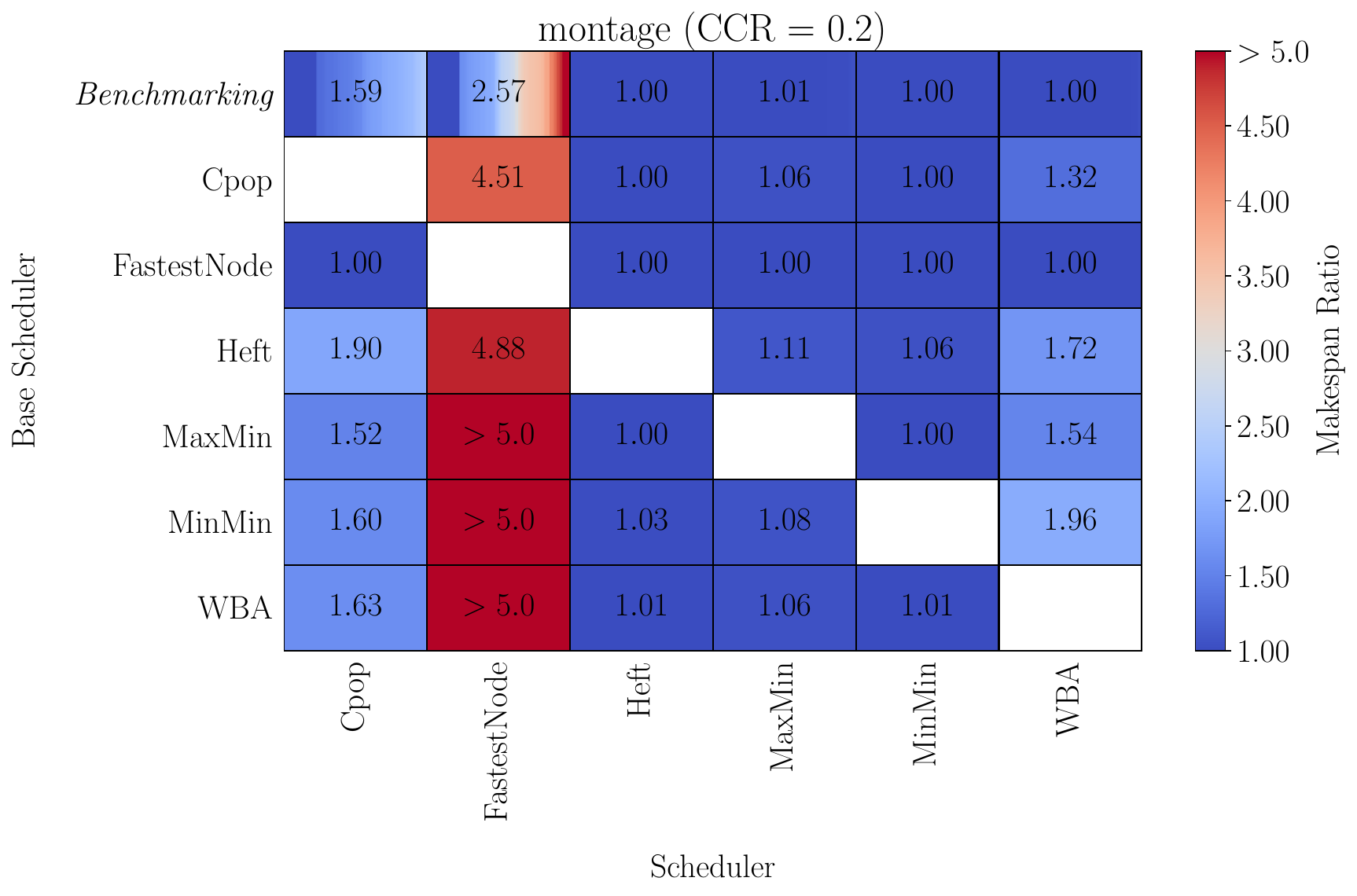}
    \end{subfigure}%
    \hfill
    \begin{subfigure}[b]{0.5\textwidth}
        \centering
        \includegraphics[width=\textwidth]{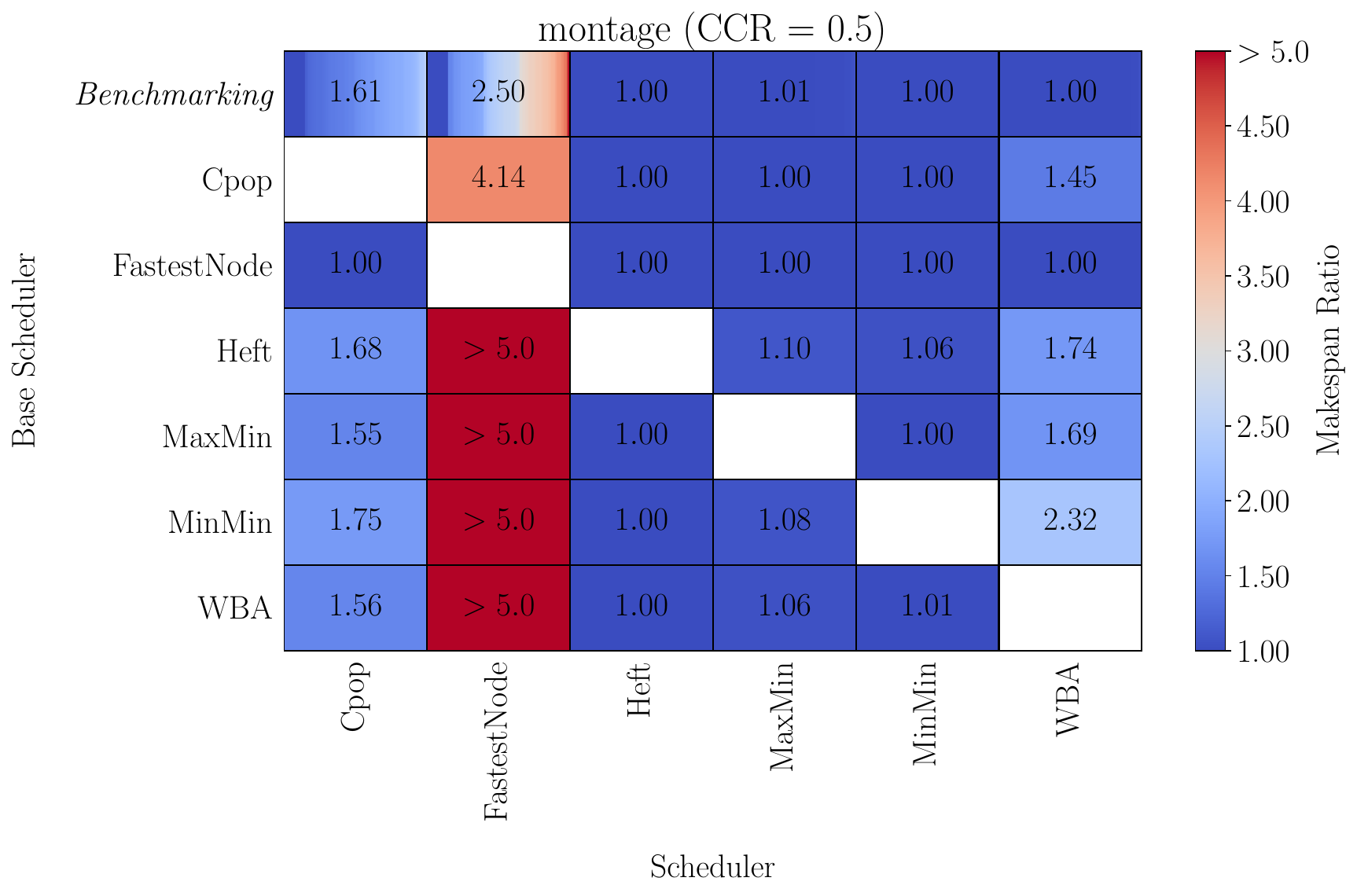}
    \end{subfigure}%
    
    \vspace{0.25cm}%
    
    \begin{subfigure}[b]{0.5\textwidth}
        \centering
        \includegraphics[width=\textwidth]{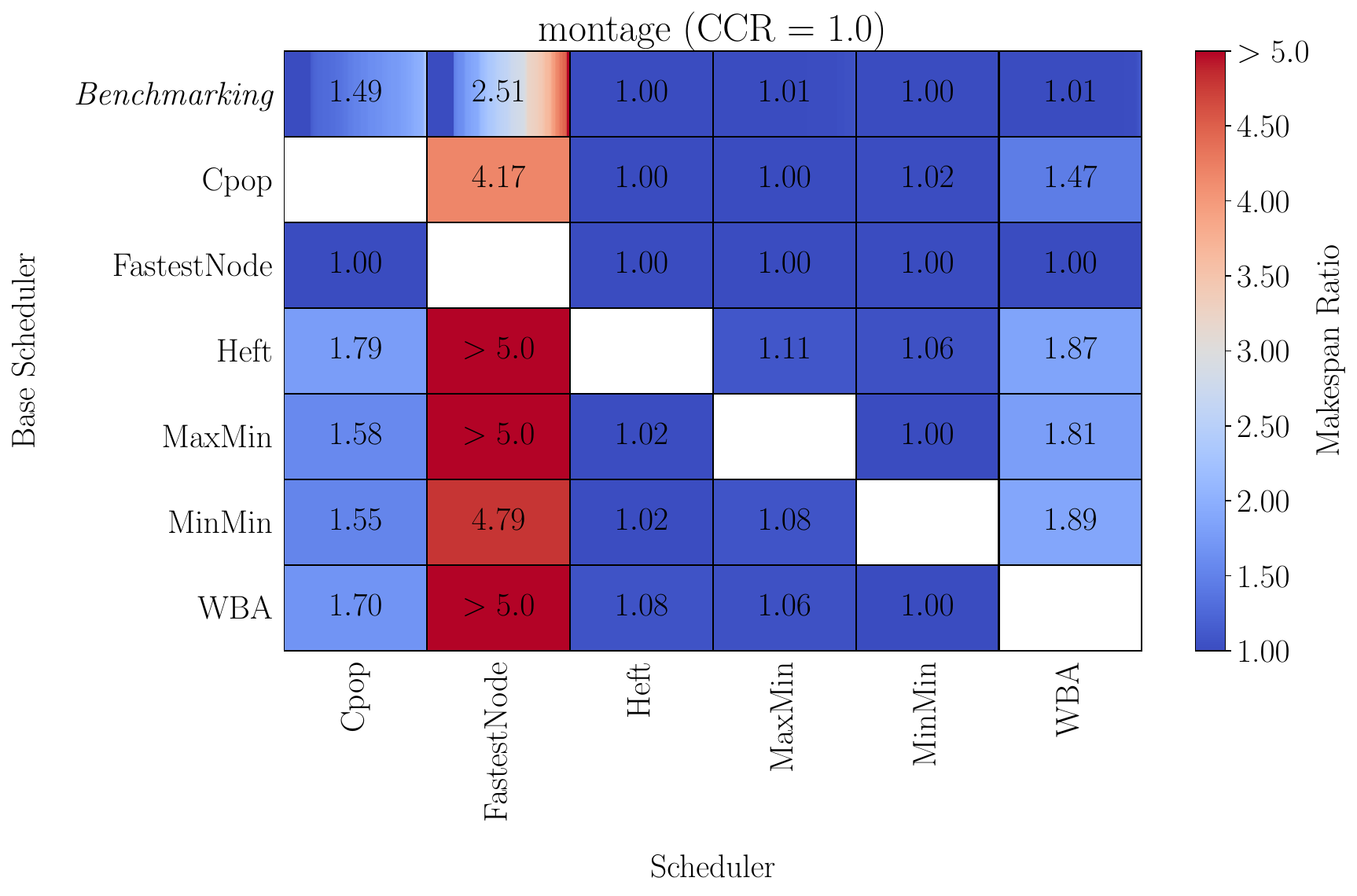}
    \end{subfigure}%
    \hfill
    \begin{subfigure}[b]{0.5\textwidth}
        \centering
        \includegraphics[width=\textwidth]{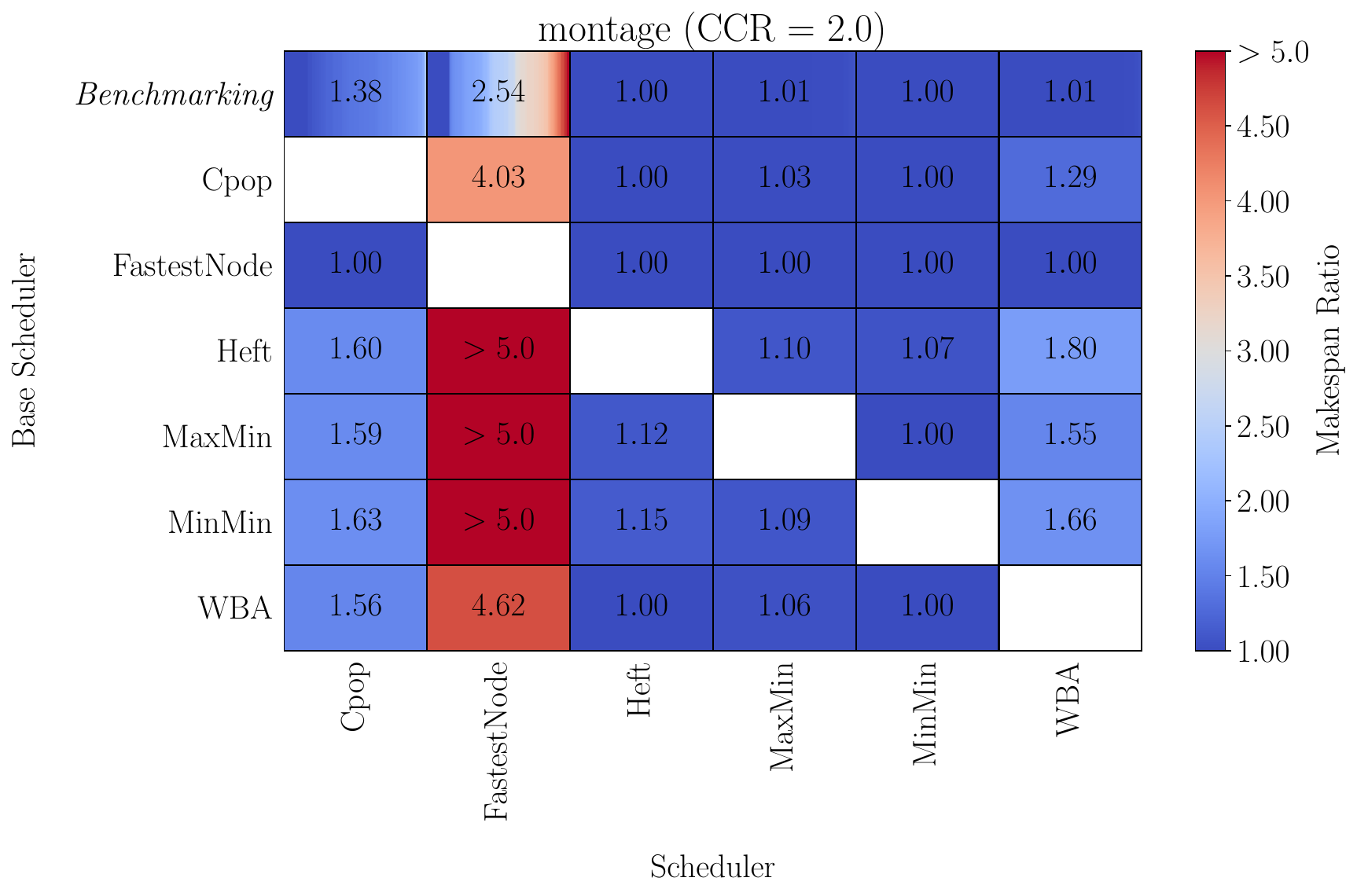}
    \end{subfigure}%
    
    \vspace{0.25cm}%
    
    \begin{subfigure}[b]{0.5\textwidth}
        \centering
        \includegraphics[width=\textwidth]{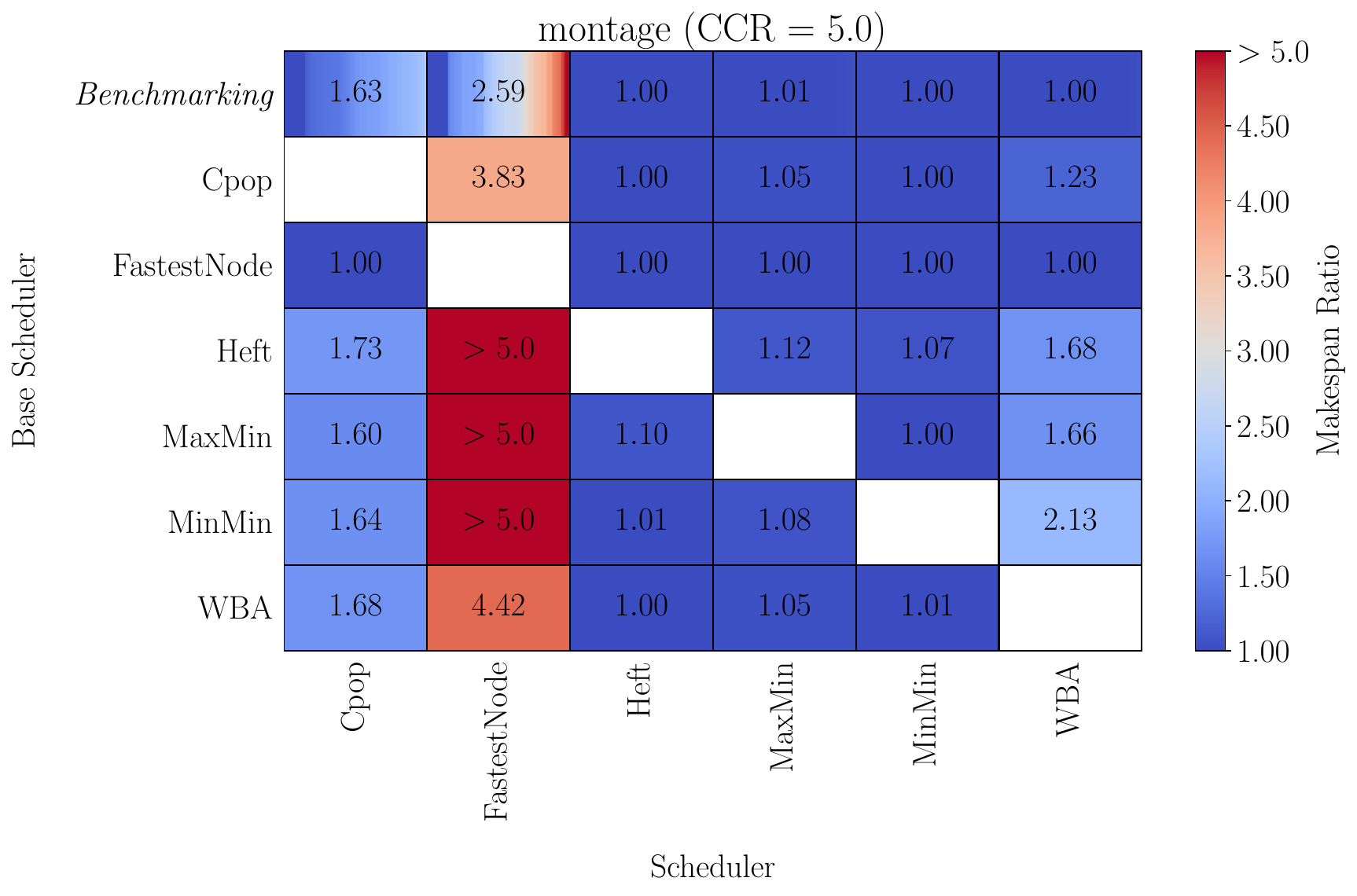}
    \end{subfigure}%
    \caption{Results for the montage scientific workflow.}
    \label{fig:apx-app:montage}
\end{figure*}

\begin{figure*}[!htb]
    \centering
    
    \begin{subfigure}[b]{0.5\textwidth}
        \centering
        \includegraphics[width=\textwidth]{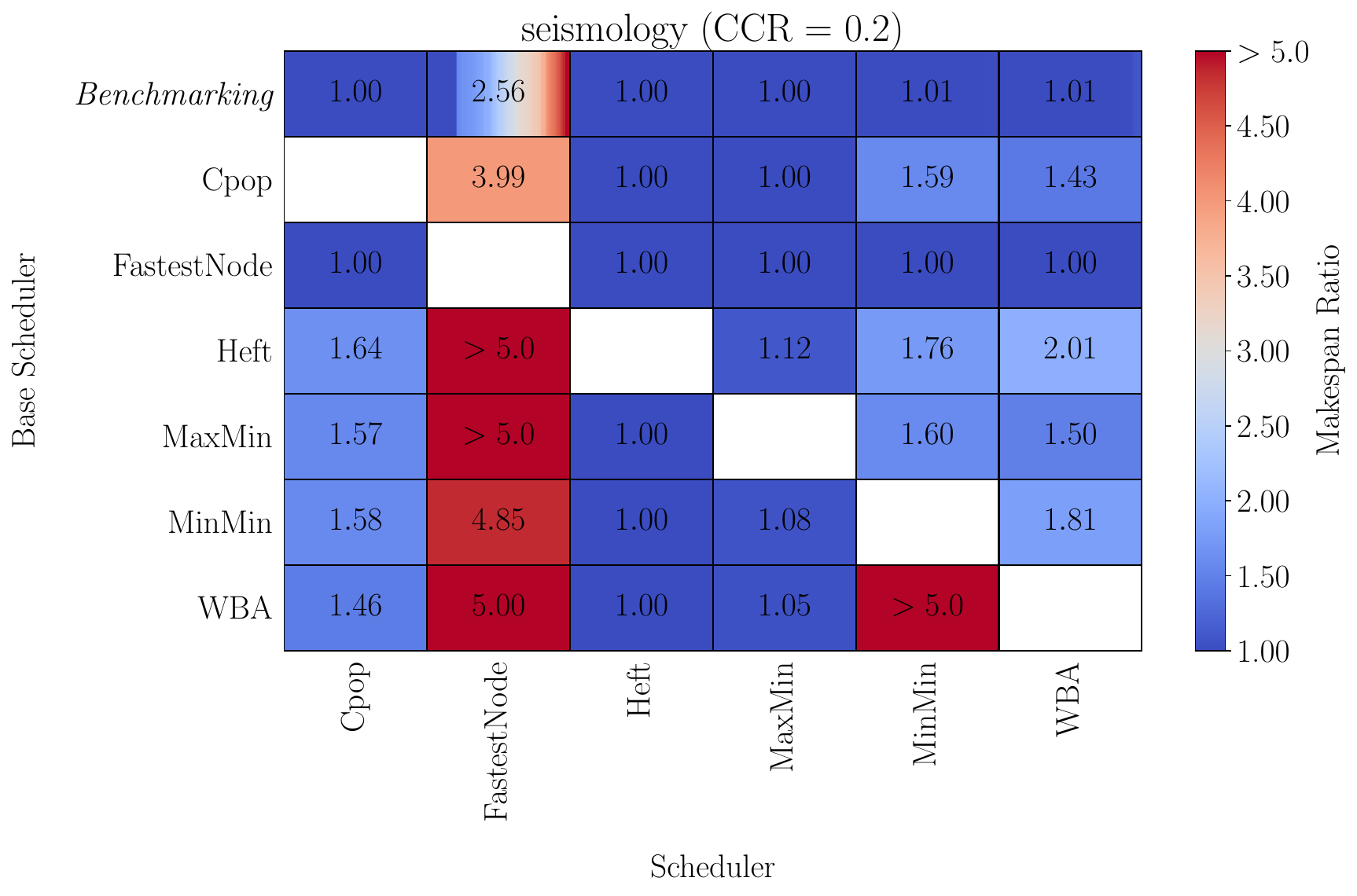}
    \end{subfigure}%
    \hfill
    \begin{subfigure}[b]{0.5\textwidth}
        \centering
        \includegraphics[width=\textwidth]{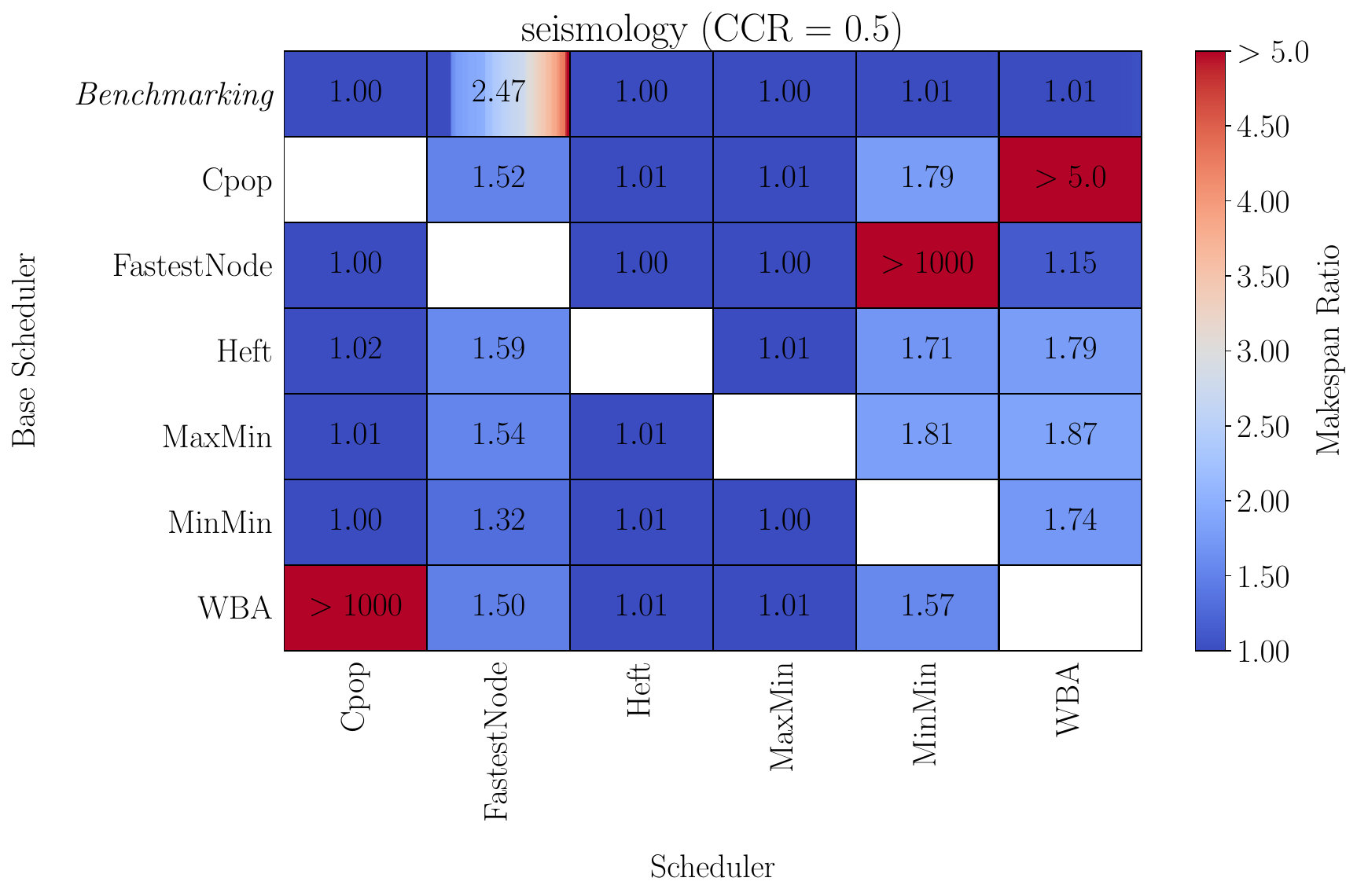}
    \end{subfigure}%
    
    \vspace{0.25cm}%
    
    \begin{subfigure}[b]{0.5\textwidth}
        \centering
        \includegraphics[width=\textwidth]{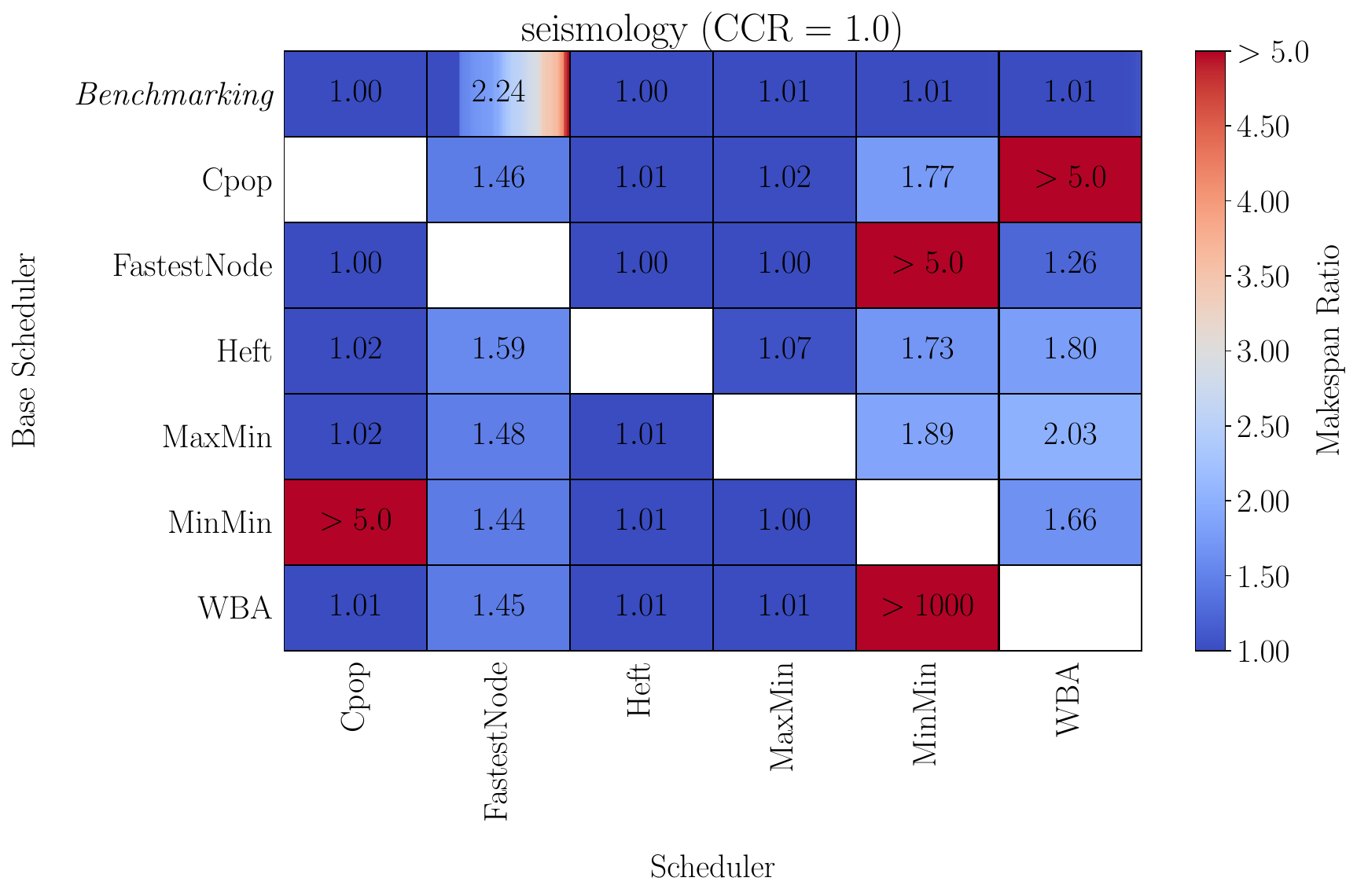}
    \end{subfigure}%
    \hfill
    \begin{subfigure}[b]{0.5\textwidth}
        \centering
        \includegraphics[width=\textwidth]{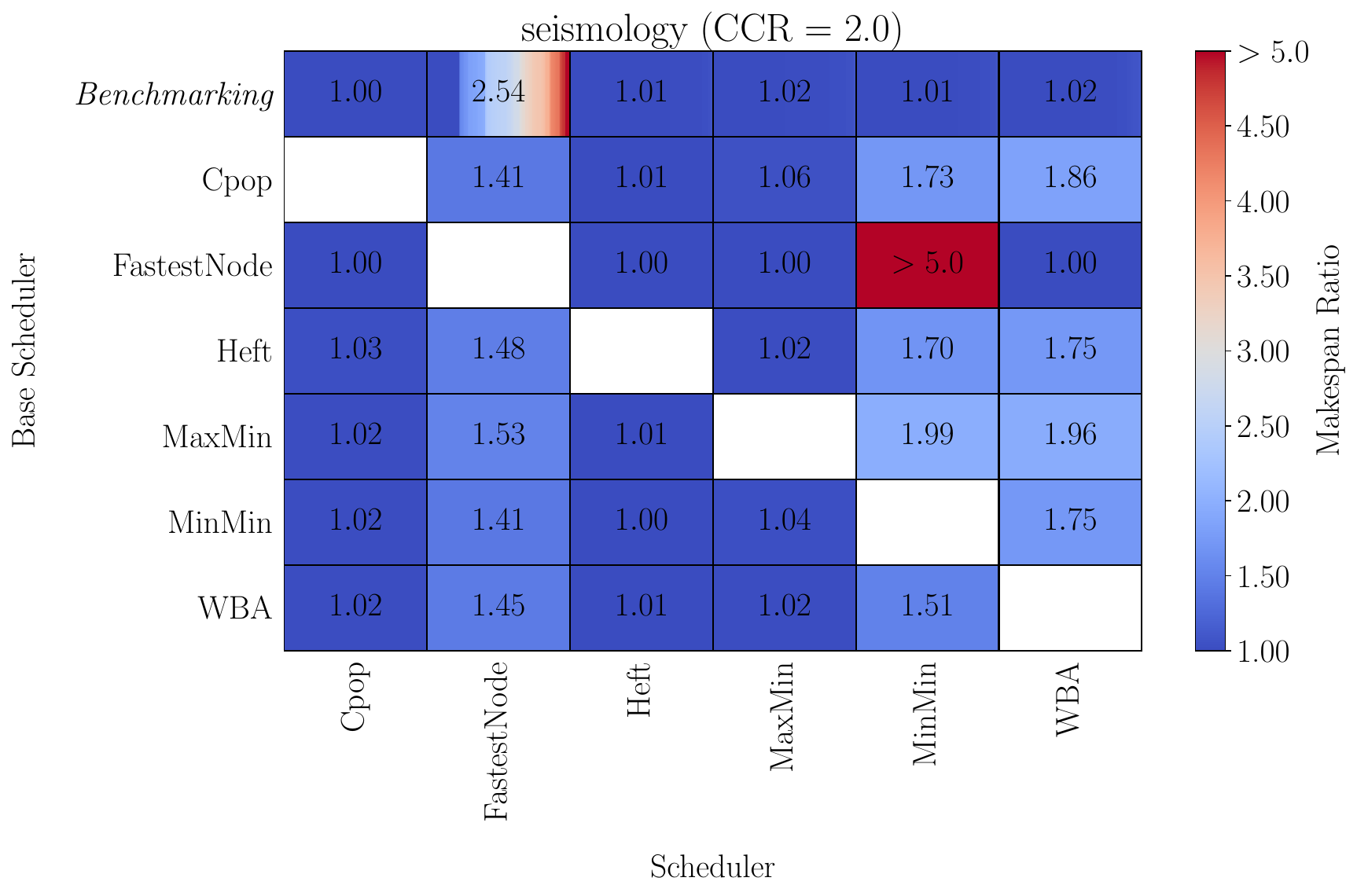}
    \end{subfigure}%
    
    \vspace{0.25cm}%
    
    \begin{subfigure}[b]{0.5\textwidth}
        \centering
        \includegraphics[width=\textwidth]{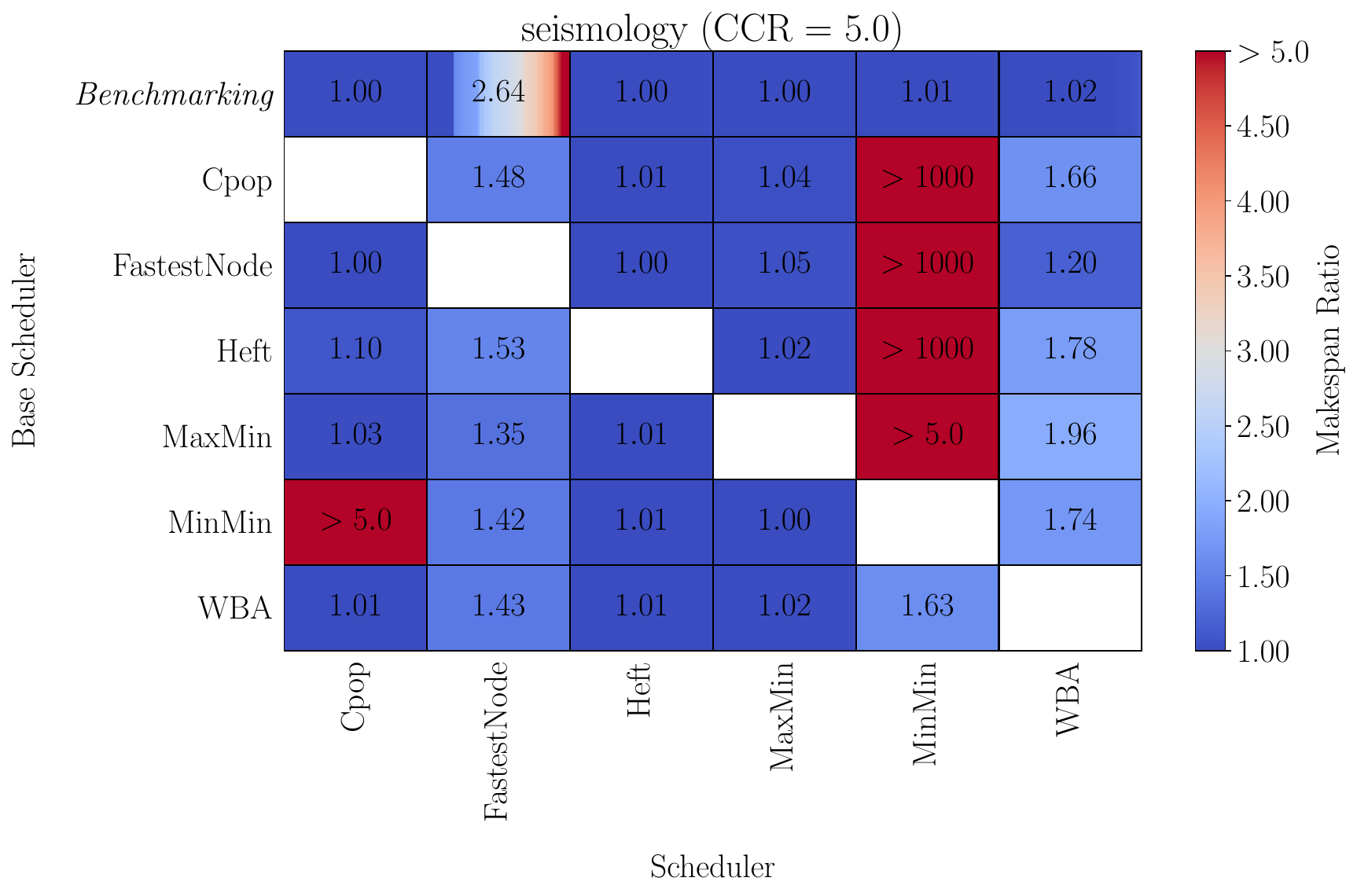}
    \end{subfigure}%
    \caption{Results for the seismology scientific workflow.}
    \label{fig:apx-app:seismology}
\end{figure*}

\begin{figure*}[!htb]
    \centering
    
    \begin{subfigure}[b]{0.5\textwidth}
        \centering
        \includegraphics[width=\textwidth]{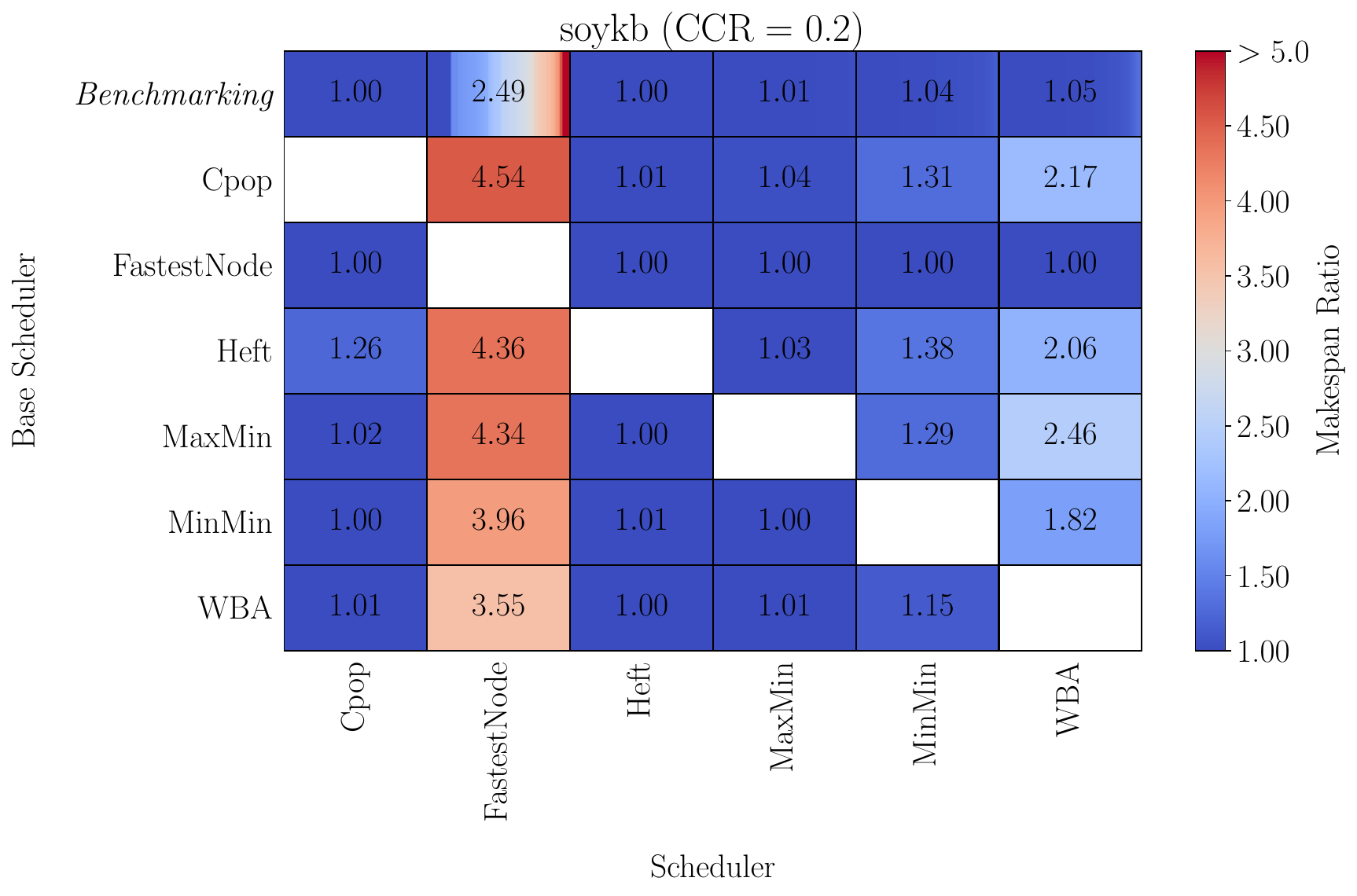}
    \end{subfigure}%
    \hfill
    \begin{subfigure}[b]{0.5\textwidth}
        \centering
        \includegraphics[width=\textwidth]{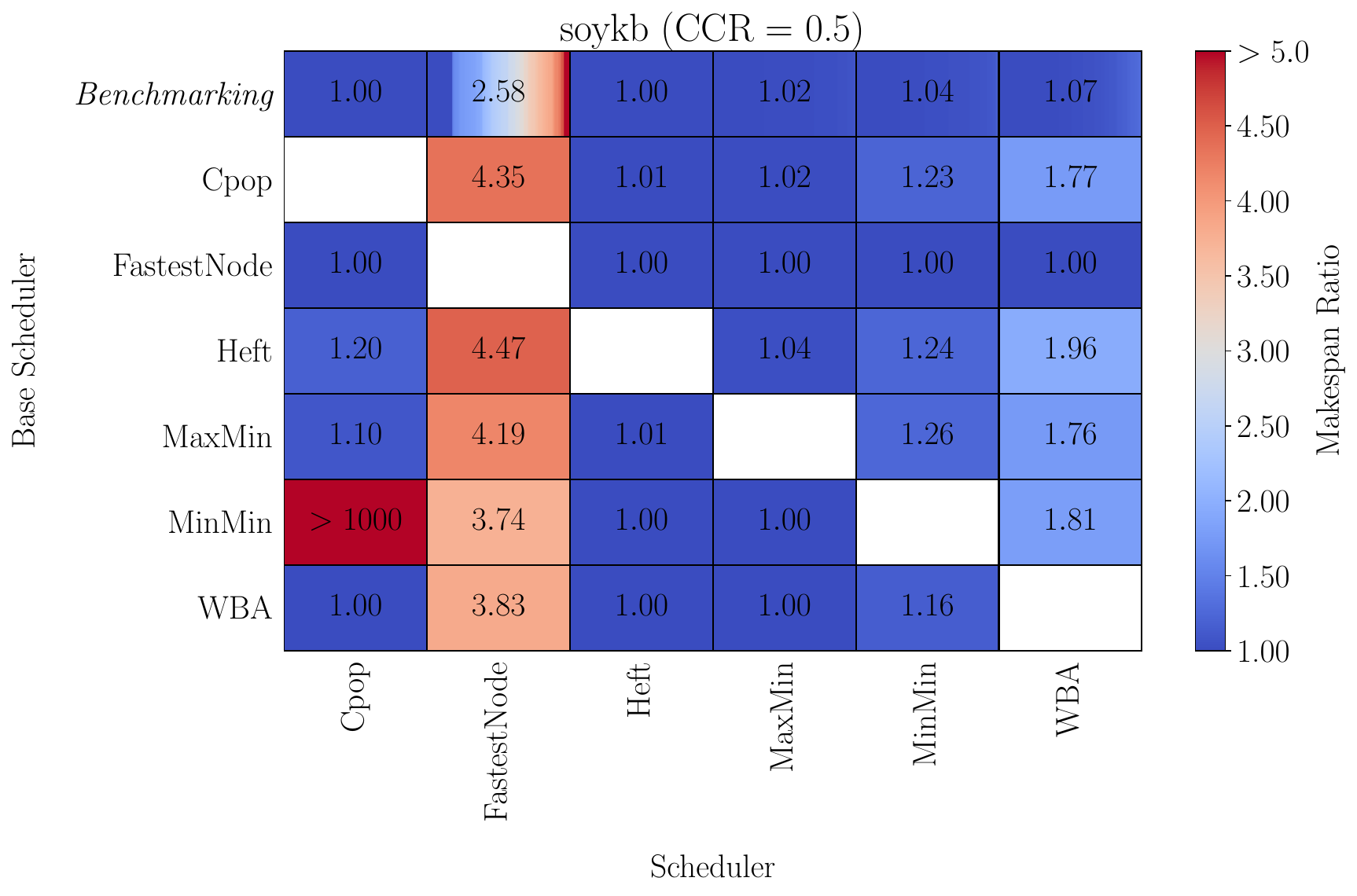}
    \end{subfigure}%
    
    \vspace{0.25cm}%
    
    \begin{subfigure}[b]{0.5\textwidth}
        \centering
        \includegraphics[width=\textwidth]{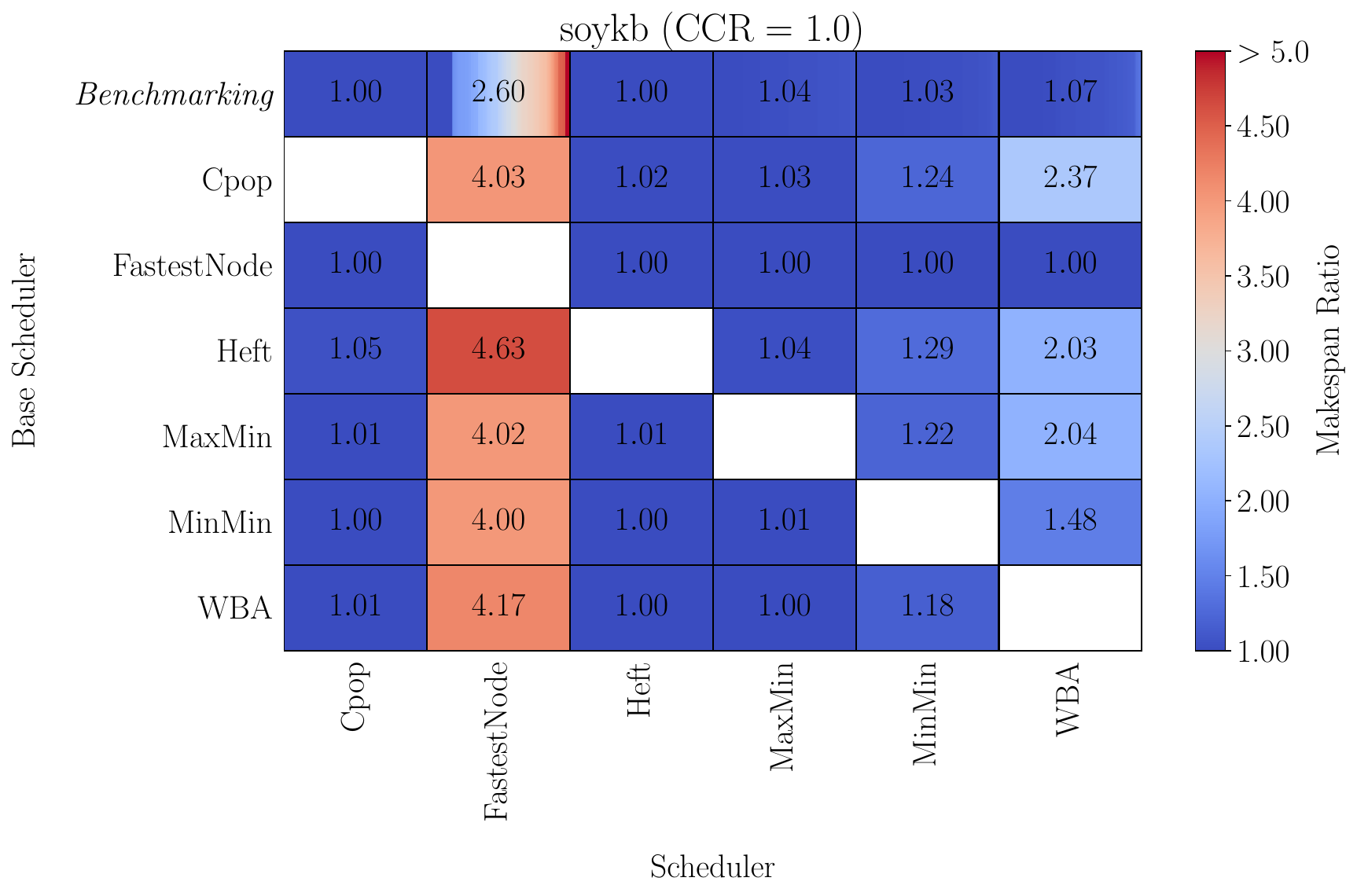}
    \end{subfigure}%
    \hfill
    \begin{subfigure}[b]{0.5\textwidth}
        \centering
        \includegraphics[width=\textwidth]{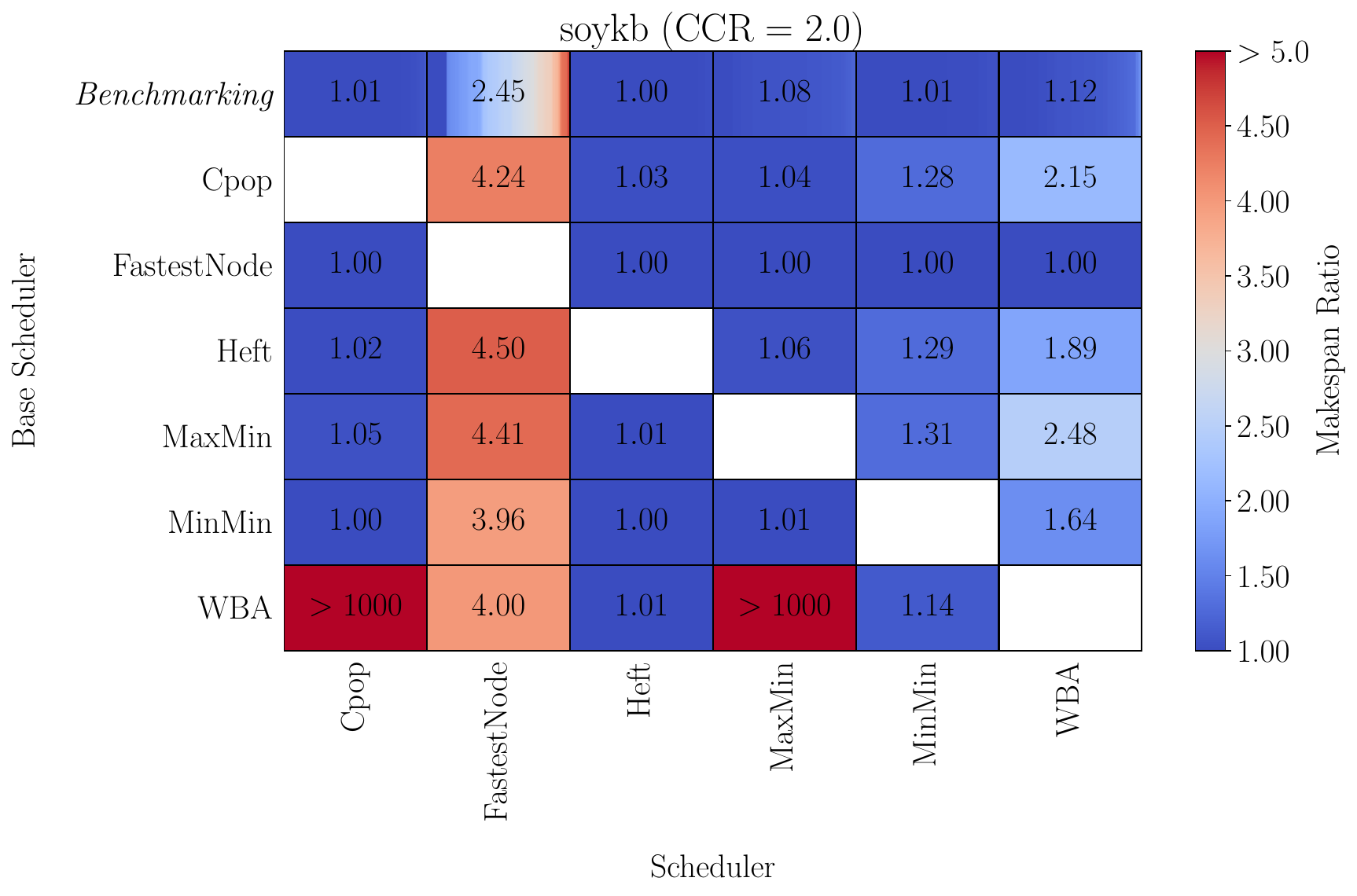}
    \end{subfigure}%
    
    \vspace{0.25cm}%
    
    \begin{subfigure}[b]{0.5\textwidth}
        \centering
        \includegraphics[width=\textwidth]{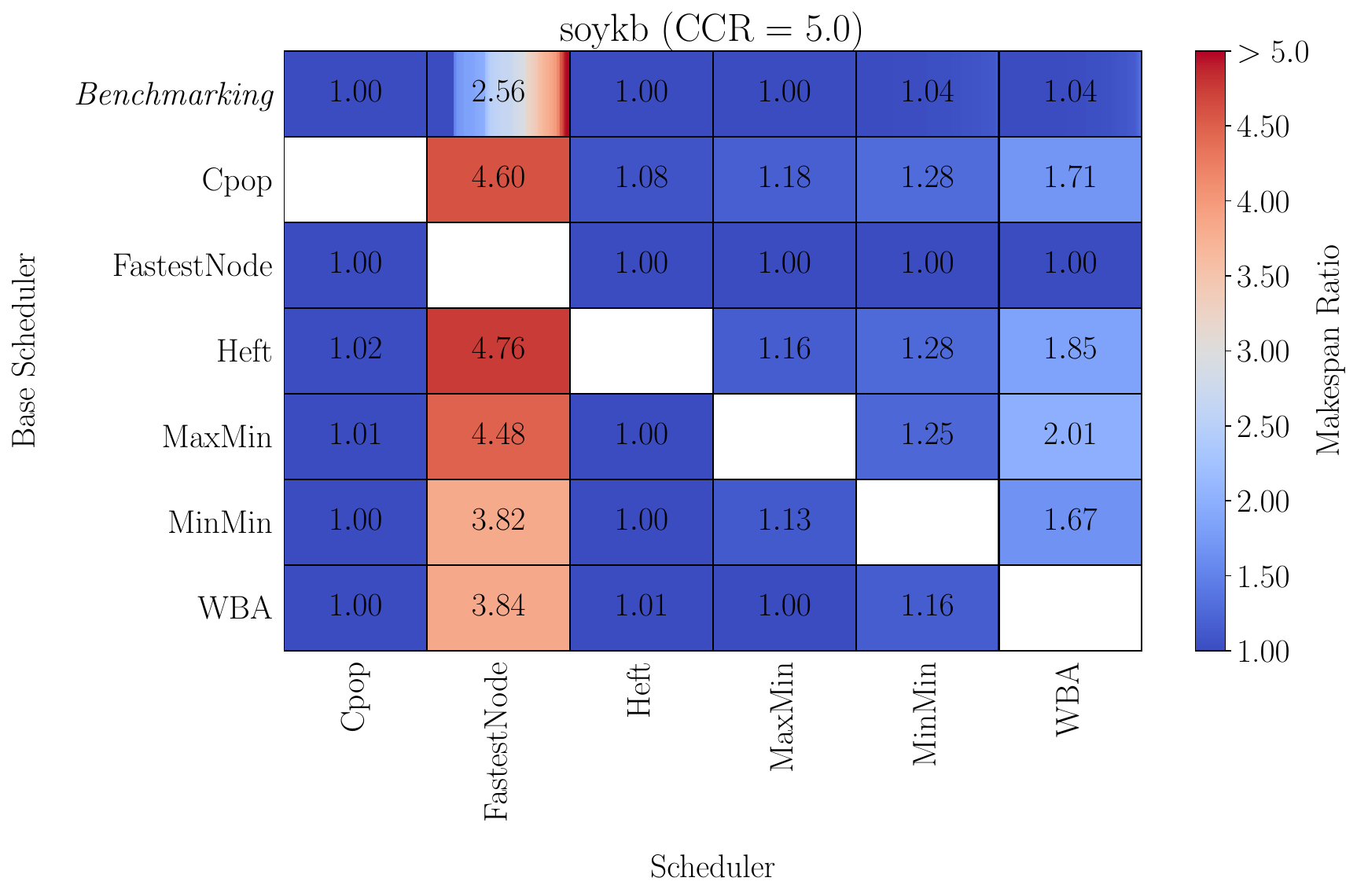}
    \end{subfigure}%
    \caption{Results for the soykb scientific workflow.}
    \label{fig:apx-app:soykb}
\end{figure*}

\end{document}